\documentclass[journal=mamobx,super=true,manuscript=note]{achemso}
\usepackage[version=3]{mhchem} 
\usepackage{amsmath,multirow,amssymb,amsbsy,relsize,comment,color,mathtools,mathrsfs,hyperref}
\usepackage[normalem]{ulem}

\setkeys{acs}{articletitle=true}

\def\D{\mathrm{d}}

\newcommand{\vectornorm}[1]{\left\|#1\right\|}

\newcommand{\pderv}[2]{\frac{\partial#1}{\partial#2}}
\newcommand{\parnths}[1]{\left(#1\right)}
\newcommand{\bracks}[1]{\left[#1\right]}
\newcommand{\cbraces}[1]{\left\{ #1 \right \}}
\newcommand{\angles}[1]{\left \langle #1 \right \rangle}
\newcommand{\tderv}[2]{\left. #1\right)_{#2}}
\def\albe{{\alpha \beta}}

\title{Equation of State-based Slip-spring Model for Entangled Polymer Dynamics}
\author{Georgios G. Vogiatzis}
\author{Grigorios Megariotis}
\author{Doros N. Theodorou}
\affiliation[NTUA]{School of Chemical Engineering, National Technical University of Athens, 
9 Heroon Polytechniou Street, Zografou Campus, GR-15780 Athens, Greece}

\makeatletter
\let\thetitle\@title
\let\theauthor\@author
\makeatother

\makeatletter
\renewcommand\section{\@startsection{section}{1}{\z@}%
                                  {-3.5ex \@plus -1ex \@minus -.2ex}%
                                  {2.3ex \@plus.2ex}%
                                  {\normalfont\small\bfseries}}                                  
\makeatother

\makeatletter
\renewcommand\subsection{\@startsection{subsection}{1}{\z@}%
                                  {-3.5ex \@plus -1ex \@minus -.2ex}%
                                  {2.3ex \@plus.2ex}%
                                  {\normalfont\small\small\bfseries}}
\makeatother

\begin{document}

\begin{abstract}  

\begin{figure} 
\begin{center}
  \includegraphics[clip,width=0.7\linewidth] {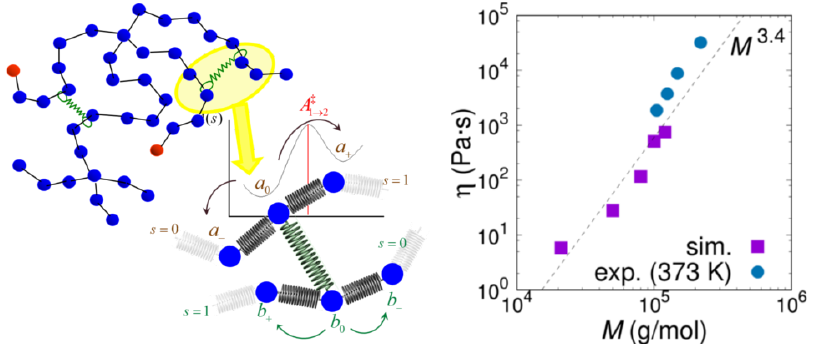}
\end{center}
\end{figure}

A mesoscopic, mixed particle- and field-based Brownian Dynamics methodology for the simulation of entangled polymer 
melts has been developed. Polymeric beads consist of several Kuhn segments and their motion is dictated by the 
Helmholtz energy of the sample, which is a sum of the entropic elasticity of chain strands between beads; slip-springs; 
and non-bonded interactions.
Following earlier works in the field (\textit{Phys. Rev. Lett.} \textbf{2012}, \textit{109}, 148302)
the entanglement effect is introduced by the slip-springs, which are springs connecting either non successive beads on 
the same chain, or beads on different 
polymer chains. The terminal positions of slip-springs are altered during the simulation through a kinetic Monte Carlo 
hopping scheme, with rate-controlled creation/destruction processes for the slip-springs at chain ends.
The rate constants are consistent with the free energy function 
employed and satisfy microscopic reversibility at equilibrium. 
The free energy of nonbonded interactions is derived from an appropriate equation of state 
and it is computed as a functional of the local density by passing an orthogonal grid through the simulation box; 
accounting for it is necessary for reproducing the correct compressibility of the polymeric material. Parameters 
invoked by the mesoscopic model are derived from experimental volumetric and viscosity data or from atomistic 
Molecular Dynamics simulations, establishing a ``bottom-up'' predictive framework for conducting slip-spring simulations of
polymeric systems of specific chemistry.
Initial configurations for the mesoscopic simulations are obtained by further coarse-graining 
of well-equilibrated structures represented at a greater level of detail. The mesoscopic simulation methodology is 
implemented for the case of cis-1,4 polyisoprene, whose structure, dynamics, thermodynamics and 
linear rheology in the melt state are quantitatively predicted  
and validated without a posteriori fitting the results to experimental measurements.
\end{abstract}

\section{Introduction}

One of the fundamental concepts in the molecular description of structure - property relations of polymer melts is
chain entanglement. When macromolecules interpentrate, the term entanglements intends to describe the
topological interactions resulting from the uncrossability of chains.
The fact that two polymer chains cannot go though each other in the course of their motion changes their 
dynamical behavior dramatically, without altering their equilibrium properties. 
Anogiannakis et al.\cite{Macromolecules_45_9475} have examined microscopically at what level topological constraints can
be described as a collective entanglement effect, as in tube model theories, or as certain pairwise uncrossability 
interactions, as in slip-link models. They employed a novel methodology, which analyzes entanglement constraints into a
complete set of pairwise interactions (links), characterized by a spectrum of confinement strengths. As a measure of the 
entanglement strength, these authors used the fraction of time for which the links are active. The confinement was 
found to be mainly imposed by the strongest links. The weak, trapped, uncrossability interactions cannot contribute to
the low frequency modulus of an elastomer, or the plateau modulus of a melt. 

In tube model theories,\cite{ProcPhysSoc_92_9,DeGennes_ScalingConcepts,Doi_Edwards_TheoryOfPolymerDynamics,RepProgPhys_51_243} 
it is postulated that the entanglements generate a confining mean field potential, which restricts the lateral monomer motion
to a tube-like region surrounding each chain. In polymer melts the confinement is not permanent, but leads to a one-
dimensional diffusion of the chain along its tube, called reptation.\cite{JChemPhys_55_572}
An alternative, discrete, localized version of the tube constraint is utilized in models employing slip-links.
\cite{JChemSocFaradayTrans2_74_1802,JChemPhys_66_3363,Polymer_22_1010,Polymer_27_483,
JChemPhys_109_10018,JChemPhys_109_10028,Macromolecules_35_6670}
The tube is replaced by a set of slip-links along the chain, which restrict lateral motion but permit chain sliding 
through them. The real chain is represented by its primitive path, which is a series of strands of average molar 
mass $M_{\rm e}$ connecting the links.

Hua and Schieber\cite{JChemPhys_109_10018,JChemPhys_109_10028} considered that the molecular details on the monomer or 
Kuhn-length level are smeared out in the slip-link model, while the segmental network of generic polymers is 
directly modeled, by introducing links between chains, which constrain the motion of segments of each chain into a 
tubular region. The motion of segments is updated stochastically, and the positions of slip-links can either be
fixed in space, or mobile. When either of the constrained segments slithers out of a slip-link constraint, they are 
considered to be disentangled, and the slip-link is destroyed. Conversely, the end of one segment can hop towards 
another segment and create another new entanglement or slip-link. 
The governing equations in the slip-link model can be split into two parts: the chain motion inside its tube is 
governed by Langevin equations and the tube motion is governed by deterministic convection and stochastic constraint 
release processes. Based on the tube model,\cite{Doi_Edwards_TheoryOfPolymerDynamics} it is assumed that the motion of 
the primitive path makes the primary contribution to the rheological properties of entangled polymer melts. Therefore, from
the microscopic information given by the slip-link model, these authors could precisely access the longest polymer
chain relaxation time. Moreover, by employing an elegant formulation, the macroscopic properties of polymer melts, 
i.e., stress and dielectric relaxation can be extracted from the ordering, spatial location and aging of the 
entanglements or slip-links in the simulations.

Later, Schieber and co-workers studied the fluctuation effect on the chain entanglement and viscosity using a 
mean-field model.\cite{RheolActa_51_1021}
Shanbhag et al.\cite{PhysRevLett_87_195502} developed a dual slip-link model with chain-end fluctuations for entangled 
star polymers, which explained the observed deviations from the ``dynamic dilution'' equation in dielectric and 
stress relaxation data. Doi and Takimoto\cite{PhilosTranRSocA_361_641} adopted the dual slip-link model to 
study the nonlinear rheology of linear and star polymers with arbitrary molecular weight distribution. 

Likhtman\cite{Macromolecules_38_6128} has shown that the standard tube model cannot be applied to neutron
spin-echo measurements because the statistics of a one-dimensional chain in a three-dimensional random-walk tube become 
wrong on the length scale of the tube diameter. He then introduced a new single-chain dynamic slip-link model to 
describe the experimental results for neutron spin echo, linear rheology and diffusion of monodisperse polymer melts. 
All the parameters in this model were obtained from one type of
experiment and were applied to predict other experimental results. 
The model was formulated in terms of stochastic differential equations, suitable for Brownian Dynamics (BD) simulations. 
The results were characterized by some systematic discrepancies, suggesting possible inadequacy of the Gaussian chain 
model for some of the polymers considered, and possible inadequacy of the time - temperature superposition 
(TTS) principle.

Del Biondo et al.\cite{JChemPhys_138_194902} extended the slip link model of Likhtman\cite{Macromolecules_38_6128} 
in order to study an inhomogeneous system. These authors studied the dependence of the relaxation modulus on the 
slip link density and stiffness, and explored the nonlinear rheological properties of the model 
for a set of its parameters. The crucial part of their work was the introduction of excluded volume interactions in 
a mean field manner in order to describe inhomogeneous systems, and application of this description to a simple 
nanocomposite model. They concentrated on an idealized situation where the fillers were well dispersed, with a simple 
hardcore interaction between the fillers and the polymer matrix. However, addressing real situations where the fillers 
are poorly dispersed and partially aggregated was clearly possible within their framework.

Masubuchi et al.\cite{JChemPhys_115_4387} performed several multichain simulations for entangled polymer melts by 
utilizing slip links to model the entanglements. These authors proposed a primitive chain network model, in which the 
polymer chain is coarse-grained into a set of segments (strands) going through entanglement points. 
Different segments are coupled together 
through the force balance at the entanglement node. The Langevin equation is applied to update the positions of 
these entanglement nodes, by incorporating the tension force from chain segments and an osmotic force caused by 
density fluctuations. The entanglement points, modeled as slip-links, can also fluctuate spatially (or three dimensionally). 
The creation and annihilation of entanglements are controlled by the number of monomers at chain ends. 
The longest relaxation time and the self-diffusion coefficient scaling, as predicted from the model, were
found in good agreement with experimental results.
Later on, the primitive chain network model was extended to study the relationship between entanglement length and 
plateau modulus.\cite{JChemPhys_119_6925,ModellSimulMaterSciEng_12_S91,JNonNewtonianFluidMech_149_87,JChemPhys_121_12650}
It was
also extended to study star and branched polymers,\cite{RheolActa_46_297} nonlinear rheology,\cite{JChemPhys_128_154901,
Macromolecules_44_9675,Macromolecules_45_2773} phase separation in polymer blends,\cite{Macromolecules_41_8275}
block copolymers,\cite{JNonCrystSolids_352_5001} and the dynamics of confined polymers.\cite{JChemPhys_130_214907}
Recently, Masubuchi has compiled an excellent review of simulations of entangled polymers.\cite{AnnuRevChemBiomolEng_5_11}

Coarse graining from an atomistic Molecular Dynamics (MD) simulation to a mesoscopic one results in soft repulsive 
potentials between the coarse-grained blobs, causing unphysical bond crossings influencing the dynamics of the system. 
The Twentanglement uncrossability scheme proposed by Padding and Briels,\cite{JChemPhys_115_2846, JChemPhys_117_925} 
was introduced in order to prevent these bond crossings. 
Their method is based on considering the bonds as elastic bands between the bonded 
blobs. As soon as two of these elastic bands make contact, an ``entanglement'' is created which prevents the elastic 
bands from crossing. In their method, entanglements were defined as the objects which prevent the crossing of chains. 
Only a few of them were expected to be entanglements in the usual sense of long-lasting topological constraints, 
slowing down the chain movement. Simulations of a polyethylene melt\cite{JChemPhys_115_2846} showed that the rheology 
can be described correctly, deducing the blob-blob interactions and the friction parameters from MD simulations.

Chappa et al.\cite{PhysRevLett_109_148302} proposed a model in which the topological effect of 
noncrossability of long flexible macromolecules is effectively taken into account by 
slip-springs, which are
local, pairwise, translationally and rotatationally invariant interactions between polymer beads that do not affect 
the equilibrium properties of the melt.
The slip-springs introduce an effective attractive potential between the beads; in order to eliminate this effect the 
authors introduced an analytically derived repulsive compensating potential.
The conformations of polymers and slip springs are updated by a hybrid Monte Carlo (MC) scheme. 
At every step, either the positions of the beads are evolved via a short Dissipative Particle Dynamics (DPD) run or the 
configuration of the slip-springs is modified by MC moves,
involving discrete jumps of the slip-springs along the chain contour and creation or deletion at the chain ends.
The number of slip-springs can vary during the simulation, obeying a prescribed chemical potential.
That model can correctly describe many aspects of the 
dynamical and rheological behavior of entangled polymer liquids in a computationally efficient manner, since 
everything is cast into only pairwise-additive interactions between beads.
The mean-square displacement of the beads evolving according to this model was found to be in favorable agreement with 
the tube model predictions. 
Moreover, the model exhibited realistic shear thinning, deformation of
conformations, and a decrease of the number of entanglements at high shear rates.

M\"uller and Daoulas were the first to introduce the hopping of slip-springs by means of discrete MC moves.\cite{JChemPhys_129_164906}
These authors thoroughly investigated the ability of MC algorithms to describe the single-chain dynamics in 
a dense homogeneous melt and a lamellar phase of a symmetric diblock copolymer. 
Three different types of single-chain dynamics were considered: local, unconstrained dynamics, slithering-snake dynamics,
and slip-link dynamics.
Ram\'irez-Hern\'andez et al.\cite{SoftMatter_9_2030,JChemPhys_146_014903} replaced the discrete hopping of a slip-spring along the chain 
contour by a one-dimensional continuous Langevin equation.
In their method, slip-springs consist of two rings that slip along different chain contours, and are connected by a harmonic spring.
The rings move in straight lines between beads belonging to different chains, scanning the whole contour of the chains 
in a continuous way.
Comparison with experimental results was made possible by rescaling the frequency and the modulus, in order to match 
the intersection point of storage and loss moduli curves of polystyrene melts. 
More recently, Ram\'irez-Hern\'andez et al.\cite{Macromolecules_46_6287} presented a more general formalism in order to 
qualitatively capture the linear rheology of pure homopolymers and their blends, as well as the nonlinear rheology 
of highly entangled polymers and the dynamics of diblock copolymers. The number of slip-springs in their approach 
remained constant throughout the simulation, albeit their connectivity changes. These authors have later presented a 
theoretically informed entangled polymer simulation approach wherein the total number of slip-springs is not 
preserved but, instead, it is controlled through a chemical potential that determines the average molecular 
weight between entanglements.\cite{JChemPhys_143_243147} 

The idea of slip-springs was in parallel used by Uneyama and Masubuchi,\cite{JChemPhys_137_154902} who proposed a
multi-chain slip-spring model inspired by the single chain slip-spring model of Likhtman.\cite{Macromolecules_38_6128}
Differently from the primitive chain network model of the same authors, they defined the total free energy for the
new model, and employed a time evolution equation and stochastic processes for describing its dynamical evolution.
All dynamic ingredients satisfy the detailed balance condition, and are thus capable of reproducing the thermal equilibrium 
which is characterized by the free energy. The number of slip-springs varies. Later, Langeloth 
et al.\cite{JChemPhys_183_104907,Macromolecules_49_9186} presented a simplified version of the slip-spring model of Uneyama and 
Masubuchi,\cite{JChemPhys_137_154902} where the number of slip-springs remains constant throughout the simulation.

Working on a different problem than melt rheology, Terzis et al.\cite{Macromolecules_33_1385,Macromolecules_33_1397,
Macromolecules_35_508} have invoked the microscopic description of entanglements and the associated processes envisioned
in slip-link models, in order to generate entanglement network specimens of interfacial polymeric systems and study their
deformation to fracture. The specimens were created by sampling the configurational distribution functions derived 
from a Self Consistent Field (SCF) lattice model. The specimens generated were not in detailed mechanical equilibrium. 
To this end, these authors developed a method for relaxing the network with respect to its density distribution, 
and thereby imposing the condition of mechanical equilibrium, without changing the network topology.\cite{Macromolecules_35_508}
The free energy function of the network was minimized with respect to the coordinates of all entanglement points and 
chain ends. Contributions to the free energy included (a) the elastic energy due to stretching of the chain strands 
and (b) the free energy due to the repulsive and attractive (cohesive) interactions between segments. The latter was
calculated by superimposing a simple cubic grid on the network and taking into account contributions between cells and 
within each cell.
A density functional was used for non-bonded interactions depending on a segment density field derived from 
coarse-grained particle positions tracked in a simulation.

Laradji et al.\cite{PhysRevE_49_3199} originally introduced the concept of conducting particle-based simulations with interactions 
described via collective variables and also two ways of calculating these collective variables: grid-based and 
continuum weighting-function-based. Along the same lines, Soga et al.\cite{EurophysLett_29_531} investigated the 
structure of an end-grafted polymer layer immersed in poor solvent through MC simulations based on Edward's Hamiltonian,
incorporating a third-order density functional. Pagonabarraga and Frenkel\cite{JChemPhys_115_5015} carried out 
Dissipative Particle Dynamics (DPD) simulations of nonideal fluids. In their work the conservative forces needed for the 
DPD scheme were derived from a free energy density obtained from an equation of state (EOS) (e.g. the van der Waals EOS). 
Later on, Daoulas and M\"uller\cite{JChemPhys_125_184904} have also employed the density functional representation of 
nonbonded interactions 
in order to study the single chain structure and intermolecular correlations in polymer melts, and fluctuation 
effects on the order-disorder transition of symmetric diblock copolymers.
In the present work we introduce a grid-based density field-based scheme for dealing with nonbonded interactions in 
coarse grained simulations which takes into account a specific equation of state.

The purpose of this work is twofold. First, our main goal is to develop a consistent and rigorous methodology capable 
of simulating polymeric melts over realistic (i.e., $10^{-2} {\rm s}$) time scales.
In our previous works,\cite{EurPolymJ_47_699,Macromolecules_46_4670} we have developed a methodology in order to 
generate and equilibrate (nanocomposite) polymer melts at large length scales (on the order of 100 nm). This 
coarse-grained representation is based on the idea that the polymer chains can be described as random flights at length 
scales larger than that defined by the Kuhn length of the polymer.
Now, we develop a methodology to track the dynamics of the system at a coarse-grained level, 
by invoking a free energy functional motivated by an EOS capable of describing real polymeric materials, where both 
conformational and nonbonded contributions are taken into account. 
The EOS-based nonbonded free energy allows us to simulate a compressible model which, in principle, can deal with 
equation of state effects, phase transitions (e.g., cavitation) and interfaces. 
It is computed as a functional of the local density by 
passing an orthogonal grid through the simulation box. The voxels of the grid employed are large enough that their 
local thermodynamic properties can be described well by the chosen macroscopic EOS. The validity of this 
assumption is tested. At our level of description, i.e. beads consisting of 50 chemical monomers, forces among beads 
are not pairwise-additive and many-body effects are likely to be important.\cite{EurophysLett_30_191,PhysRevE_53_1572,JChemPhys_135_114103}
Pairwise additive effective potentials between beads\cite{Macromolecules_42_7474,Macromolecules_42_7485}
may not work well under conditions of high stress, where phase separation phenomena such as cavitation take place,
or near interfaces, where long-range attractions are important in shaping structure.
Thus, resorting to drastically coarse-grained models based on collective variables is a promising 
route.\cite{JChemPhys_125_184904} Introducing a nonbonded compensating potential like Chappa et 
al.,\cite{PhysRevLett_109_148302} eliminates the problem of chain conformations, caused by the artificial attraction 
of the slip-springs. Here, we refrain from explicitly including a term of this kind, in order to check whether the use 
of our non-bonded energy functional can provide an alternative way for avoiding unphysical agglomerations.

Once the free energy is known, BD simulations driven by the free energy functional can be used in 
order to obtain thermodynamic averages and correlation functions.
When macromolecules interpenetrate, the term entanglement intends to describe
the interactions resulting from the uncrossability of chains. At this level of description, we introduce the 
entanglements as slip-springs connecting beads belonging to different chains.
In the course of a simulation, the topology of the entanglement network changes through the introduction of elementary
kinetic Monte Carlo (kMC) events governed by rate expressions which are based on the reptation picture of polymer dynamics and 
the free energy defined.

Our second objective concerns parameterizing this methodology in a bottom-up approach, avoiding the introduction
of non-meaningful parameters and ad-hoc fitting of the results. All ingredients of our model are based on either 
more detailed (atomistic) simulations or experimental findings. For every parameter we introduce, we thoroughly 
explain its physical meaning and provide rigorous guidelines for estimating its exact value or its range of variation.

Our work is different from the extensive current literature on the subject in a number of ways. 
All observables stem out of an explicit Helmholtz energy functional.
The stress tensor of the model, a necessary ingredient for rheology calculations, is 
rigorously derived from the free energy functional including all contributions (bonded and nonbonded). Its special
characteristics are presented and discussed. In contrast to previous relevant studies, a rigorous kinetic MC 
scheme (with rate constants defined in terms of the free energy) is used for tracking slip-spring hopping dynamics in 
the system. A hybrid integration scheme allows for rate-controlled discrete slip-spring processes to be tracked in 
parallel with the integration of equations of motion of the polymeric beads. A new density field-based scheme for 
dealing with nonbonded interactions in coarse grained simulations,
reproducing a specific equation of state, is presented and its numerical application is thoroughly reviewed. 
All necessary links to more detailed levels of 
simulation and experimental observables are established, without introducing a posteriori tuning of the parameters. 
Finally, extensive quantitative validation against existing experimental findings is performed.

\section{Model and Method}

\subsection{Polymer Description}

\begin{figure}
   \centering
   \includegraphics[width=0.4\textwidth]{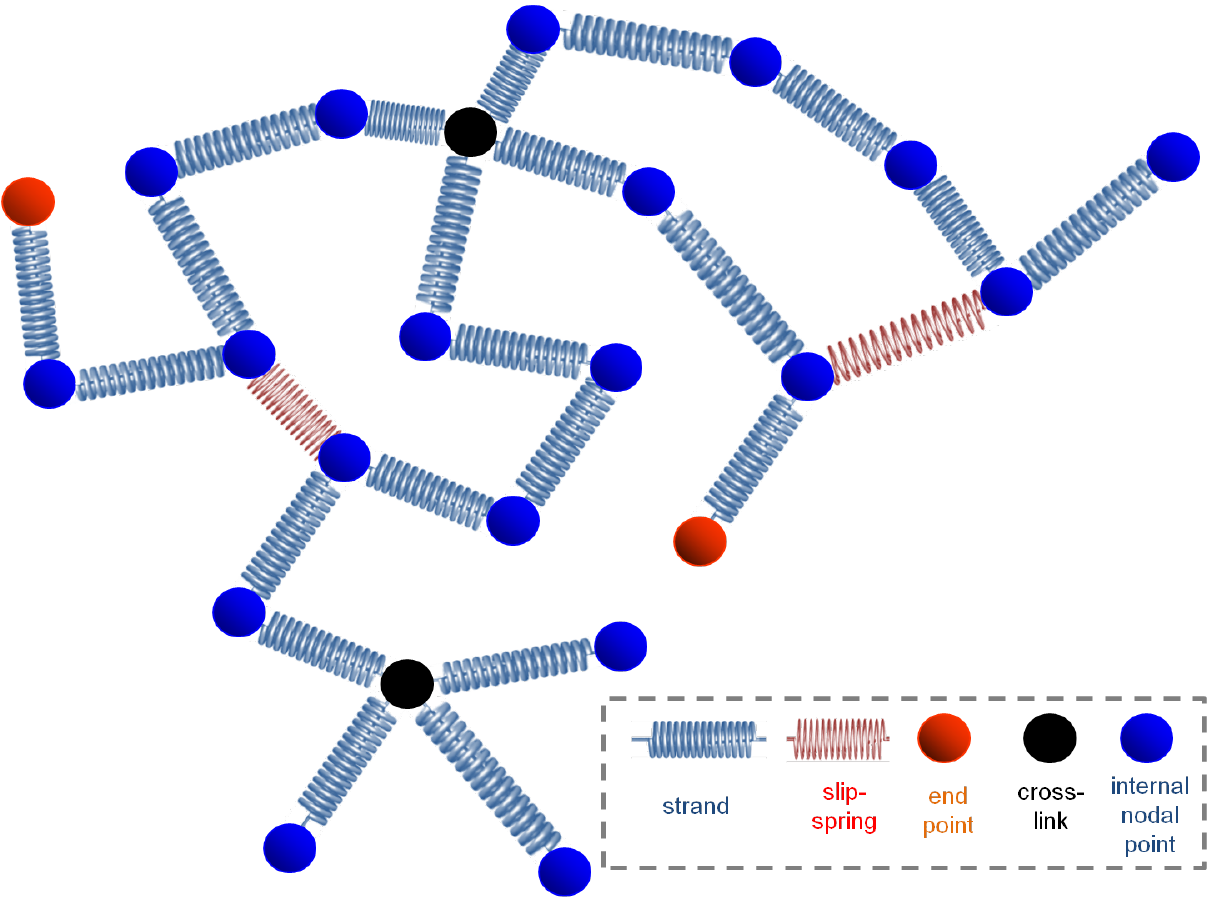}
   \caption[Network representation of a polymer melt or rubber.]
   {Network representation of a polymer melt or rubber. End beads, internal beads, crosslink beads (if 
   present) are shown as red, blue and black spheres, respectively. Entropic springs along the chain contour are shown
   in blue, while slip-springs are shown in red.}
   \label{fig:b3DkMC_polymer_schematic}
\end{figure}

Our melt consists of linear chains, represented by specific points or beads (i.e. internal nodal points, end points, or 
permanent crosslink points) along their contour, connected by entropy springs. Each coarse-grained bead represents 
several Kuhn segments of the polymer under consideration. 
Our construction results in a set of nodes for each chain, where each node $i$ has a specific contour position along the 
chain, positional vector in three-dimensional space, $\mathbf{r}_i$, and pairing with other nodes, as shown
in Figure \ref{fig:b3DkMC_polymer_schematic}. The piece of 
chain between two nodes (blue spring in Figure \ref{fig:b3DkMC_polymer_schematic}) is referred to as a strand.
This paper will be concerned with melts of linear chains, so permanent crosslink points will not be considered further.

We introduce the effect of entanglement by dispersing slip-springs,\cite{PhysRevLett_109_148302} which are designed to 
bring about reptational motion of chains along their contours, as envisioned in theories and simulations of dynamics 
in entangled polymer melts and as observed by topological analysis of molecular configurations evolving in the course 
of MD simulations.\cite{Macromolecules_45_9475}
In more detail, a slip-spring connects two internal nodal points or chain ends on two different
polymer chains and is stochastically destroyed when one of the nodes it connects is a chain-
end. To compensate for slip-spring destruction, new slip springs are created stochastically by
free chain ends in the polymer network. Along with the BD simulation of the bead motion in the periodic 
simulation box, a parallel simulation of the evolution of the slip-springs present is undertaken. The rates used for the 
kMC procedure are described in detail in the following paragraphs.
The initial contour length between consecutive entanglement points (slip-spring anchoring beads) on the same chain 
is commensurate with the entanglement molecular weight, $M_{\rm e}$, of the simulated polymer.

With the above mesoscopic model, the relative importance of reptation, constraint release (CR) and contour-length 
fluctuation (CLF) mechanisms depends on the specific melt of interest. As expected, in a monodisperse sample of long 
molecules, reptation plays a dominant role. In contrast, in a bidisperse sample composed of short and long 
macromolecules, CR and CLF may dominate the relaxation process.
\cite{Macromolecules_17_2316,Macromolecules_37_1641,Macromolecules_41_4945,Macromolecules_47_7653}

\subsection{Model Free Energy}
We postulate that the entangled melt specimen of given spatial extent, defined by the edge vectors of our periodic 
simulation box ($L_x$, $L_y$, $L_z$), under temperature $T$, is governed by a Helmholtz energy function, $A$, which 
has a direct dependence on the set of local densities $\cbraces{\rho\parnths{\mathbf{r}}}$, the temperature, $T$, and 
the separation vectors between pairs of connected polymer beads, 
$\cbraces{\mathbf{r}_{ij} \equiv \mathbf{r}_j - \mathbf{r}_i}$:
\begin{equation}
   A \big(\cbraces{\mathbf{r}_{ij}}, \cbraces{ \rho \parnths{\mathbf{r}}}, T \big) = 
   A_{\rm b} \big( \cbraces{\mathbf{r}_{ij}}, T \big) +
   A_{\rm nb} \big(\cbraces{\rho \parnths{\mathbf{r}}}, T  \big)
   \label{eq:helmholtz_energy_def}
\end{equation}
The first term on the right-hand side of eq \ref{eq:helmholtz_energy_def}, $A_{\rm b}$, is the contribution of 
bonded interactions, whereas the second one, $A_{\rm nb}$, is the contribution of the nonbonded
interactions to the Helmholtz energy.  
It should be noted that we employ the full free energy (including ideal gas contribution), 
and not the excess free energy in our formulation.
In the discussion of the nonbonded contribution
to the Helmholtz energy, $A_{\rm nb}$, we will elaborate further on our choice of employing the full Helmholtz energy,
rather than the excess one (with reference to an ideal gas of beads).

We start by considering the bonded contribution to the Helmholtz energy, which can be written as a sum over 
all bonded pairs, $\parnths{i,j}$, where $i$ is connected to $j$ with either intramolecular springs or slip-springs:
\begin{equation}
   A_{\rm b} \big(\cbraces{\mathbf{r}_{ij}}, T \big) = 
   A_{\rm b} \big(\cbraces{r_{ij}} , T \big) =
   \sum_{\parnths{i,j}} A_{\rm pair} \parnths{r_{ij}, T}
   \label{eq:helmholtz_strand_contribution}
\end{equation}
The sum runs over all pairs, where each pair is thought to interact via a distance-dependent Helmholtz energy, 
$A_{\rm pair} \parnths{r_{ij}, T}$.
The elastic force depends on the coordinates of the nodes, but it does not depend on the local network density.
The elastic force between connected beads arises due to the retractive force acting to resist the stretching of a 
strand. This force originates in the decrease in entropy of a stretched polymer strand. 
In this approximation, the force on bead $i$, due to its connection to bead $j$ is: 
\begin{equation}
   \mathbf{F}_{ij}^{\rm b} = - \nabla_{\mathbf{r}_{ji}} A_{\rm pair} \parnths{r_{ij}, T}
   \label{eq:bonded_force}
\end{equation}

The Gaussian approximation to $A_{\rm pair}\left(r_{ij}\right)$ can be used for most 
extensions.\cite{JChemSocFaradayTrans_93_4355} 
The conformational entropy of strands is taken into account via a simple harmonic expression:
\begin{equation}
   A_{\rm pair}^{\rm intra} \parnths{r_{ij}, T} =  \frac{3}{2} k_{\rm B} T 
   \frac{r_{ij}^2}{N_{ij} b_{\rm K}^2}
\end{equation}
where $N_{ij}$ is the number of Kuhn segments assigned to the strand, $b_{\rm K}$ the Kuhn length of the polymer, 
and $k_{\rm B}$ the Boltzmann constant.
The summation is carried over all pairs $\parnths{i,j}$ which lie along the contour of the chains.
The Helmholtz energy of slip-springs, which are included to account for the entanglement effect, is described by the
following equation:
\begin{equation}
   A_{\rm pair}^{\rm sl} \big( r_{ij}, T \big) = \frac{3}{2} k_{\rm B} T 
   \frac{r_{ij}^2}{l_{\rm ss}^2}
   \label{eq:sls_helmholtz_energy}
\end{equation}
where $l_{\rm ss}$ is an adjustable parameter (i.e. 
root mean square equilibrium slip-spring length) which should be larger than 
the Kuhn length, $b_{\rm K}$, and smaller than or equal to the tube diameter of the polymer under consideration, 
$a_{\rm pp}$.

In order to account for nonbonded (excluded volume and van der Waals attractive) interactions in the network 
representation, we introduce a 
Helmholtz energy functional:
\begin{equation}
   A_{\rm nb} \parnths{\cbraces{\rho\parnths{\mathbf{r}}}, T} = \int_{\rm box} \D^ 3 r \; 
   a_{\rm vol} \left[\rho\left(\mathbf{r}\right), T \right]
   \label{eq:nb_helmholtz_energ_def}
\end{equation}
where $a_{\rm vol}$ is a free energy density (free energy per unit volume) and $\rho \parnths{\mathbf{r}}$ 
is the local mass density at position $\mathbf{r}$. 
Expressions for $a_{\rm vol} \left(\rho, T\right)$ may be extracted from equations of state. 
Local density, $\rho \left(\mathbf{r}\right)$, will be resolved only at the level of entire cells,
defined by passing an orthogonal grid through the simulation domain. 
Thus, eq \ref{eq:nb_helmholtz_energ_def} will be approximated by a discrete sum over all cells of the orthogonal grid:
\begin{equation}
   A_{\rm nb} \parnths{\cbraces{\rho\parnths{\mathbf{r}}}, T } = \sum_{k=1}^{N_{\rm cells}} V_{{\rm cell}, k}^{\rm acc} 
   \: a_{\rm vol}\left(\rho_{{\rm cell}, k}, T\right)
   \label{eq:nb_helmholtz_energy_approx}
\end{equation}
where $V_{{\rm cell}, k}^{\rm acc}$ is the accessible volume of cell $k$. The cell density, $\rho_{{\rm cell}, k}$,
must be defined based on the beads in and around cell $k$. 
The spatial discretization scheme employed for nonbonded interactions is thoroughly presented in Appendix A.

In deriving nonbonded forces (see eq \ref{eq:nonbonded_force} below), it would have been more correct to use the excess
Helmholtz energy density (relative to an ideal gas of beads), rather than the total Helmholtz energy density and to
rely on the particle-based simulation for effectively including the contribution of translational entropy. Our
formulation uses the total, rather the excess Helmholtz energy density, thereby missing the translational entropy 
associated with allowing the beads which reside in one 
voxel to visit other voxels. However, we have validated our results against those obtained by employing the excess Helhmoltz 
energy. They are practically identical, because, in our implementation, the cells of the discretization grid are so large that 
they behave almost as macroscopic domains. The translational entropy associated with exchange of a bead between voxels
is negligible in comparison with the entropy associated with roaming of the bead within a voxel.

In this work we invoke the Sanchez-Lacombe equation of state,\cite{JPhysChem_80_2352} which gives for the Helmholtz 
energy:
\begin{align}
   A^{\rm SL} \left(\rho, T \right) = & -n r k_{\rm B}T^* \tilde{\rho} + n r k_{\rm B} T \left[
   \left(\frac{1}{\tilde{\rho}}-1\right)\ln{\left(1-\tilde{\rho}\right)} 
   + \frac{1}{r}\ln{\parnths{\tilde{\rho}}}\right] \nonumber \\
   & - n k_{\rm B} T \ln{\parnths{w}}
   \label{eq:sl_helmholtz}
\end{align}
where $\tilde{\rho} =\rho / \rho*$ is the reduced density with $\rho$ being the mass density of the melt and $\rho^*$ 
the close packed mass density. The temperature of the melt is denoted by $T$ and $T^* = \epsilon/ k_{\rm B}$ corresponds 
to a equivalent temperature defined in terms of $\epsilon$ which is the nonbonded, mer-mer interaction energy ($k_{\rm B}$ 
being Boltzmann's constant). The 
$n$ chains of the melt, whose molecular weight per chain is $M$, are thought to be composed of 
$r$ Sanchez Lacombe segments each (``$r$-mers'').
Finally, $w$ is a parameter quantifying the number of configurations available to the system. 
The last term in eq \ref{eq:sl_helmholtz} does not depend on the density of the system, 
being an ideal gas contribution (please see discussion in Appendix B).
All Sanchez-Lacombe parameters are presented in Table B1 of Appendix B.

The Helmholtz energy density, $a_{\rm vol}\left(\rho, T\right)$, can be calculated as:
\begin{equation}
   a_{\rm vol} \left(\rho, T \right) = \frac{A\left(\rho, T \right)}{V} 
   = \frac{A\left( \rho, T\right)}{n M/\parnths{\rho N_{\rm A}}} = \rho N_{\rm A} \frac{A\left(\rho,T\right)}{n M} 
   = \rho \; a_{\rm mass}\left(\rho, T\right)
\end{equation}
where the $a_{\rm mass} \left(\rho, T\right)$ denotes the Helmholtz energy per unit mass of the system.
Based on eq \ref{eq:sl_helmholtz}, we can calculate the above:
\begin{equation}
   a_{\rm mass} \left(\rho, T\right) = - \frac{r R T^* \tilde{\rho}}{M} + \frac{r R T}{M}
   \left[\left(\tilde{v}-1\right) \ln{\left(1-\tilde{\rho} \right)}
   +\frac{1}{r}\ln{\left( \tilde{\rho} \right)} \right]
   -\frac{R T}{M}\ln{\left(w\right)}
\end{equation}
and by using $M = r \rho^* v^*$ and $r = \left(M p^* \right) / \left(\rho^* R T^* \right)$ 
the above expression becomes:
\begin{equation}
   a_{\rm mass} \left(\rho, T\right) = 
   - \frac{p^*}{\rho^{*2}} \rho + \frac{p^* \tilde{T}}{\rho^*} 
   \left[ \left(\frac{\rho^*}{\rho} - 1\right) \ln{\left(1 - \frac{\rho}{\rho^*}\right)}
   + \frac{\rho^* R T^*}{M p^*} \ln{\left( \frac{\rho}{\rho^* w} \right)} 
   \right]
\end{equation}
The Helmholtz energy density (free energy per unit volume), $a_{\rm vol} \left(\rho, T\right)$ is:
\begin{align}
   a_{\rm vol} \left(\rho, T\right) =  & \rho \; a_{\rm mass}\left(\rho, T\right) \nonumber \\
   = & -p^* \tilde{\rho}^2  + p^* \tilde{T} \tilde{\rho} \left[ 
   \left(\frac{1}{\tilde{\rho}} - 1\right)\ln{\left(1 - \tilde{\rho} \right)}
   + \frac{\rho^* R T^*}{M p^*} \ln{\parnths{\tilde{\rho}}} -\frac{\rho^* R T^*}{M p^*} \ln{\parnths{w}} \right]  
   \label{eq:helhmoltz_energy_density}
\end{align}
where everything is cast in terms of the reduced variables $T^*$, $p^*$, $\rho^*$ and the molecular weight of a chain, 
$M$. 
Upon integration over the domain of a system with given mass, the last term within the brackets of eq 
\ref{eq:helhmoltz_energy_density}, which is strictly proportional to 
density, yields a density-independent contribution. 
Thus, it does not affect the equations of motion or thermodynamic properties.
All needed parameters (i.e. $T^*$, $p^*$, $\rho^*$) can be obtained from experimental studies.\cite{PolymBull_34_109}

The force due to non-bonded interactions on a bead $i$ is given by:
\begin{align}
   \mathbf{F}^{\rm nb}_{i} = & - \nabla_{\mathbf{r}_i} A_{\rm nb} \parnths{\cbraces{\rho_{\rm cell}}, T} 
   =  - \frac{\partial}{\partial \mathbf{r}_i} \left [ \sum_{k \in {\rm cells}} V_{{\rm cell},k}^{\rm acc} 
   a_{\rm vol} \left(\rho_{{\rm cell}, k}, T\right) \right ] \nonumber \\
   = & - \sum_{k \;\in \;{\rm cells}} V_{{\rm cell},k}^{\rm acc} 
   \left . \frac{\partial a_{\rm vol}\left(\rho, T\right)}{\partial \rho}\right |_{\rho = \rho_{{\rm cell},k}}
   \frac{\partial \rho_{{\rm cell},k}}{\partial \mathbf{r}_i}
   \label{eq:nonbonded_force}
\end{align}
with the derivative of $a_{\rm vol}\left(\rho, T\right)$ with respect to density being:
\begin{equation}
   \frac{\partial a_{\rm vol}\left(\rho, T\right)}{\partial \rho} =  - \frac{2 p^*}{\rho^{*2}}\rho
   - \frac{p^* T}{\rho^* T^*} \left[1 + \ln{\left(1-\frac{\rho}{\rho^*} \right)} \right] 
   + \frac{RT}{M}\left[ 1 + \ln{\left(\frac{\rho}{\rho^* w} \right)} \right]
\end{equation}
and the derivative $\partial \rho_{{\rm cell},k} / \partial \mathbf{r}_i$ given by eq 
\ref{eq:discrete_nonbonded_derivative} in Appendix A.
The terms of the Helmholtz energy density containing the parameter $w$ (or the thermal wavelength as
those derived in Appendix B), which are linear in the density, do not contribute to the forces. Upon integration over
the entire domain of the primary simulation box, these terms give a constant contribution to the Helmholtz energy 
and their gradient with respect to any bead position is zero.

\subsection{Generation of Initial Configurations}
Initial configurations for the linear melt are obtained by field theory - inspired Monte Carlo (FT-i MC) equilibration 
of a coarse-grained melt, wherein chains are represented as freely jointed sequences of Kuhn segments subject to a 
coarse-grained Helfand Hamiltonian which prevents the density from departing from its mean value anywhere in the 
system.\cite{EurPolymJ_47_699, Macromolecules_46_4670}
The coarse-graining from the freely jointed chain model to the bead-spring model involves placement of beads at 
regular intervals along the contour of the chains obtained after the MC equilibration.
As already discussed in the previous section, at the new (coarser, mesoscopic) level of description, the polymer is 
envisioned as a network of strands connecting internal nodal points and end points.

Starting from a well equilibrated configuration $\mathscr{R}$, obtained from a FT-i MC simulation, we determine 
the box size and shape for which the Gibbs energy function:\cite{MolPhys_111_3430}
\begin{equation}
   G \parnths{T, \boldsymbol{\tau}} = A
   \parnths{\cbraces{\mathbf{r}_{ij}}, \left\{ \rho \left(\mathbf{r} \right) \right\}, T } 
   - V_\mathscr{R} \frac{1}{3}{\rm Tr}\parnths{\boldsymbol{\tau}}
   - V_\mathscr{R} \sum_{\alpha \beta}\tau_{\alpha \beta} \epsilon_{\alpha \beta}
   = G \parnths{\cbraces{\mathbf{r}_i}, \boldsymbol{\tau}}
   \label{eq:gibbs_energy_for_minimization}
\end{equation}
becomes minimal under the given, externally imposed $\boldsymbol{\tau}$. By prescribing the stress tensor,
$\boldsymbol{\tau}$, we minimize the free energy with respect to the positions of the nodal points, 
$\cbraces{\mathbf{r}_i}$, which we assume to follow the macroscopic strain in an affine way.
The presence of any symmetry element 
in the stress tensor reduces the number of minimization parameters, and in the special case where only hydrostatic 
pressure is applied on the system, the sum of the last two terms in eq 
\ref{eq:gibbs_energy_for_minimization} is equivalent to $-p V$, letting the volume $V$ of the deformed configuration 
be the only parameter for the Gibbs energy minimization.
In that case, the volume is expressed as $V = V_\mathscr{R} \parnths{1 + \epsilon_{xx} + \epsilon_{yy} + \epsilon_{zz}}$.
The Gibbs energy function, eq \ref{eq:gibbs_energy_for_minimization}, is, of course, valid only for small deformations
away from the reference state.
In our case, we let the system relax under atmospheric hydrostatic pressure, starting from an equilibrated 
configuration at the average density for the temperature under consideration. 
Slip-springs have not been introduced yet at this point.

Slip-springs can either be placed randomly
in the initial configuration, or they can be allowed to be created, following the fluctuating kMC scheme described
in Appendix E. 
In the former way, the number of slip-springs is chosen so as to be consistent with the molar mass between 
entanglements, $M_{\rm e}$, of the polymer under consideration. Every possible pair of beads in the melt can be coupled
by a slip-spring, given that their separation distance is less than $l_{\rm ss}$ (eq \ref{eq:sls_helmholtz_energy}).
Otherwise, the initially slip-spring-free configuration is used as an input and slip-springs are generated on the go by 
utilizing the fluctuating kMC scheme (Appendix E) with a hopping pre-exponential factor, $\nu_{\rm hop}$, at the 
upper limit of the allowed range (Appendix D). In this way, slip-springs are introduced in an unbiased and rigorous way 
while the melt is in parallel equilibrated. The procedure is interrupted at the point where the number of entanglements 
(slip-springs) reaches the desired value, and the frequency factor $\nu_{\rm hop}$ is fixed at the value that ensures
conservation of the number of slip-springs. Once the slip-springs have been introduced, the Gibbs energy of the system
(eq \ref{eq:gibbs_energy_for_minimization}) can be minimized again in their presence, at fixed topology of the 
entanglement network, in order to ensure the proper simulation box dimensions. However, slip-springs contribute 
little (on the order of 1\%) to the stress tensor of the system.

\subsection{Brownian Dynamics}
When simulating a system of coarse-grained particles, some degrees of freedom are treated explicitly, whereas other 
are represented only by their stochastic influence on the former ones. In our model the effect of the surrounding melt
on the motion of the coarse-grained beads is mimicked by introducing a stochastic force plus a frictional force into 
the equations of motion of the beads. When the stochastic force contains no correlations in space or time, one obtains
the simplest form of stochastic dynamics, called Brownian Dynamics (BD).\cite{RevModPhys_15_1,ChemPhysLett_24_243,JChemPhys_69_1352}
The theory of Brownian motion was developed to describe the dynamic behavior of particles whose mass and size are much 
larger than those of the host medium particles. In this case the position Langevin equation becomes:
\begin{equation}
   \mathbf{v}_i \parnths{t} = \frac{1}{\zeta_i} \mathbf{F}_i \parnths{\cbraces{\mathbf{r}_i\parnths{t}}} +
   \frac{1}{\zeta_i} \boldsymbol{\mathcal{F}}_i\parnths{t}
   \label{eq:method_simulations_brownian_equation_of_motion}
\end{equation}
The systematic force $\mathbf{F}_{i}\parnths{t}$ is the explicit mutual force between the $N$ particles and 
$\boldsymbol{\mathcal{F}}_{i} \parnths{t}$ represents the effect of the medium on the particles. Each particle is 
characterized by its mass $m_i$ and the friction coefficient $\zeta_i$, measured in kg/s. 
The systematic force, $\mathbf{F}_i$, is to be derived from the free energy following eqs \ref{eq:bonded_force} and
\ref{eq:nonbonded_force}:
\begin{equation}
   \mathbf{F}_i \parnths{\cbraces{\mathbf{r}_i\parnths{t}}} = - \nabla_{\mathbf{r}_i\parnths{t}}  
   A \parnths{\left\{ \mathbf{r}_{ij}\parnths{t} \right\}, \left\{ \rho \left(\mathbf{r},t\right) \right\}, T }
\end{equation} 
with the expression of the Helmholtz energy as given in eq \ref{eq:helmholtz_energy_def}.
The stochastic force $\boldsymbol{\mathcal{F}}_{i}$ is assumed to be stationary, Markovian and Gaussian with zero mean and
to have no correlation with prior velocities nor with the systematic force.
For large values of $(\zeta_i / m_i) \Delta t$ in the diffusive regime, when the friction is so strong that the velocities
relax within $\Delta t$, the BD algorithm of van Gunsteren and Berendsen\cite{MolPhys_45_637}, eq 
\ref{eq:method_simulations_brownian_integration_scheme} can be used:
\begin{equation}
   r_{i, \alpha} \parnths{t_n + \Delta t} = r_{i,\alpha}\parnths{t_n} 
   + \frac{1}{\zeta_i}\bracks{F_{i,\alpha} \parnths{t_n} \Delta t + \frac{1}{2} \dot{F}_{i,\alpha} \parnths{t_n}
   \parnths{\parnths{\Delta t}^2} } + \mathcal{R}_{i,\alpha} \parnths{\Delta t}
   \label{eq:method_simulations_brownian_integration_scheme}
\end{equation}
where the random variable $\mathcal{R}_{i,\alpha,n} \left(\Delta t\right)$ is sampled from a Gausssian distribution with
zero mean and width:
\begin{equation}
   \left \langle \mathcal{R}_{i,\alpha}^2 \parnths{\Delta t} \right \rangle 
   = \frac{2 k_{\rm B} T_{\rm ref}}{\zeta_i} \Delta t
   \label{eq:method_simulations_brownian_integration_scheme_ran_displacement}
\end{equation}
The time derivatives of the forces, $\dot{F}_{i,\alpha}$, are estimated by finite differences. 
However, more sophisticated schemes, like Smart Monte Carlo, have been developed.\cite{JChemPhys_129_164906}

\subsection{Slip-spring Kinetic Monte Carlo}

\begin{figure}
   \centering
   \includegraphics[width=0.4\textwidth]{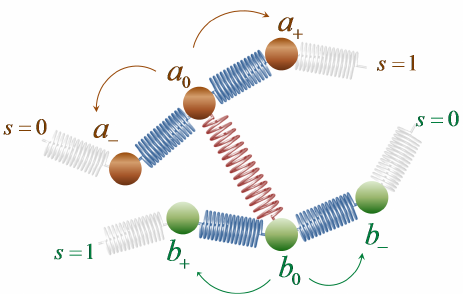}
   \caption{Illustration of the ``hopping'' scheme of slip-springs along the chain contour.}
   \label{fig:kMC_hopping_contour}
\end{figure}

Chain slippage through entanglements is modeled as hopping of slip-springs along the contours of the chains they
connect, without actual displacements of their beads. Let the ends of a slip-spring connect beads $a_0$ and $b_0$ 
along chains $a$ and $b$ (see Figure \ref{fig:kMC_hopping_contour}).
In order to track hopping events in the system, we use a discretized version of the kinetic Monte Carlo (kMC) 
algorithm, which runs in parallel with the BD integration of beads' equations of motion. 
The dynamics of slip-spring jumps is envisioned as consisting of infrequent transitions from one state to another, with long 
periods of relative inactivity between these transitions. We envision that each state corresponds to a single free 
energy basin, and the long time between transitions arises because the system must surmount an energy barrier, 
$A^\ddagger$, to get from one basin to another.
Then, for each possible escape pathway from an original to a destination basin (e.g. $\mathscr{O} \to \mathscr{N}$), there is a 
rate constant $k_{\mathscr{O} \to \mathscr{N}}$ that characterizes the probability, per unit time, 
that an escape from state $\mathscr{O}$ to a new one, $\mathscr{N}$, will occur. These rate constants are independent of
what state preceded state $\mathscr{O}$.

We sample slip-spring transitions at regular pre-defined time intervals $\Delta t_{\rm kMC}$ which are multiples of 
the timestep used for the BD integration, $\Delta t_{\rm kMC} = n_{\rm kMC} \Delta t$. Every $n_{\rm kMC}$ steps of the
BD integration along the trajectory of the system, we freeze the coordinates of all beads and calculate the transition 
probabilities,  $k_{\rm hop}\Delta t_{\rm kMC}$ for all possible transitions of the slip-springs. 
The kMC time interval, $\Delta t_{\rm kMC}$, is chosen such that the number of transitions performed at every kMC
step is much smaller than the number of slip-springs present in the system and preferably on the order of unity 
(i.e., infrequent events governed by Poisson process statistics). 
The kMC step consists in visiting all slip-springs 
of the system and computing the probabilities for them to perform a jump, considering all possible states accessible 
from their current one. A probability is assigned to every possible transition, 
$p_{\mathscr{O} \to \mathscr{N}} = k_{\mathscr{O} \to \mathscr{N}} \Delta t_{\rm kMC}$. The transition is undertaken only if 
$p_{\mathscr{O} \to \mathscr{N}}$ exceeds a pseudo-random number sampled from a uniform distribution in the interval
$[0,1)$. The use of small kMC timesteps, $\Delta t_{\rm kMC}$,
ensures that $\sum_{\mathscr{N}} p_{\mathscr{O} \to \mathscr{N}} << 1$ at every step kMC is attempted. 
In the following we will present a microscopically reversible and physically meaningful formulation for calculating 
all rates connected with slip-spring jumps, slip-spring destruction and formation. Our formulation is further 
elaborated to Appendices D, E and F, where detailed considerations are presented.

\begin{figure}
   \centering
   \includegraphics[width=0.4\textwidth]{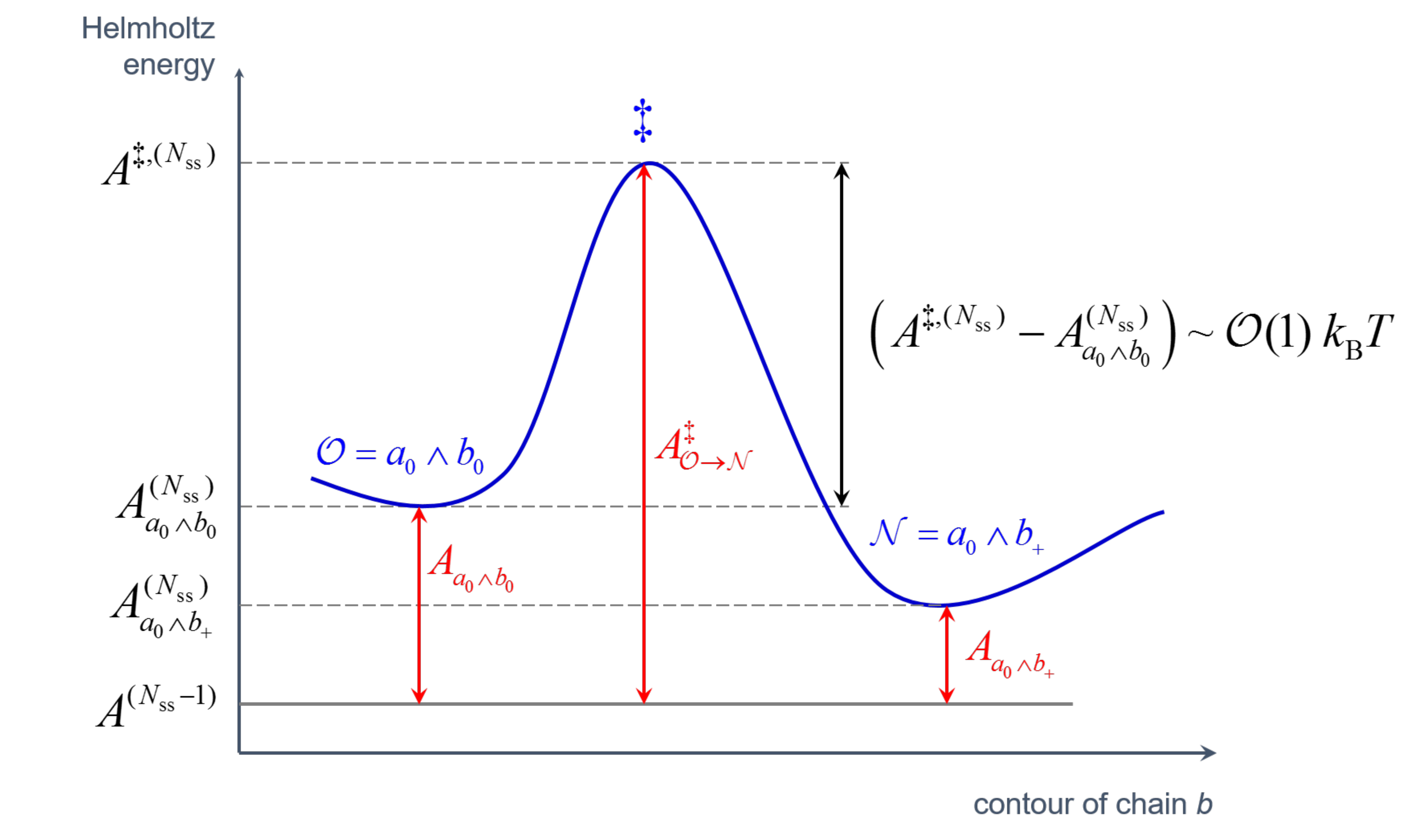}
   \caption{Schematic representation of free energy levels defined in the slip-spring hopping formulation.
   The transition considered is the slippage of the slip-spring initially connecting the beads $a_0$ and $b_0$,
   along chain $b$, i.e., $b_0 \to b_+$. 
   The height of the free energy barrier in our simulations was in the range of $3$ to $8$ $k_{\rm B}T$,
   as indicated in the figure. Please see text for details.}
   \label{fig:free_energy_levels}
\end{figure}

In order to develop a formalism for elementary events of slip-spring hopping, creation or destruction, we need 
expressions for the rate of slippage along the chain backbone. We concentrate on the slip-spring connecting beads 
$a_0$ and $b_0$. The Helmholtz energy of the system is denoted as $A^{\parnths{N_{\rm ss}}}_{a_0 \land b_0}$, 
where in our notation we have incorporated both the number of slip-springs present and the connectivity the particular
slip-spring finds itself at.
An individual jump of one end of a slip-spring along the chain backbone, e.g. from bead $b_0$ to bead $b_+$, 
takes place with rate:
\begin{equation}
   k_{\rm hop} =  \nu_0 \exp{\parnths{-\frac{A^{\ddagger,\parnths{N_{\rm ss}}} 
   - A^{\parnths{N_{\rm ss}}}_{a_0 \land b_0}}{k_{\rm B} T}}} 
   \label{eq:hopping_rate_definition_nu_0}
\end{equation}
conforming to a transition state theory (TST) picture of the slippage along the backbone as an infrequent event, 
which involves a transition 
from state $\mathscr{O} \equiv a_0 \land b_0$ to state $\mathscr{N} \equiv a_0 \land b_+$ over a barrier of free energy, 
$A^{\ddagger,\parnths{N_{\rm ss}}} - A^{\parnths{N_{\rm ss}}}_{a_0 \land b_0}$ (c.f. Figure \ref{fig:free_energy_levels}). 
Since the connectivity of the $N_{\rm ss}-1$ slip-springs does not change during the transition, we can define 
$A_{\mathscr{O} \to \mathscr{N}}^\ddagger$ as the difference between the total Helmholtz energy of the configuration 
with a slip spring at the transition state, $A^{\ddagger,\parnths{N_{\rm ss}}}$, and the total Helmholtz energy of a 
configuration that is missing the particular slip spring but is otherwise identical, $A^{\parnths{N_{\rm ss}-1}}$, i.e., 
$A^{\ddagger,\parnths{N_{\rm ss}}} = A^{\parnths{N_{\rm ss}-1}} + A_{\mathscr{O} \to \mathscr{N}}^\ddagger$.
Accordingly, the free energy of the starting configuration can be split as 
$A_{a_0 \land b_0}^{\parnths{N_{\rm ss}}} = A^{\parnths{N_{\rm ss}-1}} + A_{a_0 \land b_0}$.
All configurations depicted in Figure \ref{fig:free_energy_levels} are identical but for the presence and end point 
positions of the $N_{\rm ss}$-th slip-spring.
We use the logical conjunction operator ``$\land$'' to denote the connectivity of the system.
As depicted in Figure \ref{fig:kMC_hopping_contour}, in the case considered none of the nearest neighbor beads of the 
ends of the slip-spring is a chain end. Double jumps (e.g. $a_0 \to a_+$ and $b_0 \to b_+$) are disallowed, which is
a reasonable approximation for small timesteps, $\Delta t_{\rm kMC}$. 

Based on the proper (following Appendix D) selection of the pre-exponential factor $\nu_{\rm hop}$, the number of steps
between two successive kinetic Monte Carlo steps, $n_{\rm kMC}$, should be chosen such that the overall probability,
$p_{\rm hop} = k_{\rm hop} \Delta t_{\rm kMC} = k_{\rm hop} n_{\rm kMC} \Delta t$, is less than 1 for every slip-spring
in the system. Based on the selected $\nu_{\rm hop}$ and $\Delta t$ we set $n_{\rm kMC} = 100$. Any number smaller than
that can be used and any number larger than that renders hopping probabilities per kMC step of the order of unity
and thus has to be avoided. Our results are insensitive to $n_{\rm kMC}$ within the aforementioned range.

Assuming that all slip-springs attempting to jump face the same free energy barrier, $A_{\mathscr{O} \to \mathscr{N}}^\ddagger$, 
on top of $A^{\parnths{N_{\rm ss}-1}}$, eq \ref{eq:hopping_rate_definition_nu_0} can be more conveniently rewritten as:
\begin{equation}
   k_{\rm hop}  = \nu_0 \exp{\parnths{-\frac{A_{\mathscr{O} \to \mathscr{N}}^\ddagger
   - A_{a_0 \land b_0}}{k_{\rm B} T}}} 
    = \nu_{\rm hop}  \exp{\parnths{\frac{A_{a_0 \land b_0}}{k_{\rm B}T}}}
   \label{eq:hopping_rate_definition_nu_hop}
\end{equation} 
where the rate of hopping, $k_{\rm hop}$, depends directly on the energy stored in the slip-spring at its initial state (not
of the entire configuration), $A_{a_0 \land b_0}$, while the dependence on the height of the free energy at the barrier 
(i.e. $A_{\mathscr{O} \to \mathscr{N}}^\ddagger$) has been absorbed into the pre-exponential factor 
$\nu_{\rm hop} = \nu_0 \exp{\parnths{-\beta A_{\mathscr{O} \to \mathscr{N}}^\ddagger}}$, with 
$\beta = 1/\parnths{k_{\rm B}T}$.
Eq \ref{eq:hopping_rate_definition_nu_hop} leaves us with only one adjustable parameter to be determined, 
the pre-exponential frequency factor $\nu_{\rm hop}$. In Appendix D, a simple theoretical guideline in order to arrive 
to an estimate of $\nu_{\rm hop}$ is presented, given the fact that the average hopping rate $\angles{k_{\rm hop}}$,
can be accessed.

\begin{figure}
   \centering
   \includegraphics[width=0.4\textwidth]{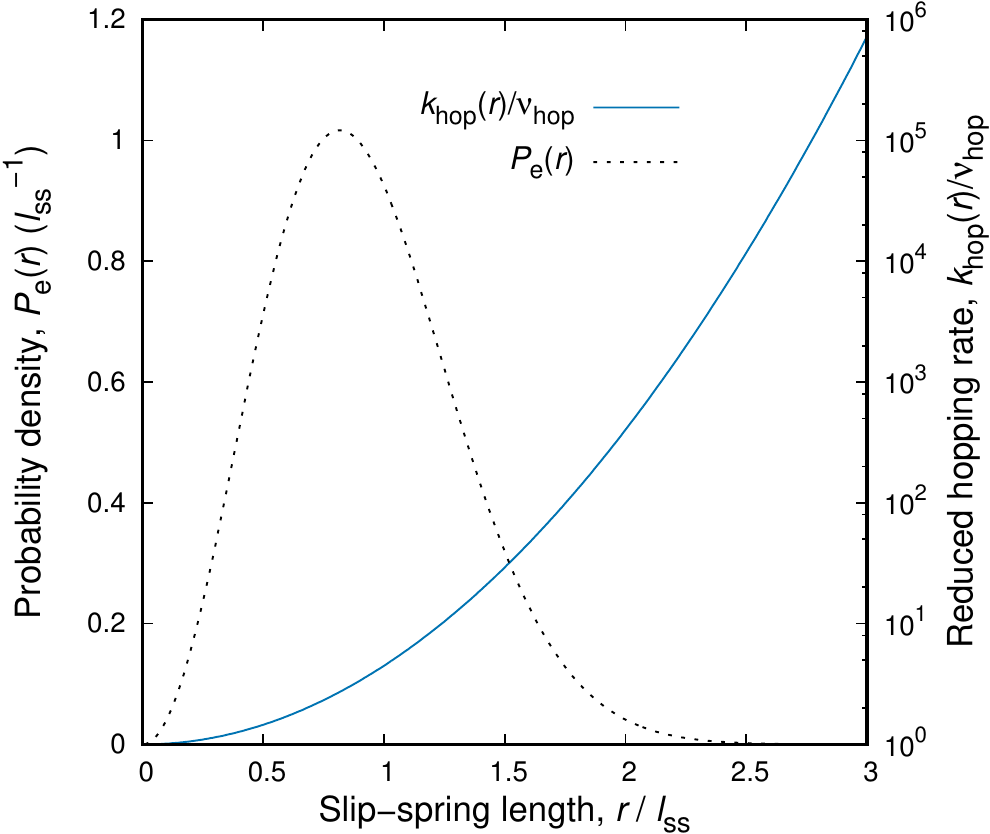}
   \caption{Reduced rate of slip-spring hopping, $k_{\rm hop}$, divided by $\nu_{\rm hop}$, as a function of 
   slip-spring length (in units of $l_{\rm ss}$, assuming Gaussian springs). The theoretical Gaussian 
   distribution of the slip-spring length is also presented in the second ordinate of the figure.} 
   \label{fig:theoretical_rate}
\end{figure}

The representation of slip-springs in terms of entropic Gaussian springs allows for a direct estimation of the 
expected hopping rate, $\angles{k_{\rm hop}}$.
The distribution of the length of the slip-springs, $P_{\rm e}\parnths{r}$, and the corresponding hopping rate,
$k_{\rm hop}\parnths{r}$, are presented in Figure \ref{fig:theoretical_rate}.
Given the probability distribution, the expected average rate of hopping per slip-spring can be estimated as 
$\angles{k_{\rm hop}} = \int_{0}^{\infty} P_{\rm e}\parnths{r} k_{\rm hop}\parnths{r} \D r 
/ \int_{0}^{\infty} P_{\rm e}\parnths{r} \D r$.
By assuming a maximum extension of slip-springs up to $3 \l_{\rm ss}$, a distance excluding less than 
$10^{-4}$ of the cumulative probability distribution, the expected hopping rate is roughly equal to 
$\angles{k_{\rm hop}} / \nu_{\rm hop} \simeq 38$.

\begin{figure}
   \centering
   \includegraphics[width=0.4\textwidth]{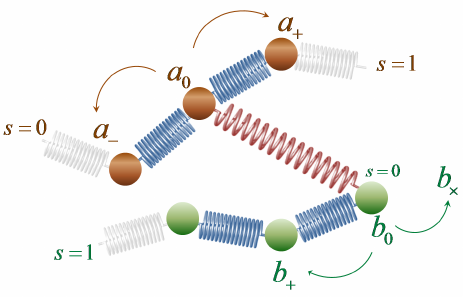}
   \caption{Illustration of the ``hopping'' scheme of a slip-springs whose one end is a chain end.}
   \label{fig:kMC_hopping_destruction}
\end{figure}

Let us assume that the ends of the slip-spring connect beads $a_0$ and $b_0$ as those are depicted in Figure
\ref{fig:kMC_hopping_destruction}. 
One of the beads connected by the slip-spring (e.g. $b_0$) is a chain end. One end of the slip spring can hop along 
the chain, while the second can either hop moving away from the chain end or be destroyed. 
The rate of destruction of a slip-spring is equal to the rate of hopping along the chain, $k_{\rm hop}$. 
This assumption keeps the necessary adjustable parameters of our model to an absolute minimum.
Through the intrinsic dynamics of the system, it is possible for a slip-spring end to leave its chain. This process 
mimics the disentanglement at the chain ends and the process of CR.
The process of slip-spring destruction is introduced in the model in order to represent the chain disentanglement,
as that is envisioned by the polymer tube theories.
In the case that both ends of a slip-spring are attached to chain ends, based on the scheme we have introduced, there
are equal probabilities for any of them sliding across any of the chain ends and getting destroyed.

As far as the event of creation of a slip-spring is concerned, two schemes have been developed; the former 
preserves the number of slip springs, while the latter allows a fluctuating number of slip springs obeying
microscopic reversibility. For the sake of clarity, we will present the constant scheme here, while the fluctuating 
scheme is presented in Appendix E. Since both schemes yield identical trajectories for the systems studied we will 
refrain from further distinction between them in the following.
In order to preserve the number of slip-springs constant throughout the simulation, we assume that a new slip-spring 
is created whenever an existing slip-spring is destroyed through slippage past a chain end. In creating a new slip-spring,
a free end of a chain can be paired to 
another bead if they are separated by a distance less than $\alpha_{\rm attempt}$. Thus, in order to ensure 
microscopic reversibility, a slip-spring can be destroyed only if its extent is less than $\alpha_{\rm attempt}$
at the time the destuction is decided to take place. 

\begin{figure}
   \centering
   \includegraphics[width=0.4\textwidth]{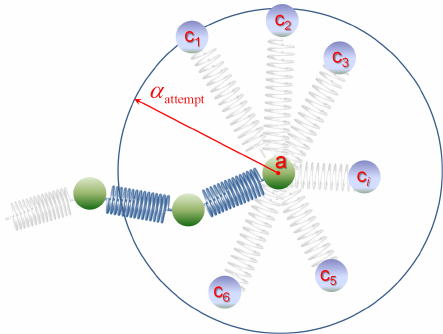}
   \caption{Illustration of the slip spring creation procedure.}
   \label{fig:kMC_hopping_creation}
\end{figure}

If a slip-spring is considered for destruction, all possible 
connections of $a_0$ (i.e., the bead that is not a candidate for slippage along its chain contour, see Figure 
\ref{fig:kMC_hopping_destruction}) with other beads $b$ lying inside a sphere of radius $\alpha_{\rm attempt}$ 
centered at $a_0$ are identified. If no such beads can be found, the destruction/creation move is rejected,
due to violation of the microscopic reversibility, since the system would not be able to access its present
state through the kMC moves. The Rosenbluth weight of the old configuration is accumulated, by iterating over
all neighboring beads $b$, $W_{\mathscr{O}} = \sum_{b} \exp{\parnths{-\beta A_{a_0 \land b}}}$. 
A chain end $a^\prime_0$ in the system is randomly selected with probability $1/\parnths{2n}$,
with $n$ being the total number of chains. All possible connections of $a^\prime_0$ with beads $b^\prime$ lying inside 
a sphere of radius $\alpha_{\rm{attempt}}$ centered at it are identified and the Rosenbluth weight 
$W_{\mathscr{N}} = \sum_{b^\prime} \exp{\parnths{-\beta A_{a^\prime_0 \land b^\prime}}}$ is accumulated.
From all candidate anchoring points, $b^\prime$, for the slip-spring emanating from chain end $a^\prime_0$, one
is chosen to serve as the end of the new slip-spring, $b^\prime_0$ with probability 
$P_{a^\prime_0 \land b^\prime_0} = \exp{\parnths{-\beta  A_{a^\prime_0 \land b^\prime_0}}} /W_{\mathscr{N}}$.
The new connectivity of the system is finally accepted with probability:
\begin{equation}
   P^{\rm accept}_{\mathscr{O}\to \mathscr{N}} = \min{\bracks{1,\exp{\parnths{-\frac{A_{a_0 \land b_0} - A_{a^\prime_0 \land b^\prime_0}}{k_{\rm B}T}}}
   \frac{W_{\mathscr{N}}}{W_{\mathscr{O}}}}} 
   \label{eq:kMC_creation_acceptance_criterion}
\end{equation}
If the creation of a new slip-spring $a^\prime_0 \land b^\prime_0$ is rejected, based on the criterion of 
eq \ref{eq:kMC_creation_acceptance_criterion}, the slip-spring $a_0 \land b_0$ remains as is, i.e. anchored to beads 
$a_0$ and $b_0$. Otherwise, the actual probability for a specific slip-spring destruction/creation event to take place 
at any point in the simulation where changes in slip-spring ends are considered is:
\begin{equation}
   P_{a_0\land b_0 \to a^\prime_0 \land b^\prime_0} = k_{{\rm hop}, a_0 \land b_0} 
   \Delta t_{\rm kMC} P_{\mathscr{O}\to \mathscr{N}}^{\rm accept} 
\end{equation}
where $k_{{\rm hop},a_0\land b_0}$ as defined in eq \ref{eq:hopping_rate_definition_nu_hop} and $\Delta t_{\rm kMC}$ 
is the elapsed time from the previous kMC step. By design the overall destruction/creation scheme satisfies 
microscopic reversibility. The corresponding proof can be found in Appendix F.

\subsection{Simulation Details} 
In this section we will discuss the parameterization of our methodology in a bottom-up approach.
During the development of the model we have tried to keep its adjustable parameters to an absolute minimum, without
introducing new parameters which cannot be mapped to physically relevant observables. We will 
elaborate on how values of the parameters are obtained with respect to structure, thermodynamics (e.g., equation of 
state), entanglement density, and friction/hopping rates, either from experimental evidence, or from more detailed 
simulation levels.
The systems considered were cis-1,4 polyisoprene (natural rubber) melts in the range of molar masses, 
$M = 21 \;{\rm kg/mol}$ to $M = 120 \; {\rm kg/mol}$. 
All simulations were carried out in the canonical statistical ensemble, at the temperature of $T=400 \text{K}$,
where rheological measurements are available.\cite{Macromolecules_27_4639}
The simulation box was cubic with varying edge length from $30 \; {\rm nm}$ to $100 \; {\rm nm}$.

The mean squared end-to-end distance of a PI chain can be either expressed in terms of the number of isoprene units or 
Kuhn lengths:
\begin{equation}
\left \langle R_{\rm e,0}^2\right \rangle = C_{\infty}4 N l^2 = N_{\rm Kuhn} b^2
\end{equation}
where $C_\infty$ is the characteristic ratio of PI, $N$ is the length of the chain measured in monomers, $l$ is the
root mean-squared average carbon-carbon bond length over an isoprene monomer and $b$ is the Kuhn length of polyisoprene. The factor 
of 4 is required because each isoprene contains four backbone bonds.
Using Mark's \cite{JAmChemSoc_88_4354} values for $C_{\infty}$ and $l$ of $4.7$ and $0.1485 \;{\rm nm}$, respectively,
we obtain $b = 0.958\;{\rm nm}$ and $2.21$ isoprene units per Kuhn length.\cite{JChemPhys_134_064906}
The effective length of an isoprene unit has been shown to have a slight temperature dependence,
\cite{JChemPhys_133_084903} however, we ignore this effect in our model.
Each bead in our representation consists of $n_{\rm Kuhns/bead} = 10$ PI Kuhn segments. Thus, its mass is 
$m_{\rm bead} = 1506 \:{\rm g/mol}$ and its characteristic mean-square end-to-end distance (if considered as a random 
walk) should be $\left( n_{\rm Kuhns/bead}b^2 \right) = 9.1776 \times 10^{-18} \:{\rm m}^2$. 

As far as the parameterization of the slip-springs is concerned, Fetters et al. \cite{Macromolecules_27_4639} have 
estimated the entanglement molecular weight of PI, $M_{\rm e}$ to be $5430 \; {\rm g/mol}$, thus allowing us 
to define the average required number of slip-springs present in the system as:
\begin{equation}
   \angles{N_{\rm ss}} = \frac{n \parnths{Z-1}}{2}
\end{equation}
with $n$ being the number of chains present and $Z = M/M_{\rm e}$.
Based on the chain discretization we have introduced above, the average distance between the slip-springs is roughly 
four coarse-grained beads. 
It is not at all clear or well defined what a slip-spring is. In this work we try to relate the parameters of the
slip-springs to quantities obtainable from atomistic simulations or experiments, so as to reduce the number of
adjustable parameters. However, one slip-spring may not correspond to exactly one entanglement. Other interpretations
can be invoked, i.e. thinking of the slip-spring as a soft constraint of the motion of the chain perpendicular to
its contour and not as a ``kink'' in the primitive path.

The stiffness of the slip-springs can be tuned by the parameter $l_{\rm ss}$, introduced 
in eq \ref{eq:sls_helmholtz_energy}. We set this equal to a reasonable estimate of the entanglement tube diameter of PI,
$l_{\rm ss} = \alpha_{\rm pp} \simeq 6.2 \;{\rm nm}$.\cite{MarkHandbook}

At this point, we should stress an important design rule for the degree of coarse-graining to be chosen. In order for the 
chains to preserve their unperturbed fractal nature in the melt, entropic springs along them should be stiffer than 
slip-springs, i.e., $l_{\rm ss} > \sqrt{n_{\rm Kuhns/bead}} b$. If this is not the case, $n_{\rm Kuhns/bead}$ should be 
tuned (i.e., the degree of coarse-graining should be adjusted) in order to ensure that entropic attraction between beads 
belonging to the same chain is at least an order of magnitude stronger than attraction due to slip-springs (connecting 
mostly beads belonging to different chains).

We have used the parameters of the Sanchez-Lacombe equation of state, given by Rudolf et al.\cite{PolymBull_34_109}
who have conducted $pVT$ measurements on several polymers under isothermal conditions, above their glass transition
temperature. 
These authors suggest $p^* = 383.0 \; {\rm MPa}$, $T^* = 631.2 \;{\rm K}$ and $\rho^{*} = 0.961 \;  {\rm g/cm}^3$
for molten polyisoprene of $M = 2594 \; {\rm g/mol}$.
Based on these values, the density of polyisoprene at $T=300 \;{\rm K}$ is estimated as
$\rho(300\:{\rm K})= 0.908 \;  {\rm g/cm}^3$.

Klopffer et al. \cite{Polymer_39_3445} have characterized the rheological behavior of a series of polybutadienes 
and polyisoprenes over a wide range of temperatures. The viscoelastic coefficients resulting from the 
time-temperature superposition principle were determined. A Rouse theory modified for undiluted 
polymers was used to calculate the monomeric friction coefficient, $ \zeta_0$.
The monomeric friction coefficient, $\zeta_0$, characterizes the resistance encountered by a monomer unit moving 
through its surroundings.
It was concluded that, within experimental error, a single set of Williams-Landel-Ferry (WLF) 
parameters\cite{JAmChemSoc_77_3701} at $ T_{\rm g}$ was adequate 
to characterize the relaxation dynamics irrespective of the vinyl content of the polybutadienes and polyisoprenes. 
These authors proposed that the variation of the monomeric friction coefficient with temperature can be given by: 
\begin{equation}
\log{\zeta_0 \left(T\right)} = \log{\zeta_\infty} + \frac{C_1^{\rm g}C_2^{\rm g}}
{T-T_{\rm g} + C_2^{\rm g}}
\end{equation}
with the parameters $C_1^{\rm g} = 13.5 \pm 0.2$, $ C_2^{\rm g} = 45 \pm 3 \;{\rm K}$, 
$ \log{\zeta_\infty} = -10.4 \;{\rm dyn\;s\;cm}^{-1}$ and $ T_{\rm g} = 211.15 \;{\rm K} $. 
At a temperature of 298 K, $\zeta_0(298\:{\rm K}) = 1.61 \times 10^{-6} \;{\rm dyn\;s\;cm}^{-1}$, while
at a temperature of 400 K, $\zeta_0(400\:{\rm K}) = 1.1508 \times 10^{-11} \;{\rm kg/s}$

If we think of the friction coefficient as being proportional to the mass of the entity it refers to, we can estimate
the friction coefficient of our coarse-grained beads as:
\begin{equation}
\zeta_{\rm bead} = \frac{m_{\rm bead}}{m_{\rm monomer}}\zeta_0
\end{equation}
where $m_{\rm monomer} = 68.12 \: {\rm g/mol}$ refers to the molar mass of a PI monomer.
The friction coefficient of a coarse-grained bead, at the temperature of $413\:{\rm K}$ is:
\begin{equation}
\zeta_{\rm bead} = 2.54 \times 10^{-10} \frac{\rm kg}{\rm s}
\end{equation}
Moreover, Doxastakis et al. \cite{JChemPhys_112_8687,JChemPhys_119_6883} have estimated the self-diffusion coefficient
of unentangled PI chains consisting of 115 carbon atoms, at $T = 413 \:{\rm K}$ to be:
\begin{equation}
D_{\ce{C115}} = 4.4 \times 10^{-11} \; \frac{{\rm m}^2}{\rm s}
\end{equation}
The length of these chains corresponds to $23$ monomers (or $11$ PI Kuhn segments). If we use the Rouse model 
\cite{JChemPhys_21_1272} to predict the diffusivity of these chains, based the parameters we have chosen above, that 
would be: 
\begin{equation}
D_{{\rm Rouse},\text{\ce{C115}}} = \frac{k_{\rm B} T}{N_{\rm monomers} \zeta_0} =  2.15 \times 10^{-11} 
\:\frac{{\rm m}^2}{\rm s}
\end{equation}
where $N_{\rm monomers}$ is the chain length measured in monomers and $\zeta_0$ the monomeric friction coefficient.
Our estimation of the self-diffusivity of Rouse chains, based on the model parameters we have
introduced coheres with what Doxastakis et al. have measured both experimentally and by all-atom MD simulations. 

Finally, as far as the implementation of the model is concerned, we integrate the equations of motion,
eq \ref{eq:method_simulations_brownian_integration_scheme}, by employing 
a timestep $\Delta t = 10^{-11} \; {\rm s}$ and up to at least $10 t_{\rm run}$ with $t_{\rm run} = 10^{-2} \; {\rm s}$. 
The kMC scheme is carried out every $n_{\rm kMC} = 100$ steps, thus resulting in $\Delta t_{\rm kMC} = 10^{-9} \; {\rm s}$. 
The frequency factor needed for the 
hopping rates of the kMC scheme is allowed to vary in the range $10^{-1} \le \nu_{\rm hop} \le 10 \; {\rm s}^{-1}$ 
and is kept constant for the whole range of molecular weights studied. In the following, its optimal value will be
thoroughly discussed.

\section{Results and Discussion}

\subsection{Hopping Dynamics}

\begin{figure}
   \centering
   \includegraphics[width=0.4\textwidth]{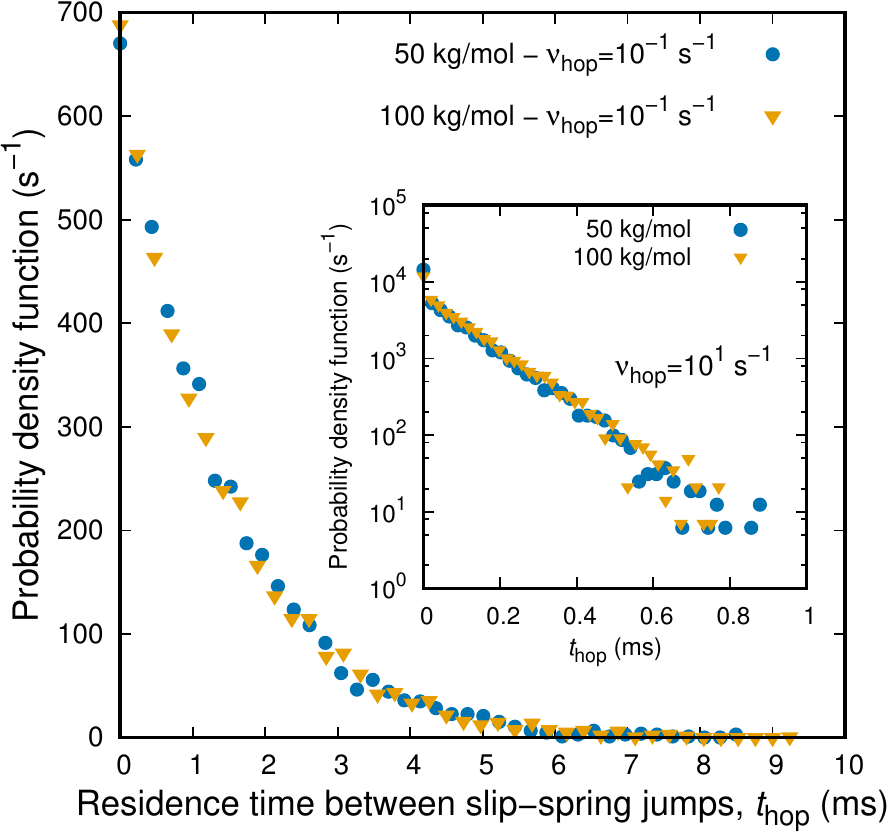}
   \caption{Probability density functions of the residence time between two successive slip-spring jumps, for two 
   different molar masses of the chains ($50$ and $100$ ${\rm kg/mol}$). Two different values of the hopping 
   frequency factor, $\nu_{\rm hop} = 10^{-1} \; {\rm s}^{-1}$ and $\nu_{\rm hop} = 10^{1} \; {\rm s}^{-1}$ are 
   considered in the main figure and in the inset to the figure, respectively.}
   \label{fig:residence_time_histo_vs_mw}
\end{figure}

At first we study the distribution of residence times between successive slip-spring jumps, in the course of a 
BD/kMC simulation. 
The probability density functions presented in Figure \ref{fig:residence_time_histo_vs_mw} include a number of salient 
points which should be discussed. Based on our picture of slip-spring hops as infrequent transitions, the hopping 
process should follow Poisson statistics, which is evident from the inset to the figure, where the distributions of 
residence times are clearly straight lines in semi-logarithmic axes. There are minor deviations at long time scales,
which are expected, since the population of slip-springs being stationary till that time is extremely small.
We have studied two molecular weights, namely $50 \; {\rm kg/mol}$ and $100 \; {\rm kg/mol}$. It is evident that the
distribution of residence times is independent of the molecular weight of the chains. The use of the same pre-exponential
frequency factor, $\nu_{\rm hop}$, yields indistinguishable results, as far as the individual motion of the slip-springs
is concerned. This is crucial for the methodology we have 
developed, since we will employ the same $\nu_{\rm hop}$ across different molecular weights and we expect the differences in 
chain dynamics to arise from the difference of their sizes (as they should) and not from tuning the slip-spring dynamics.

\begin{figure}[h]
   \centering
   \includegraphics[width=0.4\textwidth]{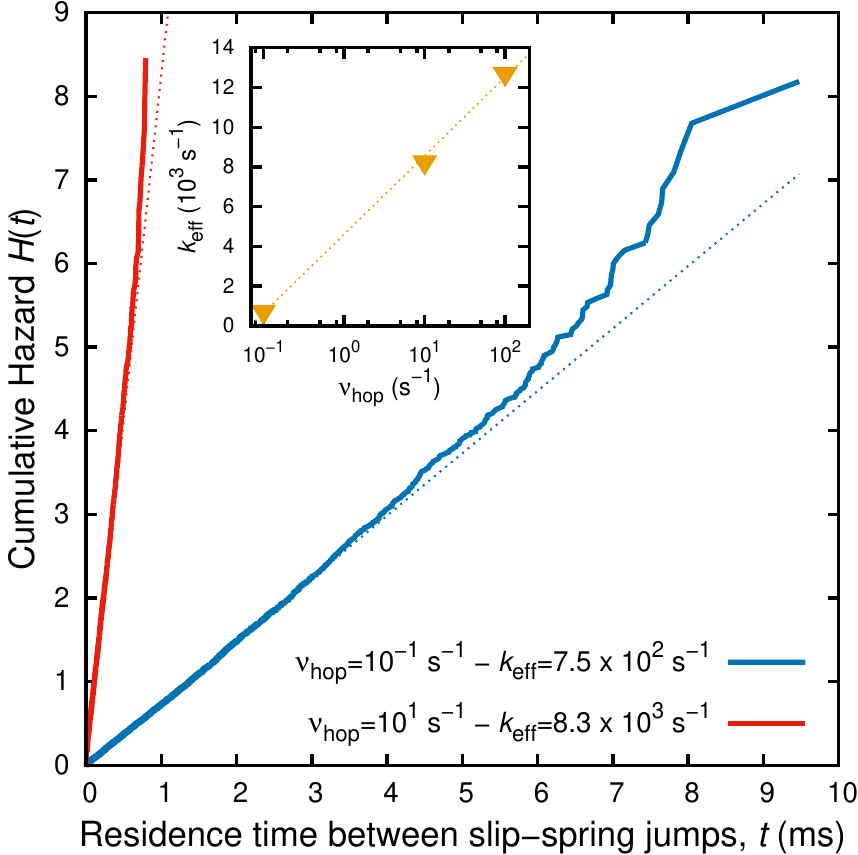}
   \caption{Hazard plot of the time between slip-spring transitions for systems consisting of two different chain lengths. 
            Linear fits are also included which provide an estimate of the slope, i.e. the effective hopping rate of the 
            slip-springs. In the inset to the figure, the dependence of the effective hopping rate,
             $k_{\rm eff}$, on the hopping pre-exponential frequency factor, $\nu_{\rm hop}$,
             is presented for the 100 kg/mol system.}
   \label{fig:hazard_plot_vs_rate_100K}
\end{figure}

In order to evaluate the effective hopping rate of the slip-springs, $k_{\rm hop}$, we invoke a hazard-plot analysis
\cite{JChemPhys_69_1010, Macromolecules_13_526, JPhysChemB_112_10619} of the hopping events which have taken place
along the BD trajectory.
The hazard rate, $h(t)$, is defined such that $h(t)\D t$ is the probability that a slip-spring which has survived a 
time $t$ in a certain state since its last transition, will undergo a transition (i.e., jump) at the time between $t$ 
and $\D t$. The cumulative hazard is defined as $H(t) = \int_0^t h\left(t^\prime\right)\D t^\prime$.
By assuming a Poisson process consisting of elementary transitions with first order kinetics from one state to 
the other, the Poisson rate can be extracted as the slope of a (linear) plot of the cumulative hazard versus the residence
time. Thus, the effective rate constant of the slip-spring hopping can be obtained from the hazard plot of the 
residence time of the slip-springs at a specific topology (i.e. the elapsed time between consecutive slip-spring jumps),
which is presented in Figure \ref{fig:hazard_plot_vs_rate_100K}.
At first, we can confirm that the hopping processes follow Poisson statistics, since the main parts of the plots are 
straight lines. However, there is a significant error related to fitting the hazard plots, due to the fact that at 
short time scales a lot of short-lived events are monitored.
Given eq \ref{eq:hopping_rate_definition_nu_0}, knowledge of $\nu_{\rm hop}$ which is a controlled parameter and the 
effective rate of slip-spring jumps from the hazard plots (i.e., $k_{\rm hop}$) allows us an estimation of the average 
free energy of the slip-springs, $A_{a_0 -b_0}$, entering eq \ref{eq:hopping_rate_definition_nu_0}.
The effective hopping rate, $k_{\rm hop}$, is barely sensitive to the frequency factor, $\nu_{\rm hop}$, as can be
observed in the inset to Figure \ref{fig:hazard_plot_vs_rate_100K}. 
The dependence of the former on the latter seems logarithmic, a rule of thumb that can be used if fine tuning of the 
dynamics of the model is needed.
This observation seems to be in apparent contradiction to the discussion connected with Figure 
\ref{fig:theoretical_rate}. However, it is fully justified, since when deriving a theoretical simplified estimate of 
the average hopping rate on the pre-exponential frequency factor, a number of assumptions were made. 
The entanglement creation and destruction processes were not taken into account. If, for example, a slip-spring to 
be destroyed does not find the proper candidates for its new anchoring points, it remains as is, perturbing the 
statistics. Moreover, slip-spring jumps are not fully independent events. The jump of a slip-spring at a specific 
timestep may help (or hinder) the remaining ones to relax in the course of the simulation or vice versa. Finally, 
the slope of the hazard plots that we have used in order to estimate the hopping rate is sensitive to the recrossing 
events taking place at short time scales (e.g. a specific slip-spring jumping back and forth). Summarizing, the 
effective hopping rate presented in Figure \ref{fig:hazard_plot_vs_rate_100K} incorporates cooperative effects 
between different slip-springs and reduction of slip-spring jumps due to slip-spring creation rules employed.
However, we expect that the linear dependence may be recovered for higher $\nu_{\rm hop}$ frequency factors.

\subsection{Structural Features}

\begin{figure}
   \centering
   \includegraphics[width=0.4\textwidth]{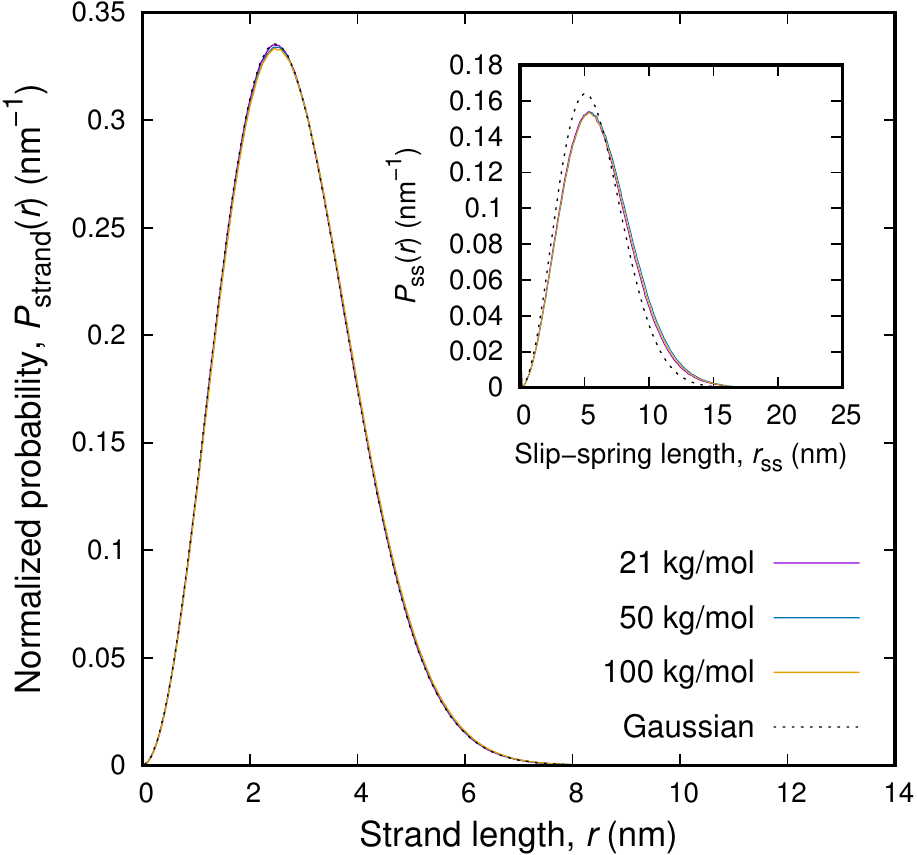}
   \caption{Distribution of the distance between successive beads (strands) along the contour of the chains, for
   three different molar masses. In the inset to the figure, the distribution of the length of the slip-springs 
   is presented.}
   \label{fig:strand_and_ss_length_distribution}
\end{figure}

We start by examining the structure of the melt chains at the level of individual strands, where we refer to a 
strand as the distance between successive beads along the contour of a chain. The distribution of the length 
of the strands (entropic springs) that connect the beads along the contour of the chains is depicted in Figure 
\ref{fig:strand_and_ss_length_distribution}. 
The entropic springs along the chain are considered Gaussian with an unperturbed length which equals the total length
of the Kuhn segments represented by a coarse-grained bead. The conformational features, at least at the strand level, 
continue to be respected during the BD simulation. Moreover, the distribution of strand lengths during the 
simulation coincides with a Gaussian distribution centered at the unperturbed strand length, as is theoretically 
expected. Our simulation scheme seems to produce trajectories of the system consistent with the imposed Helmholtz energy 
function, eq \ref{eq:helmholtz_strand_contribution}.

In the inset to Figure \ref{fig:strand_and_ss_length_distribution} we examine the distribution of slip-spring lengths. 
Slip-springs represent entanglements of a chain with its surrounding chains. The polymer tube model considers a 
tube formed around the primitive path of the chain, which fluctuates in time. Recent simulations have shown that 
the probability of finding segments of the neighboring chains inside the tube of the chain under consideration is 
Gaussian.\cite{Macromolecules_47_6077}
This is also the case in our simulations. The use of Gaussian entropic springs for describing the free energy of the 
slip-springs results in a Gaussian distribution of slip-spring lengths, conforming to the picture obtained by more 
detailed simulations and theoretical arguments. The distributions obtained from the simulations are slightly shifted
to higher distances. However, they are independent of the molecular weight of the chains under consideration.

\begin{figure}[h]
   \centering
   \includegraphics[width=0.4\textwidth]{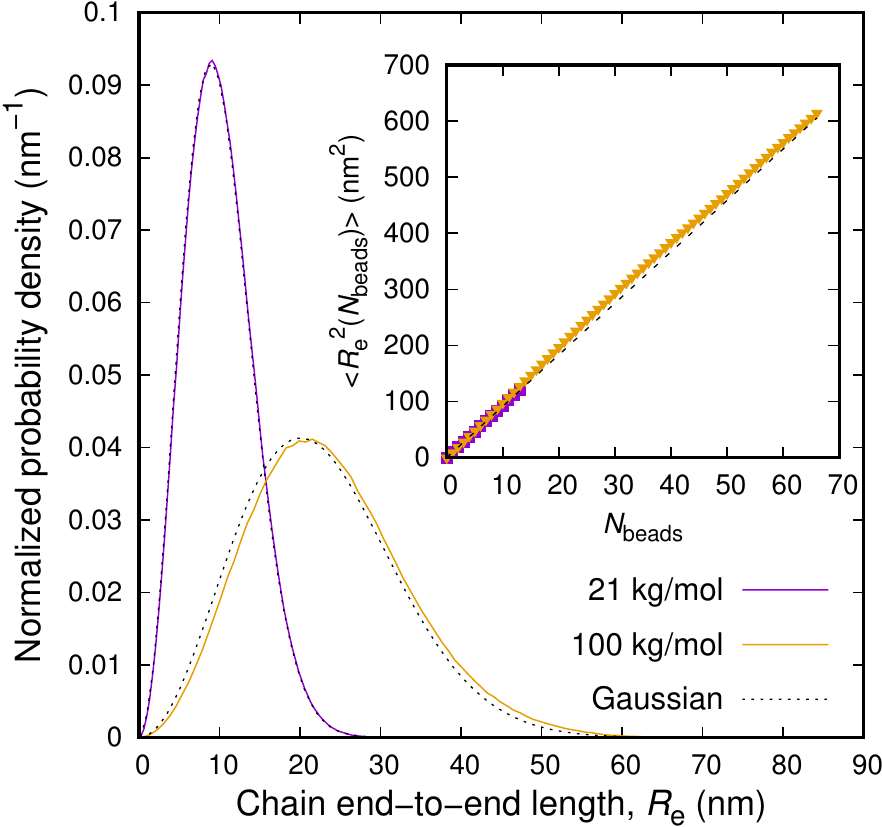}
   \caption{Distribution of the end-to-end distances of the chains for two different molecular weights. In the inset
   to the figure, the scaling of the end-to-end distance is presented, with respect to the number of beads, 
   $N_{\rm beads}$, of subchains along the original chains.}
   \label{fig:ete_vs_contour_mw}
\end{figure}

Finally, we examine the chain dimensions, as those can be quantified by the end-to-end length, $R_{\rm e}$. In Figure
\ref{fig:ete_vs_contour_mw} we examine the distribution of end-to-end distances for two molecular weights, lying at 
the extremes of our range of study. Shorter chains, of $M = 21 \; {\rm kg/mol}$, behave extremely close their 
unperturbed, Gaussian statistics. This is not the case for larger chains which 
become conformationally stiffer compared to the
theoretical prediction. End-to-end departures occur at distances commensurate or larger than the average distance 
between slip-spring anchoring beads. At this length scale, slip-springs bring up an extension of the chains. 
However, even this extension does not affect the scaling law $\sim N^{1/2}$ of the end-to-end length of the chains, as
can be observed in the inset to Figure \ref{fig:ete_vs_contour_mw}, where the end-to-end distance is calculated 
as a function of the number beads of sub-chains. Polymer chains, in both cases, behave as random walks, with a larger
``Kuhn'' step, though.
Our minimalistic approach has significantly reduced, but not eliminated, perturbations in chain 
conformations caused by the unphysical attraction introduced by the slip-springs. This attraction is not causing 
obvious problems under the considered conditions, but is always present in our simulations. The only way 
of restoring chain conformations is by introducing a compensating potential as that of Chappa et 
al.,\cite{PhysRevLett_109_148302} which fully compensates the effect of slip-springs on equilibrium properties.

\subsection{Thermodynamics}
Thermodynamics is naturally introduced in our model, via the nonbonded energy which is dictated by a suitable equation 
of state (e.g., the Sanchez-Lacombe equation of state). Several thermodynamic properties can be calculated. As a proof
of concept, the compressibility of our PI melts can be calculated. We can calculate the time evolution of local densities
at the cells of our computational grid, as presented in Appendix A. By treating each cell of the grid as a system 
capable of exchanging mass with its surroundings, compressibility can be estimated by the following fluctuation 
formula:
\begin{equation}
   \frac{\kappa_{\rm T}}{V} = \frac{\angles{\parnths{\delta n_{\rm Kuhns/cell}}^2}}{k_{\rm B}T \angles{n_{\rm Kuhns/cell}}^2}
   \label{eq:grand_canonical_compressibility}
\end{equation}
where the averages $\angles{...}$ are taken over all cells of the grid and along the trajectory of the simulation.
The averages appearing in the numerator and denominator of eq \ref{eq:grand_canonical_compressibility} are the variance
and the mean number of Kuhn segments assigned to each cell, respectively. It should be noted that local densities of 
grid cells are obtained via the smearing scheme introduced in Appendix A.
The isothermal compressibility, estimated from our simulations varies from $1.09 \times 10^{-4} {\rm bar}^{-1}$ 
to $1.04 \times 10^{-4} {\rm bar}^{-1}$ for samples from $21 \; {\rm kg/mol}$ to $120 \; {\rm kg/mol}$, respectively.
On the other hand, starting from the macroscopic equation of state, the compressibility is:
\begin{equation}
   \kappa_{\rm T} \equiv -\frac{1}{V} \tderv{\pderv{V}{p}}{T} = \tderv{\pderv{\ln{\tilde{\rho}}}{p}}{T}
\end{equation}
which is equal to $1.27 \times 10^{-4} {\rm bar}^{-1}$. 
The small difference (lower compressibility in our simulation) is probably attributable to the additional contributions
to the Helmholtz energy, i.e., entropy springs and slip-springs. Our model can faithfully reproduce the compressibility 
of polymeric melts with reasonable accuracy.
Realistic compressibility is missing from the majority of coarse-grained models and methods.

\subsection{Polymer Dynamics}

\begin{figure}[h]
   \centering
   \includegraphics[width=0.4\textwidth]{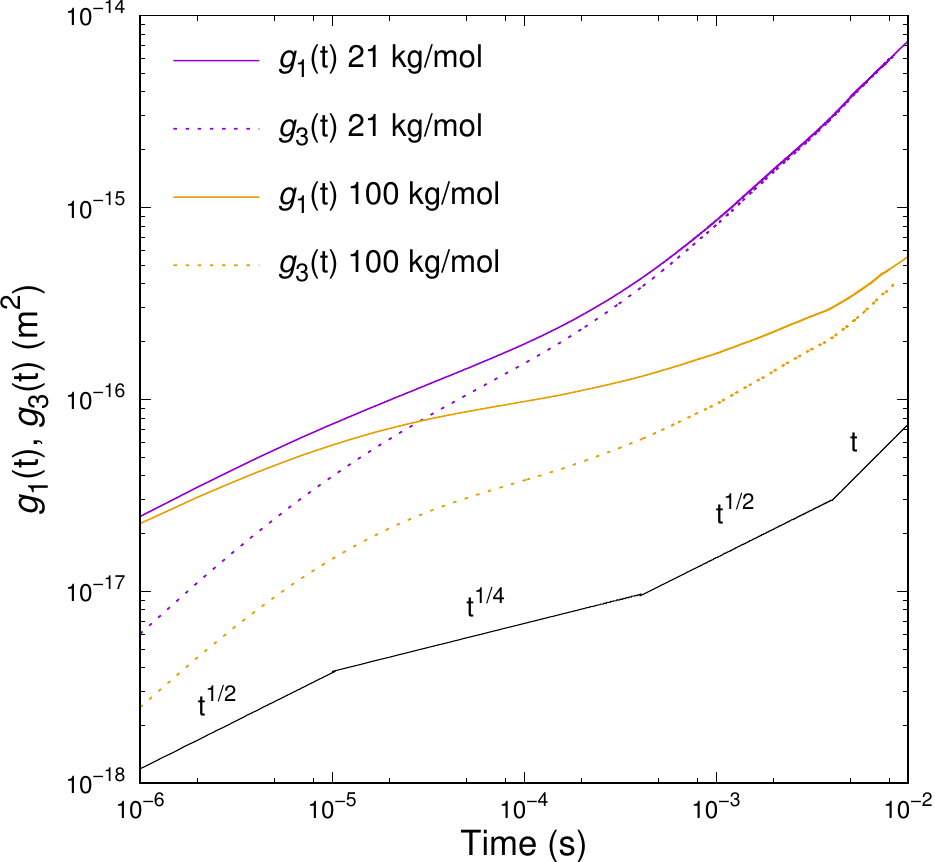}
   \caption{Time evolution of the mean-square displacements of the inner beads ($g_1 \parnths{t}$) and center of mass 
   of the chains ($g_3 \parnths{t}$) for $M = 21$ and $100 \; {\rm kg/mol}$. Slip-springs are present. Alongside, 
   the scaling regimes of the tube model are also included.}
   \label{fig:msd_vs_mw}
\end{figure}

A rigorous way of studying the mobility of a polymeric melt is to calculate the mean-squared displacement (MSD) of the
structural units of the chains. In order to avoid chain end effects,\cite{JChemPhys_116_436,Macromolecules_36_1376} 
only the innermost beads along the chain contribute to the calculations:
\begin{equation}
   g_1\parnths{t} = \frac{1}{2 n_{\rm inner} + 1} \sum_{i=N/2-n_{\rm inner}}^{i=N/2+n_{\rm inner}}
   \angles{\parnths{\mathbf{r}_{i} \parnths{t_0+t}-\mathbf{r}_{i}\parnths{t_0}}^2}
\end{equation}
with the value of the parameter $n_{\rm inner}$ quantifying the number of innermost atoms, on each side of the middle
segment of each chain that are monitored. In our case, $n_{\rm inner}$ is set in such a way that we track more than
half of the chain, excluding one-fifth of the chain close to one end and one-fifth close to the other end.
Moreover, the motion of the chains can be described in terms of the mean-squared displacement of their centers of mass,
$g_3 \parnths{t}$:
\begin{equation}
   g_3 \parnths{t} = \angles{\parnths{\mathbf{r}_{\rm cm}\parnths{t_0+t} - \mathbf{r}_{\rm cm}\parnths{t_0}}^2}
\end{equation}
Figure \ref{fig:msd_vs_mw} contains results for the time dependence of the mean-square displacements, $g_1 \parnths{t}$ 
and $g_3 \parnths{t}$ for two PI melts of different molecular weights. 

To a first approximation, the dynamics of short chains in a melt can be described by the Rouse model.\cite{JChemPhys_21_1272}
As far as the lower molecular weight chains are concerned, it can be seen in Figure \ref{fig:msd_vs_mw} that the 
mean-square center-of-mass displacement, $g_3\parnths{t}$, remains almost linear at all times; this means the intermolecular 
forces between polymers are too weak to affect diffusive behavior and play a minor role compared to the bonded 
interactions. However, small departures are fully justified due to the few slip-springs present even it the short-chain
system. The bead mean-squared displacements, $g_1\parnths{t}$, exhibit a subdiffusive 
behavior that arises from chain connectivity, and is characterized by a power law of the form $g_1\parnths{t} \sim t^{1/2}$. 
After an initial relaxation time where a change in $g_1\parnths{t}$ occurs, a regular diffusive regime is entered, 
where $g_1\parnths{t} \sim t$. This sequence of scaling trends is predicted by the Rouse theory. 
The limiting behavior of the chains' center-of-mass displacement can yield an estimate of the diffusivity of the chains:
$\lim_{t \to \infty} g_3 \parnths{t} = 6 D_{\rm cm} t$,
which has been found in excellent agreement with the diffusivity predicted by all-atom MD simulations of 
PI chains of the same molecular weight by Doxastakis et al.\cite{JChemPhys_112_8687}
The introduction of slip-springs in those short-chain systems does not seem to affect the scaling laws of the 
unentangled melt significantly. 

Moreover, Figure \ref{fig:msd_vs_mw} includes results for the mean-square displacements, $g_1\parnths{t}$ and 
$g_3\parnths{t}$, as a function of time, obtained from the simulation of a 100 kg/mol PI melt with slip-springs 
present. At short time scales, the segmental MSD curves, $g_1\parnths{t}$, for
both melts coincide. However at longer time scales, the characteristic signature of an entangled polymer system can be 
observed in the longer-chain system.
It can be seen that at short times the bead mean-square displacement, $g_1\parnths{t}$, shows a scaling 
regime with a power law $t^{1/2}$; at intermediate times, a regime with a power law  $t^{1/4}$ appears; eventually we 
observe a crossover to regular diffusion at long times. The mean-square displacement of the chain center-of-mass, 
$g_3\parnths{t}$, also exhibits subdiffusive behavior at intermediate times, with a scaling behavior $t^{1/2}$, 
as predicted by the tube model; at long times, regular diffusion is achieved. From the long time behavior of 
$g_3\parnths{t}$, we can estimate the longest relaxation time, $\tau_{\rm d}$. 
It should be noted that all transitions between scaling regimes are very smooth in our model.
Despite the fact that crossovers between the scaling regions are not accurately discerned in our results, the scaling 
of polymer melts, as expected from tube theory, is observed.

\begin{figure}[h]
   \centering
   \includegraphics[width=0.4\textwidth]{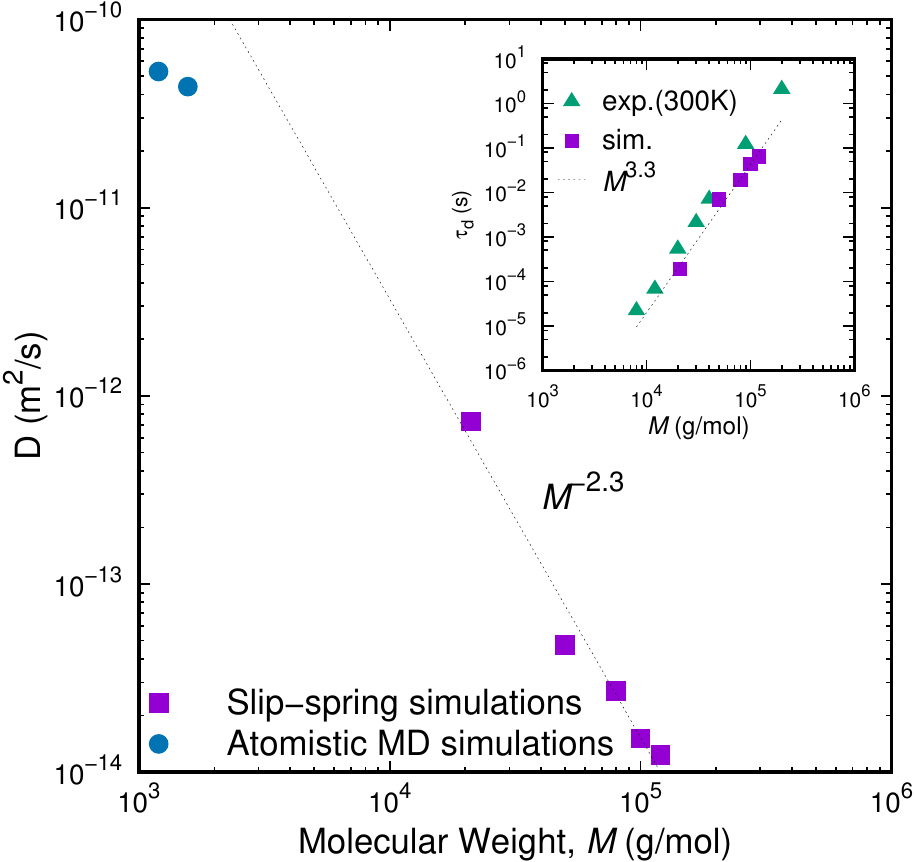}
   \caption{
      Diffusion coefficient, $D$, plotted against the molecular weight of the chains, $M$. The dashed line provides 
      a guide to the eye showing the $M^{-2.3}$ scaling law.\cite{PhysRevLett_83_3218} 
      Results from earlier atomistic simulations are also included.\cite{JChemPhys_119_6883}
      In the inset to the figure, the disengagement time, $\tau_{\rm d}$, is plotted as a function of the molecular
      weight, $M$. The dashed line is a power law fit with an exponent 3.3 and experimental estimates of $\tau_{\rm d}$
      at a lower temperature are also included.\cite{JRheol_52_801}
   }
   \label{fig:diff_coeff_vs_mw}
\end{figure}

From the long time behavior of the mean-square displacement of the center of mass of the chains, $g_3\parnths{t}$, 
we compute the diffusion coefficient and the longest relaxation time (disengagement time), $\tau_{\rm d}$,
as a function of molecular weight. The results are presented in Figure \ref{fig:diff_coeff_vs_mw}.
As expected, $D$ is a monotonically decreasing function of $M$ (or $N$) with a power scaling law of -2 (tube model)
or -2.3 (experimental observations \cite{PhysRevLett_83_3218}) for well entangled melts. 
Our slip-spring simulations can recover the aforementioned scaling behavior
of the diffusion coefficient and their predictions lie in the correct order of magnitude, cohering with the results 
from detailed atomistic simulations of Doxastakis et al.\cite{JChemPhys_119_6883} In the inset to Figure 
\ref{fig:diff_coeff_vs_mw}, the molecular weight dependence of the disengagement time, $\tau_{\rm d}$, is presented. 
For large $M$ (or $N$), the longest relaxation time should grow as $N^3$ (as predicted from the tube model) or 
with slightly higher exponents (around 3.5\cite{JRheol_52_801}). Our results fit well to a scaling law of 3.3 which
is reasonable. Experimental results at a significantly lower temperature\cite{JRheol_52_801} are also included for 
comparison. The exponent of the experimental measurements is slightly larger (around 3.5) for PI measurements at a
temperature of 300 K.

\subsection{Rheology}

Linear rheological properties can be characterized through the shear relaxation modulus, 
$G\parnths{t} = \tau_{xy} \left(t\right) / \gamma $, with $\gamma$ being a small shear deformation and $xy$ two 
orthogonal axes. In computer simulations, the most convenient way of evaluating 
$G(t)$ is to use the fluctuation-dissipation theorem:\cite{Likhtman_Viscoelasticity}
\begin{equation}
   G\parnths{t} = \frac{V}{k_{\rm B}T} \left \langle \tau_\albe \parnths{t_0 + t} \tau_\albe \parnths{t_0}
   \right \rangle
   \label{eq:goft_definition}
\end{equation}
where $\albe$ stand for any two orthogonal directions. One can show that the stress relaxation after any small 
deformation will be proportional to $G\parnths{t}$, and thus $G\parnths{t}$ fully characterizes the linear rheology of 
a polymeric system with given external parameters. However, the stress autocorrelation function (acf) is notoriously difficult to 
calculate due to huge fluctuations at long times. In order to improve the accuracy of our calculations, we average 
$\left \langle \tau_\albe \parnths{t_0 + t} \tau_\albe \parnths{t_0} \right \rangle$ 
over all possible ways of selecting a pair of perpendicular axes $\alpha$ and $\beta$.\cite{Likhtman_Viscoelasticity}
We compute the time correlation functions in eq \ref{eq:goft_definition} by using the multiple-tau correlator 
algorithm of Ram\'irez et al.\cite{JChemPhys_133_154103}

\begin{figure}
   \centering
   \includegraphics[width=0.4\textwidth]{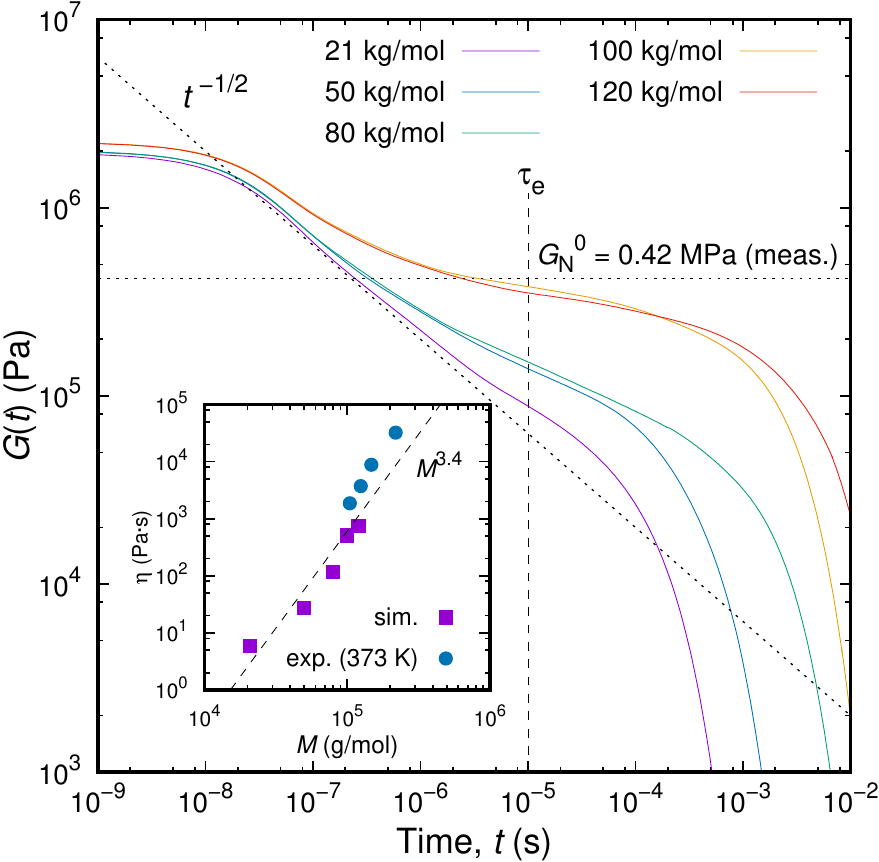}
   \caption{Temporal evolution of the shear relaxation modulus, $G\parnths{t}$, for a series of PI melts at 
   $T = 400 \;{\rm K}$. In the inset to the figure, the zero shear viscosity of the melts is presented, alongside 
   experimental results, at a slightly lower temperature ($T = 373 \;{\rm K}$).\cite{Macromolecules_17_2767}}
   \label{fig:b3kMC_goft_vs_mw}
\end{figure}

Let us first consider Figure \ref{fig:b3kMC_goft_vs_mw}, which displays the time evolution of the shear relaxation modulus,
as defined in eq \ref{eq:goft_definition}, for PI melts of different chain lengths.
The initial value of the modulus is $\rho k_{\rm B}T$, with $\rho$ being the mean density of the polymer melt (slightly
higher for longer chains). The relaxation of the shortest chains (21 kg/mol) considered is typical of unentangled polymer 
melts, and is close to the shear relaxation modulus predicted by the Rouse model. 
In particular, the intermediate time scale behavior of the stress acf is consistent with the Rouse model scaling,
$G(t) \sim t^{-1/2}$, while the long time decay is exponential, with a time constant characterizing the longest stress
relaxation time in the system. It is clear that the longest relaxation time from the stress autocorrelation function
follows the same power law scaling with $N$ as the end-to-end vector relaxation time.
Upon increasing the chain length, a plateau starts to appear; this indicates that the viscoelastic character of polymer 
melts is captured by our model. 
However, we should note the log-log scale of Figure \ref{fig:b3kMC_goft_vs_mw}, which implies that the rubbery plateau
still involves significant relaxation, even for $M = 120\;{\rm kg/mol}$.

The shear viscosity, $\eta$, can be computed from the stress relaxation function through a Green-Kubo relation:
\begin{equation}
   \eta = \int_{0}^{\infty} G_{\alpha \beta} (t) {\rm d}t
\end{equation}
In the inset to Figure \ref{fig:b3kMC_goft_vs_mw} we present the viscosity as a function of the molecular weight.
The viscosity, $\eta$, is a monotonically increasing function of $N$, which exhibits a scaling behavior $N^{3.4}$, as 
experimental data suggest.\cite{Macromolecules_17_2767} It should be noted that the tube model predicts a $N^3$ 
dependence of viscosity. Our simulation results seem to cohere with the $N^{3.4}$ scaling. Moreover, there is 
very good quantitative agreement with available experimental data for the same system. Experimental results have been
obtained at a slightly lower temperature, that being the reason for higher viscosity values for the same molecular
weight. It is very promising that our methodology, parameterized in a bottom-up approach, captures the correct values
of viscosity without any adjustment or rescaling of the shear relaxation function.

The precise scaling of viscosity of polymer melts is still elusive and remains an open question for theoretical, simulation 
and experimental studies.\cite{RheolRev_2003_197} The results presented by Auhl et al.\cite{JRheol_52_801}
imply a $M^{3.6}$ scaling law for polyisoprene melts, in contrast to the established $M^{3.4}$ scaling of the early 
experimental studies. A slip-spring simulation scheme can capture modes of entangled motion beyond pure reptation 
($M^3$ scaling). In linear response contour length fluctuation (CLF), the Brownian fluctuation of the length of the 
entanglement path through the melt, modifies early-time relaxation. Similarly, the process of constraint release 
(CR), by which the reptation of surrounding chains endows the tube constraints on a probe chain with finite lifetimes, 
contributes to the conformational relaxation of chains at longer times. Both processes of CLF and CR exist in 
slip-spring models and their relative importance contributes to the quantitative prediction of linear rheology. 
In our case, achieving a $M^{3.4}$ scaling law can be considered as a balance between all relaxation mechanisms 
involved in the rheology of linear melts.\cite{PhysRevLett_81_725,Macromolecules_35_6332}
Thus, compared to the recent experimental results, we may reasonably underestimate the exponent (slope in the 
log-log plot of $\eta$ versus $N$), given that the slope in the log-log plot of $D$ versus $N$ is reasonable. 
Furthermore, our simulation results are in good agreement with previous slip-spring simulations of other research 
groups,\cite{JChemPhys_137_154902, JChemPhys_138_194902} which exhibit the same or even lower scaling exponent.

\begin{figure}
   \includegraphics[width=0.4\textwidth]{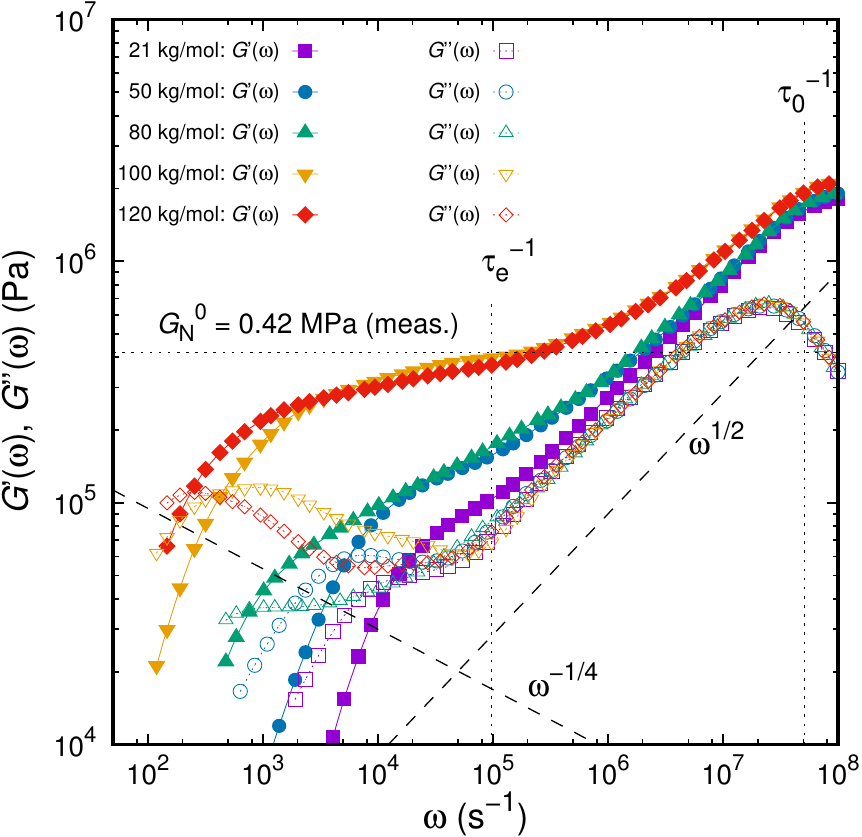}
   \caption{Loss and storage moduli for monodisperse polyisoprene samples of different molecular weights.
   Measured plateau modulus from Fetters et al.\cite{Macromolecules_27_4639} 
   is also included.}
   \label{fig:b3DkMC_gofomega_vs_mw}
\end{figure}

In experimental practice, the cosine and sine Fourier transforms of $G\parnths{t}$ can be measured:
\begin{align}
   G^\prime \parnths{\omega} & = \omega \int_0^\infty \sin{\parnths{\omega t}} G\parnths{t} \D t \\
   G^{\prime \prime} \parnths{\omega} & = \omega \int_0^\infty \cos{\parnths{\omega t}} G\parnths{t} \D t
\end{align}
by applying small oscillatory deformation and recording the in-phase and out-of-phase responses 
$G^\prime \parnths{\omega}$ and $G^{\prime \prime} \parnths{\omega}$, respectively, which are called storage and loss
moduli. A straightforward way to obtain the Fourier transform of $G\parnths{t}$ is to fit it with a sum of exponentials:
\begin{equation}
   G\parnths{t} = \sum_{i=1}^{m} G_i \exp{\parnths{-\frac{t}{\tau_i}}} 
\end{equation} 
where $G_i$ and $\tau_i$ are the amplitude and the relaxation time of the mode number $i$ (often called the Maxwell
modes). Our results for the storage and loss moduli of the PI melts under study are presented in Figure
\ref{fig:b3DkMC_gofomega_vs_mw}.

The forms of the $G^\prime \parnths{\omega}$ and $G^{\prime \prime}$ curves for PI are typical of terminal and 
plateau responses for monodisperse linear polymers in the entangled 
region.\cite{Macromolecules_14_1668,Macromolecules_17_2767}
For short chains of $M = 21 \; {\rm kg/mol}$, relaxation is rather fast; the longest relaxation time being on the 
order of $10^{-4} \; {\rm s}$. For longer chains, several important features arise.
At low frequencies, $G^\prime \sim \omega^2$ and $G^{\prime\prime} \sim \omega$, providing values of the zero-shear
viscosity $\eta$ and steady-state recoverable compliance, $J_{\rm e}^0$:\cite{FerryViscoelasticProperties}
\begin{align}
   \eta & = \lim_{\omega \to 0} \frac{G^{\prime\prime \parnths{\omega}}}{\omega} \\
   J_{\rm e}^0 & = \frac{1}{\eta^2} \lim_{\omega \to 0}\frac{G^\prime \parnths{\omega}}{\omega}
\end{align}
When moving to higher frequencies, $G^{\prime\prime} \parnths{\omega}$ passes through a shallow maximum, 
$G^{\prime\prime}_{\rm max}$ at $\omega_{\rm max}$. The terminal loss peak becomes progressively more well-defined 
(especially for $100$ and $120 \; {\rm kg/mol}$),  i.e., further separated from the transition response, with 
increasing chain length. The plateau modulus, $G_{\rm N}^{0,\rm sim}$ can be determined by an integration over a fully
resolved terminal loss peak:\cite{Macromolecules_17_2767}
\begin{equation}
   G_{\rm N}^{0,\rm sim} = \frac{2}{\pi} \int_{-\infty}^{\infty} G^{\prime\prime} \parnths{\omega} \D \ln{\omega}
   = 0.39 \; {\rm MPa}
\end{equation}
which is in excellent agreement with the measured plateau modulus, $G_{\rm N}^{0} = 0.42 \; {\rm MPa}$.
The majority of the plateau modulus values reported in the literature have been obtained by this 
method.\cite{Macromolecules_27_4639, Macromolecules_35_10096,Polymer_47_4461} 
Even for the longest chains, the loss modulus (open symbols in Figure \ref{fig:b3DkMC_gofomega_vs_mw}) shows that 
some relaxation is nevertheless present, revealing a power law close to $\omega^{-1/4}$.

It is often observed that rheological response measured at different temperatures can be reduced to a single master 
curve at the reference temperature $T_0$ if one shifts the time (or frequency) appropriately. 
According to the time-temperature superposition (TTS) principle, 
the complex relaxation modulus of rheologically simple polymers, defined as 
$G^* \parnths{\omega} = G^\prime \parnths{\omega} + i G^{\prime\prime} \parnths{\omega}$,
measured at different temperatures, obeys:
\begin{equation}
   G^* \parnths{\omega,T} = \mathcal{B}\parnths{T} G^*\parnths{\mathcal{A}\parnths{T} \omega, T_0}
\end{equation}
where $T_0$ is the reference temperature, and $\mathcal{A}\parnths{T}$ and $\mathcal{B}\parnths{T}$ are horizontal
and vertical shift factors for this particular reference temperature.\cite{FerryViscoelasticProperties} 
The range of validity of TTS include only those processes whose rates are proportional to the rate of one particular
basic process (e.g. monomer diffusion), which in turn depends on temperature. In polymer
melts, the shift factors are usually given by empirical equations:
\begin{align}
   \log_{10}{\mathcal{A}\parnths{T}} &= \frac{C_1 \parnths{T-T_0 +C_2/M}}{T -T_{\infty} + C_2/M} \\
   \mathcal{B}\parnths{T} & = \frac{\rho\parnths{T} T}{\rho\parnths{T_0}T_0}
\end{align}
with the former being the Williams-Landel-Ferry (WLF)\cite{JAmChemSoc_77_3701} equation (with two material 
constants $C_1$ and $C_2$), and the latter incorporating the dependence of polymer density on temperature.
In our case, we use the experimentally derived curves of Auhl et al.\cite{JRheol_52_801}, horizontally and vertically
shifted to our temperature of interest ($400 \; {\rm K}$). The values of the relevant parameters are: 
$T_0 = 25 \; ^\circ{\rm C}$, $T_\infty = -114.03 \; ^\circ{\rm C}$, $C_1 = 4.986$ and $C_2 = 14.65$ with $M$ measured
in ${\rm kg/mol}$.\cite{JRheol_52_801}
The relation for the temperature dependence of the PI density was obtained from the Sanchez-Lacombe equation of state
used for describing the nonbonded interactions of the our coarse-grained model.

\begin{figure}
   \includegraphics[width=0.4\textwidth]{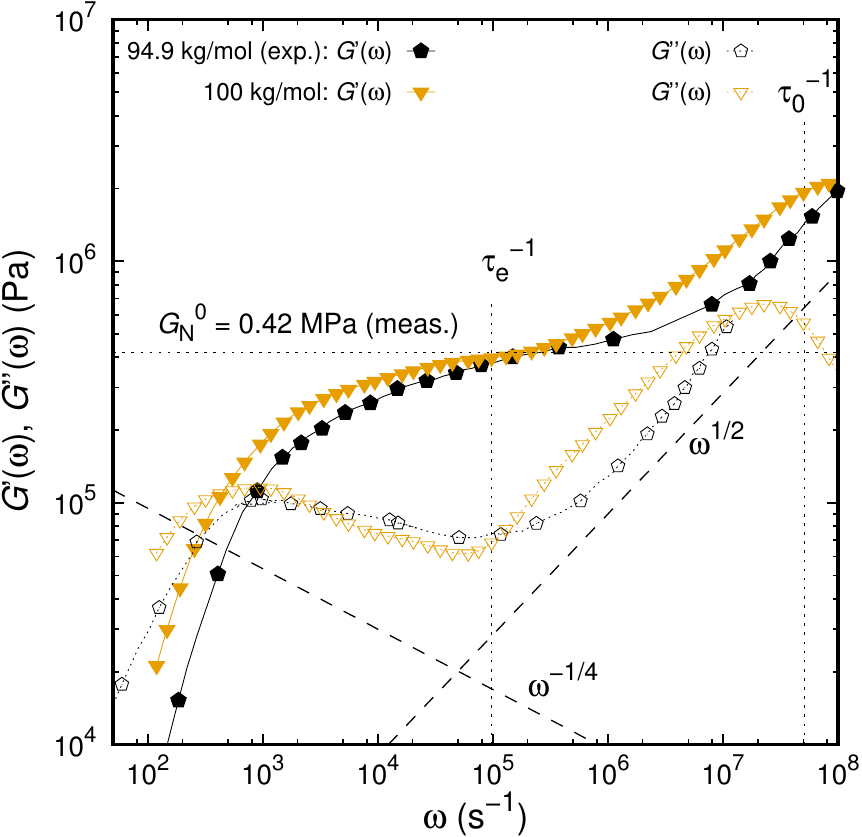}
   \caption{
   Loss and storage moduli for monodisperse polyisoprene samples of 100 kg/mol.
   Experimental results have been taken from Auhl et al.\cite{JRheol_52_801} and have been properly shifted 
   to the simulation temperature. 
   Measured plateau modulus from Fetters et al.\cite{Macromolecules_27_4639} 
   is also included.}
   \label{fig:b3DkMC_gofomega_vs_mw_experimental}
\end{figure}

In Figure \ref{fig:b3DkMC_gofomega_vs_mw_experimental}, the predicted moduli of the polyisoprene
sample of 100 kg/mol are compared against the experimental measurements of Auhl et al\cite{JRheol_52_801}
shifted to the temperature of the simulations following the TTS principle described in the previous paragraph. 
Despite the fact that experimental measurements have been conducted for a slightly lower molecular weight, 
$M=94.9 \; {\rm kg/mol}$, the predicted moduli are in quantitative agreement with the experimental observations.
The minor discrepancies observed can be attributed to several reasons. In the case of our simulations, 
polyisoprene melts are strictly monodisperse, which does not hold for the experimental studies, where a finite 
polydispersity exists. 
The plateau value obtained from the simulation curve around $1/\tau_{\rm e}$ is the same as the one measured by
rheological experiments, as well as the shape of the whole storage modulus curve in the frequency range of 
$10^2$ to $10^6 \; {\rm s}^{-1}$. 
The use of the kMC scheme for the slip-spring hopping introduces a relaxation time scale on the order
of $10^{-6} \;{\rm s}$, which accounts for the different slope of the simulated storage modulus curve with respect 
to the experimental in the range $\omega > 10^6 \; {\rm s}^{-1}$. 
As far as the loss modulus is concerned, discrepancies start to appear at high frequencies (close to the shortest 
relaxation time, $\tau_0 \simeq \zeta b_{\rm K}^2 / \parnths{3 \pi^2 k_{\rm B}T}$), which correspond to glassy 
behavior that cannot be captured by a coarse grained model which omits truly microscopic, atomistic processes. 
Experimental loss modulus curves are monotonically increasing with a power law $\omega^{1/2}$ at high frequencies, 
which is not the case for our soft, coarse-grained model.

\section{Summary and Conclusions}
The first steps towards a consistent coarse-grained model capable of reproducing the rheological properties of 
polymer melts have been presented.
The methodology and the corresponding computer code have been developed for the case of a pure polymer melt.
Chains are modeled as sequences of beads, each bead encompassing approximately 10 Kuhn segments. The Helmholtz
energy of the system is written as a sum of three contributions: entropy springs, representing the entropic elasticity 
of chain strands between beads; slip-springs, representing entanglements; and non-bonded interactions. The Helmholtz 
energy of non-bonded interactions is estimated by invoking an arbitrary equation of state and is computed as a 
functional of the local density by passing an orthogonal grid through the simulation box. 
Slip-springs are envisioned as connecting nodes on different polymer chains or on the same chain. 
Equations for a stochastic description of the dynamics are derived from the coarse-grained Helmholtz energy function. 
All beads execute Brownian motion in the high friction limit. The ends of the slip-springs execute
thermally activated hops between adjacent beads along chain backbones, these hops
being tracked by kMC simulation. In addition, creation/destruction processes are included for the slip-springs. 
A slip spring is destroyed when one of its ends slips past the free end of a polymer chain. A new slip spring is formed 
when a chain end captures a bead of another chain lying within a certain radius from it,
according to a prescribed rate constant. Parameters needed in the model are derived
from experimental volumetric and melt viscosity data or from atomistic molecular
dynamics simulations. Initial configurations for the network are obtained from  MC simulations of linear melts. 
Tests of the simulation code on molten linear (non-crosslinked) cis-1,4 polyisprene of high molar mass at
equilibrium have given satisfactory results for the mean square displacement of beads and for the shear relaxation 
modulus. 

By following the rigorous procedures described before, a consistent methodology has been set in place.
To implement this methodology we have developed a general purpose in-house software program to simulate 
polymer melts of abstract geometry. 
Predicting the rheology of binary blends of chains of different geometries poses formidable challenges for tube and slip-link models. 
\cite{Macromolecules_49_4964} 
Nonlinear rheology can also be studied with the formulation developed in this work and strain rate experiments
can be conducted.\cite{Macromolecules_50_386}
The methodology has already been implemented to study tensile deformations of crosslinked cis-14 
polyisoprene, with very encouraging results.\cite{JPhysConfSer_738_012063}
Morover, it can be readily extended to systems containing solid surfaces, particles or 
cavities.\cite{ARCME_2017}
Therefore, when all relevant relaxation processes are correctly treated, the viscoelastic response of the melt should 
be fully obtained from the simulation findings.

\section{Appendix A: Discrete Nonbonded Interactions Scheme for Slip-Spring Simulations}

\begin{figure}
   \centering
   \includegraphics[width=0.4\textwidth]{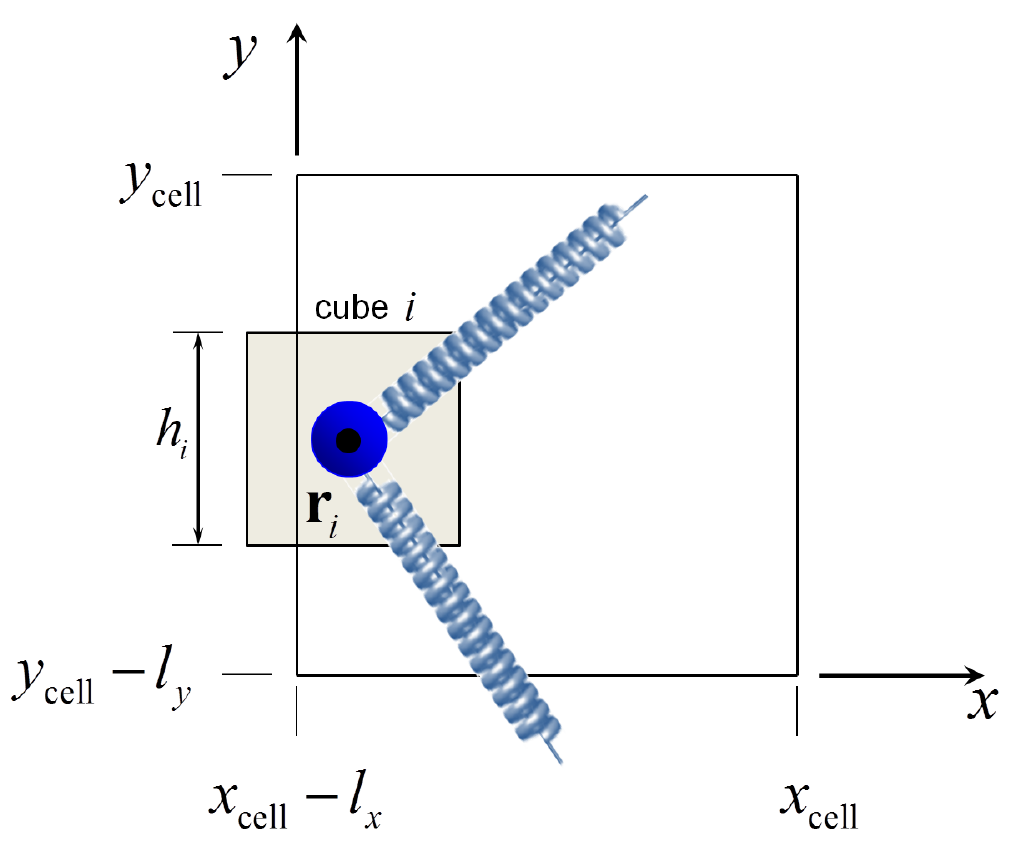}
   \caption{Schematic representation of a grid cell and a nodal point with its surrounding cube.}
   \label{fig:discrete_scheme_grid}
\end{figure}

The simplest option for relating the positions and masses of the beads to $\rho_{\rm cell}$ is to envision each 
bead $i$ as a cube containing mass $m_i$, of edge length $h_i$, centered at position 
$\mathbf{r}_i = \left(x_i, y_i, z_i \right)$, as shown in Figure \ref{fig:discrete_scheme_grid}.
Our scheme was designed to permit fast analytical calculation of nonbonded forces. It ensures 
continuity of the nonbonded forces as functions of bead position, but not of the derivatives of these forces (which
are not needed in our simulation setup). It has proven satisfactory for use in the Brownian Dynamics simulations
conducted in this work and can be readily extended to systems containing solid surfaces, particles or cavities.
Other schemes for determining the instantaneous density field from the particle positions,\cite{Macromolecules_46_4670}
including gridless ones,\cite{JChemPhys_123_144102} seem not to be straightforwardly usable in the aforementioned
cases.

The cell dimensions along the $x$, $y$ and $z$ directions will be 
denoted as $l_x$, $l_y$, and $l_z$, respectively. We will focus on a cell extending between $x_{\rm cell} - l_x$ and
$x_{\rm cell}$ along the $x$-direction, between $y_{\rm cell} - l_y$ and $y_{\rm cell}$ along the $y$-direction, and
between $z_{\rm cell} - l_z$ and $z_{\rm cell}$ along the $z$-direction. In the regular grid considered, if 
$\left(0,0,0\right)$ is taken as one of the grid points, $x_{\rm cell}$, $y_{\rm cell}$, and $z_{\rm cell}$ will be
integer multiples of $l_x$, $l_y$ and $l_z$, respectively.
In the following, we assume that $h_i < \min{\left(l_x, l_y, l_z \right)}$.

The mass contributed by the node to the cell is:
\begin{equation}
   m_{i,{\rm cell}} = m_i \frac{V_{{\rm cube}\;i\;\cap\;{\rm cell}}}{V_{{\rm cube }\;i}}
   \tag{A1}
   \label{eq:mass_cell_def}
\end{equation}
with $V_{{\rm cube}\;i\;\cap\;{\rm cell}}$ being the volume of the intersection of cube $i$, associated with bead $i$,
and the considered cell, while $V_{{\rm cube }\;i} = h_i^3$ is the volume of cube $i$.

Under the condition $h_i < \min{\left(l_x, l_y, l_z \right)}$, $V_{{\rm cube}\;i\;\cap\;{\rm cell}}$ is obtainable as:
\begin{align}
   V_{{\rm cube}\;i\;\cap\;{\rm cell}} = & 
   \max{\left\{ \left[  \min{\left(x_i + \frac{h_i}{2}, x_{\rm cell} \right)}
                      - \max{\left(x_i - \frac{h_i}{2}, x_{\rm cell} - l_x \right)} 
   \right],0 \right\}} 
   \nonumber \\
   \times & \max{\left\{ \left[  \min{\left(y_i + \frac{h_i}{2}, y_{\rm cell} \right)}
                      - \max{\left(y_i - \frac{h_i}{2}, y_{\rm cell} - l_y \right)} 
   \right],0 \right\}} 
   \nonumber \\
   \times & \max{\left\{ \left[  \min{\left(z_i + \frac{h_i}{2}, z_{\rm cell} \right)}
                      - \max{\left(z_i - \frac{h_i}{2}, z_{\rm cell} - l_z \right)} 
   \right],0 \right\}} 
   \tag{A2}
   \label{eq:vcube_cap_cell_def}
\end{align}
As defined by eq \ref{eq:vcube_cap_cell_def}, $V_{{\rm cube}\;i\;\cap\;{\rm cell}}$ is a linear function of the bead
coordinates. Clearly, if cube $i$ lies entirely within the cell, 
$V_{{\rm cube}\;i\;\cap\;{\rm cell}} = V_{{\rm cube}\;i}$ and, consequently, $m_{i,{\rm cell}}=m_i$.
If however, the borders of the cube $i$ intersect the borders of the considered cell, then node $i$ will contribute a 
mass $m_{i,{\rm cell}} < m_i$ to the cell. The total mass contributed by bead $i$ to all cells in which it participates
will always be $m_i$.

The density $\rho_{\rm cell}$ in the considered cell is estimated as:
\begin{equation}
   \rho_{\rm cell} = \frac{1}{V_{\rm cell}^{\rm acc}} \sum_{i} m_{i,{\rm cell}}
   \tag{A3}
   \label{eq:cell_density_def}
\end{equation}
Clearly, only beads $i$ whose cubes have a nonzero overlap with the considered cell will contribute to the summation of
eq \ref{eq:cell_density_def}. The position vectors $\mathbf{r}_i$ of these nodal points will necessarily lie within
the considered cell or its immediate neighbors.

The precise conditions for cube $i$ to have common points with the considered cell are:
\begin{align}
   & x_{\rm cell} - l_x < x_i + \frac{h_i}{2} < x_{\rm cell} + h_i \nonumber \\
   & y_{\rm cell} - l_y < y_i + \frac{h_i}{2} < y_{\rm cell} + h_i \nonumber \\
   & z_{\rm cell} - l_z < z_i + \frac{h_i}{2} < z_{\rm cell} + h_i 
   \tag{A4}
\end{align}
According to the above approach, the force on node $i$ due to nonbonded interactions is:
\begin{equation}
   \mathbf{F}_i = - \nabla_{\mathbf{r}_i} A_{\rm nb} = - \sum_{{\rm cells}\;\cap{\rm cube}\;i}
   V_{{\rm cell}}^{\rm acc} 
   \left. \frac{\partial a_{\rm vol}\left(\rho, T\right)}{\partial \rho}\right|_{\rho = \rho_{{\rm cell}}}
   \nabla_{\mathbf{r}_i} \rho_{{\rm cell}}
   \tag{A5}
\end{equation}

From eqs \ref{eq:mass_cell_def}, \ref{eq:vcube_cap_cell_def} and \ref{eq:cell_density_def} one can obtain the 
components of $\nabla_{\mathbf{r}_i} \rho_{\rm cell}$. Along the $x$ direction,
\begin{equation}
   \small
   \frac{\partial}{\partial x_i} \rho_{\rm cell} = 
   \begin{dcases}
      0 & \mbox{if } x_i \le x_{\rm cell}- l_x - \frac{h_i}{2} \\
      \frac{m_i}{V_{\rm cell}^{\rm acc}} \frac{V_{{\rm cube}\;i\;\cap\;{\rm cell}}}{V_{{\rm cube}\;i}}
      \frac{1}{x_i + \frac{h_i}{2} - x_{\rm cell} + l_x}  
      & \mbox{if } x_{\rm cell} - l_x - \frac{h_i}{2} < x_i < x_{\rm cell} - l_x + \frac{h_i}{2} \\
      0 & \mbox{if } x_{\rm cell}-l_x + \frac{h_i}{2} \le x_{i} \le x_{\rm cell}-\frac{h_i}{2} \\
      -\frac{m_i}{V_{\rm cell}^{\rm acc}} \frac{V_{{\rm cube}\;i\;\cap\;{\rm cell}}}{V_{{\rm cube}\;i}}
      \frac{1}{x_{\rm cell}-x_i+\frac{h_i}{2}} 
      & \mbox{if } x_{\rm cell}-\frac{h_i}{2} < x_i < x_{\rm cell}+\frac{h_i}{2} \\
      0 & \mbox{if } x_i \ge x_{\rm cell} + \frac{h_i}{2} 
   \end{dcases}
   \tag{A6}
   \label{eq:discrete_nonbonded_derivative}
\end{equation}
and similarly for $\partial \rho_{\rm cell} / \partial y_i$ and  $\partial \rho_{\rm cell} / \partial z_i$

The derivatives are bounded but not continuous. To make them continuous, an extension of the ``smearing scheme''
for beads which uses a continuous density distribution for the contribution of each bead would be required.
Two nodes whose cubes lie entirely within a cell experience the same nonbonded force (zero). This is not true,
however, of nodes whose cubes intersect cell borders.

The edge length of the cube assigned to node $i$, $h_i$, can be set approximately equal to the root mean square
end-to-end distance of the strands assigned to a node:
\begin{equation}
   h_i = \left(\frac{m_i}{m_{\rm K}} \; b_{\rm K}^2 \right)^{1/2} 
   \tag{A7}
\end{equation}
where $m_{\rm K}$ and $b_{\rm K}$ are the mass and the length of a Kuhn segment of the polymer under consideration.

\begin{figure}
   \centering
   \includegraphics[width=0.4\textwidth]{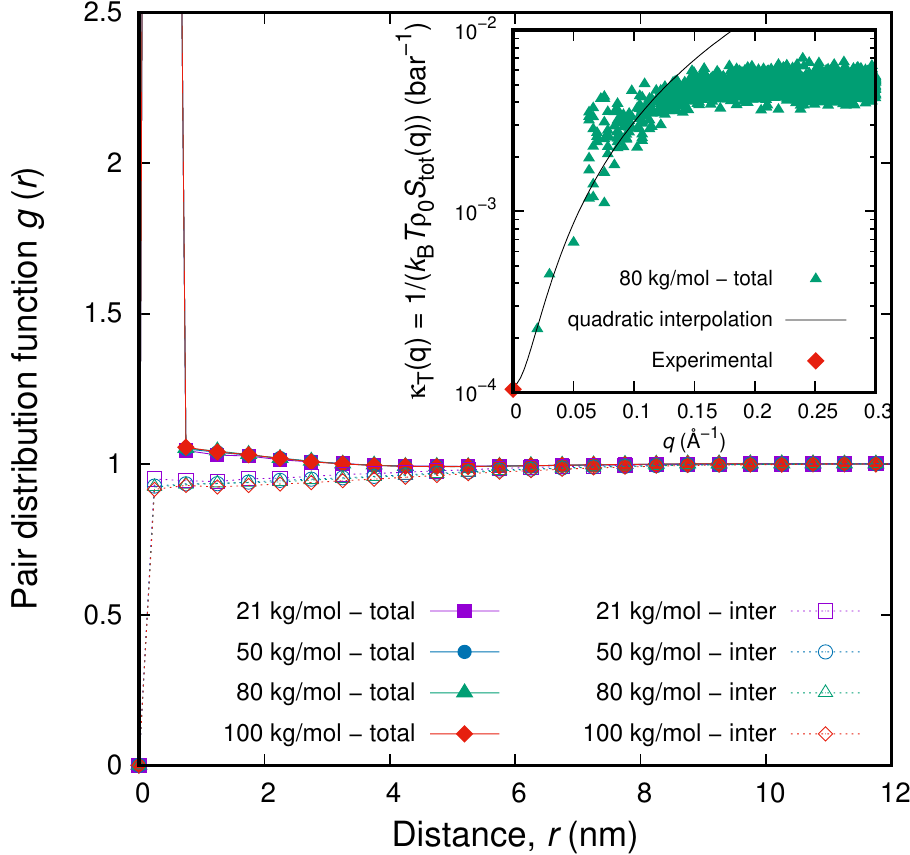}
   \caption{
   Total and intermolecular pair distribution functions of polymeric beads accumulated in 0.5-nm-thick bins. 
   The edge length of the non-bonded discretization cells, $l_x=l_y=l_z$, is 5 nm. In the inset to the figure, the compressibility
   is presented as a function of the observation length scale. Compressibility is estimated from the 
   static structure factor, $S_{\rm tot}(\mathbf{q})$ of a 80 kg/mol polyisoprene melt. The experimental value, 
   $\kappa_{\rm T} = 1.27 \times 10^{-4} {\rm bar}^{-1}$, predicted from the Sanchez-Lacombe EOS, is also included.\cite{PolymBull_34_109}}
   \label{fig:gofr}
\end{figure}

Figure \ref{fig:gofr} shows the variation of the total and the intermolecular pair distribution functions with the distance 
between polymeric beads. On very short length scales, the total distribution function diverges as $1/r$ because of the 
intramolecular contribution dictated by the Gaussian chain model. This spike formed is an artificial effect, caused by 
the fact that chain self-intersection is not prevented in the Gaussian chains used in our calculations. If an atomistic 
description were used, this spike would be replaced by a series of intramolecular peaks reflecting the bonded geometry 
and conformational preferences of polyisoprene molecules. In the regime of small distances, segments of other molecules 
are expelled from the volume of a reference chain and this gives rise to a slight ``correlation hole'' effect in the
intermolecular $g\parnths{r}$.
For large length scales, the intermolecular distribution function approaches unity, as the intramolecular distribution 
function approaches zero. The use of the non-bonded scheme employed in this work leads to a uniform density profile 
throughout the volume of the simulation box (the non-bonded discretization length employed for Figure \ref{fig:gofr} was 5 nm).

In a dense polymer melt, density fluctuations are strongly suppressed on large length scales, resulting in the 
screening of excluded volume interactions among the polymer beads. The correlations of local segment density in
homopolymer melts are characterized by the structure factor $S_{\rm tot}(\textbf{q})$, i.e. the Fourier transform 
with respect to distance ${\bf r} - {\bf r}^\prime$ of the density correlation 
function $\left \langle \rho\left(\mathbf{r}\right) \rho\left(\mathbf{r}^\prime\right)\right\rangle$, where brackets denote
averaging in the canonical ensemble. Since our simulations deal with explicit particle coordinates, 
$S_{\rm tot}(\textbf{q})$ can be calculated according to: 
\begin{equation}
S_{\rm tot}(\textbf{q}) = \frac{1}{n N} \left\langle
\left|
\sum_{j=1}^{nN} e^{-i\mathbf{q}\cdot\mathbf{r}_j}
\right|^2 \right\rangle
\tag{A8}
\end{equation}
where $\mathbf{r}_j$ stands for the position vector of bead $j$.
The limiting value of the total structure factor at large length scales, $q\to 0$, is directly connected to the 
compressibility of the system:
\begin{equation}
\lim_{|\mathbf{q}|\to 0}S_{\rm tot}(\mathbf{q}) = \frac{1}{k_{\rm B}T\rho_0 \kappa_{\rm T}}
\tag{A8}
\end{equation}
where $\rho_0$ is the mean segment density and $\kappa_T$ the isothermal compressibility.
By calculating $S_{\rm tot}(\textbf{q})$ for a range of $q$-vectors, the compressibility of the system can be estimated 
as a function of the observation length scale. 
In the inset to Figure \ref{fig:gofr} the compressibility of a polyisoprene system is plotted as a function of 
the scattering vector $\mathbf{q}$ used for the calculation of the total structure factor. 
Coarse-grained simulations with soft interactions are performed with potential expressions and parameters which yield 
more compressible systems than their atomistically simulated counterparts. This applies to our simulations, too. It is 
evident that the compressibility of the system deviates from the macroscopic one ($\sim 10^{-4}\;{\rm bar}^{-1}$), as 
$q$ increases (i.e., as one moves to smaller length scales). However, at large length scales compressibility values 
lie in the range expected for experimental systems.

\section*{Appendix B: Equation of State Considerations}
One of the most widely-used equations of state (EOS) for polymer fluids is the one derived by Sanchez and Lacombe.\cite{JPhysChem_80_2352}
These authors have employed a lattice formulation, wherein the polymer chains are occupying discrete lattice sites,
while there also exist vacant lattice sites (holes).
The Gibbs energy, based on the Sanchez-Lacombe equation of state can be expressed in dimensionless variables as:
\begin{equation}
   \tilde{G} \equiv \frac{G}{n r \varepsilon^*} 
   = -\tilde{\rho} + \tilde{p} \tilde{v} + \tilde{T} \bracks{\parnths{\tilde{v}-1} \ln{\parnths{1 - \tilde{\rho}}}
   + \frac{1}{r} \ln{\parnths{\frac{\tilde{\rho}}{w}}}}
   \tag{B1}
   \label{eq:method_polymer_sl_gibbs_energy}
\end{equation}
where $\tilde{T}$, $\tilde{p}$, $\tilde{v}$ and $\tilde{\rho}$ are the reduced temperature, pressure, volume and
density. 
The parameter $w$ is connected to the number of different configurations available to a system of $n$ $r$-mers. It will
be examined in detail later.
The Sanchez-Lacombe parameters are presented in Table B1.%
The corresponding equation of state (EoS) can be extracted, by minimizing $G$:
\begin{equation}
   \tderv{\pderv{\tilde{G}}{\tilde{v}}}{\tilde{T}, \tilde{p}} = 0 
   \tag{B2}
\end{equation}
which yields the equation of state for the system:
\begin{equation}
   \tilde{\rho}^2 + \tilde{p} + \tilde{T} \bracks{\ln{\parnths{1 - \tilde{\rho}}} + \parnths{1-\frac{1}{r}} \tilde{\rho}} = 0
   \tag{B3}
   \label{eq:method_polymer_sl_equation_of_state}
\end{equation}
It should be noted that, in the isothermal-isobaric ensemble, $\tilde{p}$ and $\tilde{T}$ are the independent variables, 
while $\tilde{\rho}$ is the dependent one. Therefore, eq \ref{eq:method_polymer_sl_equation_of_state} defines the 
value of $\tilde{\rho}$ at given $\parnths{\tilde{T}, \tilde{p}}$ that minimizes the free energy.
Equations \ref{eq:method_polymer_sl_gibbs_energy} and \ref{eq:method_polymer_sl_equation_of_state} contain the complete
thermodynamic description of the model fluid.

\begin{table*}
   \centering 
   \caption*{Table B1: Sanchez-Lacombe Notation}
   \label{tab:method_polymer_sl_notation}
   \begin{tabular}{ccc}
   \hline
   Symbol & Explanation & Units \\ \hline
   $n$ & number of molecules & - \\
   $M$ & molecular weight & kg \\
   $r$ & Sanchez-Lacombe segments per molecule & - \\
   $V$ & volume of the system & m$^3$ \\
   $\rho$ & mass density of the system & ${\rm kg/m}^3$ \\
   $\rho_{\rm mol} = \frac{n}{V}$ & molecular density & ${\rm m}^{-3}$ \\[10pt]
   $v^*$ & close packed volume of a mer & ${\rm m}^3$ \\
   $r v^*$ & close packed volume of an $r$-mer molecule & ${\rm m}^3$ \\
   $V^* = n\parnths{r v^*}$ & close packed volume of the $n$ $r$-mers & ${\rm m}^3$ \\
   $\rho^* = \frac{M}{r v^*}$ & close packed mass density & ${\rm kg/m}^3$ \\
   $T^* = \varepsilon^* / k_{\rm B}$ & characteristic temperature & K \\
   $p^* = \varepsilon^* / v^*$ & characteristic pressure & ${\rm kg}/\parnths{{\rm m s}^2}$ \\[10pt]
   $\tilde{T} = T /T^*$ & reduced temperature & - \\
   $\tilde{p} = p / p^*$ & reduced pressure & - \\
   $\tilde{\rho} = \rho/\rho^* = \rho_{\rm mol} M /\rho^*$ & reduced density & \\
   $\tilde{v} = 1/\tilde{\rho} = V/V^*$ & reduced volume & - \\ \hline
   \end{tabular}
\end{table*}

Since the Sanchez-Lacombe theory is not a corresponding states theory, the state parameters are predicted to vary
with chain length. It has been shown that thermal expansion coefficients, compressibilities and free volumes are predicted by 
the Sanchez-Lacombe EoS to decrease with increasing degree of polymerization, in agreement with 
experiment. The decrease of the free volume with increasing molecular weight is supported by the observation that
the glass temperature increases with increasing molecular weight.\cite{EurPolymJ_11_297}

If we wish to obtain some physical insight into the parameter $w$, we can compare the Sanchez - Lacombe EoS
with the ideal gas EoS. The former should reduce to the latter, in the limit of zero molecular density, 
$\rho_{\rm mol} = n/V$:
\begin{equation}
   \lim_{\rho_{\rm mol} \to 0^+} A^{\rm SL} = \lim_{\rho_{\rm mol} \to 0^+} A^{\rm id.g.}
   \tag{B4}
   \label{eq:method_polymer_sl_idg_limits_equation}
\end{equation}
where we have used $A^{\rm SL}$ and $A^{\rm id.g.}$ to denote the Helmholtz energy obtained from the Sanchez-Lacombe
and the ideal gas EoS, respectively.

The Helmholtz energy can be obtained from eq \ref{eq:method_polymer_sl_gibbs_energy}:
\begin{align}
   A^{\rm SL} \left(\rho, T \right) 
   = & G - p V = G - \tilde{p} \frac{n r \epsilon^*}{\tilde{\rho}} \nonumber \\
   = & -n r k_{\rm B}T^* \tilde{\rho} + n r k_{\rm B} T \left[
   \left(\tilde{v}-1\right)\ln{\left(1-\tilde{\rho}\right)} 
   + \frac{1}{r}\ln{\left(\frac{\tilde{\rho}}{w} \right)}\right]
   \tag{B5}
   \label{eq:method_polymer_sanchez_lacombe_helmholtz}
\end{align}
hence, the limit appearing in the left-hand side of eq \ref{eq:method_polymer_sl_idg_limits_equation} reduces to:
\begin{equation}
   \lim_{\rho_{\rm mol} \to 0^+} A^{\rm SL} = n k_{\rm B} T \bracks{ \lim_{\rho_{\rm mol} \to 0^+} 
   \ln{\parnths{\frac{\rho_{\rm mol} M}{ N_{\rm A} \rho^* w}}} -r}
   \tag{B6}
   \label{eq:method_polymer_sl_limit}
\end{equation}

On the other hand, the Helmholtz energy of an ideal gas can be written as:
\begin{align}
   A^{\rm id.g.} \parnths{\rho, T} & = G^{\rm id.g.} \parnths{\rho, T} - p^{\rm id.g.} V \nonumber \\
   & = n \mu^{\rm id.g.}\parnths{\rho, T} - n k_{\rm B}T \nonumber \\
   & = n k_{\rm B} T \bracks{\ln{\parnths{\frac{\rho N_{\rm A} \prod_{i} \Lambda_i^3}{M \mathcal{Z}_0}}}-1}
   \tag{B7}
   \label{eq:method_polymer_ideal_gas_helmholtz}
\end{align}
with $\Lambda_i$ being the thermal wavelength of atom $i$ of a molecule of the fluid and $\mathcal{Z}_0$ 
the configurational integral of a single molecule with respect to all but three translational degrees of freedom. 
The molar mass of the molecule is denoted by $M$.
At this point we can introduce the density of molecules, $\rho_{\rm mol}$, in eq 
\ref{eq:method_polymer_ideal_gas_helmholtz} and consider its limit at $\rho_{\rm mol} \to 0^+$:
\begin{equation}
   \lim_{\rho_{\rm mol} \to 0^+}  A^{\rm id.g.} = n k_{\rm B}T \lim_{\rho_{\rm mol} \to 0^+} 
   \bracks{\ln{\parnths{\frac{\rho_{\rm mol} \prod_i \Lambda_i^3}{\mathcal{Z}_0 e}}}}
   \tag{B8}
   \label{eq:method_polymer_ideal_gas_limit}
\end{equation}

Finally, by setting the right-hand side term of eq \ref{eq:method_polymer_sl_limit} equal to the right-hand side term
of eq \ref{eq:method_polymer_ideal_gas_limit} we obtain:
\begin{equation} 
   \lim_{\rho_{\rm mol} \to 0^+} \bracks{\ln{\parnths{\frac{\rho_{\rm mol}M}{N_{\rm A} \rho^* w e^r}}}}
   = \lim_{\rho_{\rm mol} \to 0^+} \bracks{\ln{\parnths{\frac{\rho_{\rm mol} \prod_i \Lambda_i^3}{\mathcal{Z}_0 e}}}}
   \tag{B9}
\end{equation}
which results in:
\begin{equation}
   \frac{1}{w} = \frac{N_{\rm A} \rho^* \prod_i \Lambda_i^3}{M \mathcal{Z}_0} e^{\parnths{r-1}}
   \tag{B10}
   \label{eq:method_polymer_sanchez_lacombe_w_definition}
\end{equation}
that connects the parameter $w$ with the thermal wavelengths of the molecules and their intramolecular configurational
integral, $\mathcal{Z}_0$. The molecular weight, $M$, enters eq \ref{eq:method_polymer_sanchez_lacombe_w_definition}
because $\rho^*$ refers to the critical mass density of the equation of state.

\section{Appendix C: Derivation of the Model Stress Tensor}

We consider the free energy per unit mass, $A \left(T, V\right)/m$ of the system.
The thermodynamic stress tensor, $\boldsymbol{\tau}$ is:\cite{JRheol_40_69,Macromolecules_31_6310}
\begin{equation}
   \boldsymbol{\tau} = \rho \mathbb{F} \cdot \left( \pderv{\left(A/m\right)}{\mathbb{F}} \right)^{\rm T}
   \tag{C1}
   \label{eq:method_stress_deform_gradient_def}
\end{equation}
where $A$ is the Helmholtz energy, $m$ is the total mass in the system and $\mathbb{F}$ denotes the deformation 
gradient tensor.
The same equation can be written in component form as:
\begin{equation}
   \tau_{\alpha \beta} = \rho \sum_{\gamma = 1}^{3} \mathbb{F}_{\alpha \gamma} 
   \pderv{\left( A/m \right)}{\mathbb{F}_{\beta \gamma}} 
   \tag{C2}
\end{equation}
If we assume that our simulation box does not exchange mass with its environment, the stress tensor can further be
simplified:
\begin{equation}
   \tau_{\alpha \beta} = \frac{1}{V_{\mathscr{R}}} \sum_{\gamma = 1}^{3} \mathbb{F}_{\alpha \gamma} 
   \pderv{A}{\mathbb{F}_{\beta \gamma}} 
   \tag{C3}
\end{equation}
where $V_{\mathscr{R}}$ is the volume of the simulation box in the reference state $\mathscr{R}$.
In general the prefactor should be $1/V$ with $V$ being the current volume of the system. The density, $\rho$, generally
changes with $\mathbb{F}$ and $V = V_{\mathscr{R}} \parnths{1 + {\rm det}\parnths{\mathbb{F}}}$.

In our model, the free energy per unit mass, 
$A \big(\cbraces{\mathbf{r}_{ij}}, \cbraces{\rho \parnths{\mathbf{r}}}, T \big)/m$, is a function of 
the separation vectors between all connected beads, $\cbraces{\mathbf{r}_{ij}}$,
local densities, $\cbraces{\rho \parnths{\mathbf{r}}} \equiv \cbraces{\rho_{\rm cell}}$ and 
temperature, $T$. The thermodynamic stress tensor, $\pmb{\tau}$ is given by eq \ref{eq:method_stress_deform_gradient_def}:
\begin{equation}
   \boldsymbol{\tau} = \rho_\mathscr{R} \; \mathbb{F} \cdot \left(\frac{\partial 
   \left(A\big( \left\{ \mathbf{r}_{ij} \right\}, \left\{ \rho_{\rm cell} \right\}, T \big)/m\right)}
   {\partial \mathbb{F}} \right)^{\rm T}
   \tag{C4}
   \label{eq:b3kMC_stress_matrix_form}
\end{equation}
where $A$ is the total Helmholtz energy (eq \ref{eq:helmholtz_energy_def}) and $m$ is the total mass in the system.
It should be noted that density, $\rho_\mathscr{R}$, refers to the reference, $\mathscr{R}$, configuration of the system.
$\mathbb{F}$ denotes the deformation gradient tensor defined through a mapping of an infinitesimal vector 
$\D \mathbf{x}$ of the initial configuration onto the infinitesimal vector $\D \mathbf{x}^\prime$ after the deformation.
Equation \ref{eq:b3kMC_stress_matrix_form} can be written in component form as:
\begin{equation}
   \tau_{\alpha \beta} = \rho_{\mathscr{R}} \; \sum_{\gamma = 1}^{3} \mathbb{F}_{\alpha \gamma} \frac{\partial 
   \left( A \big(\left\{ \mathbf{r}_{ij} \right\}, \left\{ \rho_{\rm cell} \right\}, T \big)/m\right)}
   {\partial \mathbb{F}_{\beta \gamma}} 
   \tag{C5}
   \label{eq:b3kMC_stress_component_form}
\end{equation}

Invoking the functional dependence of our Helmholtz energy (eq \ref{eq:helmholtz_energy_def}):
\begin{align}
   \tau_{\alpha \beta} & = \frac{\rho_\mathscr{R}}{m} \sum_{\gamma = 1}^3 \mathbb{F}_{\alpha \gamma} \left[
   \frac{\partial A_{\rm b} \left(\left\{ r_{ij} \right\}, T\right)}
   {\partial \mathbb{F}_{\beta \gamma}} +
   \frac{\partial A_{\rm nb} \left(\left\{ \rho_{\rm cell} \right\}, T \right)}
   {\partial \mathbb{F}_{\beta \gamma}} \right] \nonumber \\
   & = \tau_{\alpha \beta}^{\rm b} \big(\left\{r_{ij} \right\}, T \big) 
   + \tau_{\alpha \beta}^{\rm nb} \big(\left\{\rho_{\rm cell} \right\}, T \big) 
   \tag{C6}
\end{align}
where
\begin{equation}
   \tau_{\alpha \beta}^{\rm b} \big(\left\{ r_{ij} \right\}, T \big) = 
   \frac{1}{V_{\mathscr{R}}} \sum_{\gamma = 1}^3 \mathbb{F}_{\alpha \gamma} \left\{
   \frac{\partial}{\partial \mathbb{F}_{\beta \gamma}} \left[ \sum_{\left(i,j\right)} A_{\rm pair} 
   \parnths{r_{ij} , T} \right] \right \}
   \tag{C7}
\end{equation}
is the bonded contribution to the stress tensor, and 
\begin{equation}
   \tau_{\alpha \beta}^{\rm nb}\left(\left\{\rho_{\rm cell} \right\}, T \right) =
   \frac{1}{V_{\mathscr{R}}} \sum_{\gamma = 1}^3 \mathbb{F}_{\alpha \gamma} \left\{
   \frac{\partial}{\partial \mathbb{F}_{\beta \gamma}}\left[ \sum_{k\; \in\;{\rm cells}} V_{{\rm cell},k}^{\rm acc} 
   a_{\rm vol}\left(\rho_{{\rm cell},k}, T \right) \right] \right \}
   \tag{C8}
\end{equation}
is the nonbonded contribution to the stress tensor.

We start by calculating the bonded contribution to the stress tensor, which depends only on the distances between 
connected pair of atoms, $r_{ij}$, and temperature, $T$:
\begin{align}
   \tau_{\alpha \beta}^{\rm b}\left(\left\{r_{ij} \right\}, T \right) & =
   \frac{1}{V_\mathscr{R}} \sum_{\gamma = 1}^{3} \mathbb{F}_{\alpha \gamma} \left\{ \sum_{\left(i,j\right)}
   \left [ \frac{\partial A_{\rm pair} \left( r_{ij} , T\right)}{\partial \mathbb{F}_{\beta \gamma}}
   \right ] \right \} \nonumber \\
   & = \frac{1}{V_\mathscr{R}} \sum_{\gamma = 1}^{3} \mathbb{F}_{\alpha \gamma} \left\{ \sum_{\left(i,j\right)}
   \left [  
   \frac{\partial A_{\rm pair} \left( r_{ij} , T\right)}{\partial r_{ij}}
   \frac{\partial r_{ij}}{\partial \mathbb{F}_{\beta \gamma}}
   \right ] \right \}  \nonumber \\
   & = \frac{1}{V_\mathscr{R}} \sum_{\left( i, j\right)} 
   \frac{\partial A_{\rm pair} \left(r_{ij}, T\right)}{\partial r_{ij}} 
   \sum_{\gamma = 1}^{3} \frac{\partial r_{ij}}{\partial \mathbb{F}_{\beta \gamma}} \mathbb{F}_{\alpha \gamma}
   \nonumber \\
   & = \frac{1}{V_\mathscr{R}} \sum_{\left( i, j\right)} 
   \frac{\partial A_{\rm pair} \left(r_{ij}, T\right)}{\partial r_{ij}} 
   \sum_{\gamma = 1}^{3} \frac{\partial r_{ij}}{\partial r_{ij,\beta}} 
   \frac{\partial r_{ij,\beta}}{\partial \mathbb{F}_{\beta \gamma}} \mathbb{F}_{\alpha \gamma}
   \nonumber \\
   & = \frac{1}{V_\mathscr{R}} \sum_{\left( i, j\right)} 
   \frac{\partial A_{\rm pair} \left(r_{ij}, T\right)}{\partial r_{ij}} 
   \frac{r_{ij,\beta}}{r_{ij}}
   \sum_{\gamma = 1}^{3} \frac{\partial r_{ij,\beta}}{\partial \mathbb{F}_{\beta \gamma}} \mathbb{F}_{\alpha \gamma}
   \tag{C9}
   \label{eq:bonded_stress}
\end{align}
where we have made use of the fact that the partial derivative of the Euclidean norm of a vector 
$\mathbf{a} = \left(a_1,a_2, ..., a_n\right)$ with respect to one of its components, $a_j$, is:
\begin{equation}
   \frac{\partial \vectornorm{\mathbf{a}}}{\partial a_j} = 
   \frac{\partial}{\partial a_j} \left( \sqrt{\sum_{i = 1}^{n} a_i^2}\right)
   = \frac{1}{2 \vectornorm{\mathbf{a}}} \sum_{i=1}^n \left[\frac{\partial}{\partial a_j} \left(a_i^2\right) \right]
   = \frac{a_j}{\vectornorm{\mathbf{a}}}
   \tag{C10}
\end{equation}
It should be noted that eq \ref{eq:bonded_stress} is valid for any kind of deformation (both affine and nonaffine).

We envision that, at a certain time, a homogeneous deformation is applied on the polymer that displaces bond ends 
affinely, in the sense that their positions are changed in the same way as material points in a macroscopic continuum 
description. Straight parallel lines in the reference configuration map to straight parallel lines in the deformed 
configuration
Let $\mathbf{r}_i$ and $\mathbf{r}_j$ be the positions of the start and the end of the bonded beads before the 
deformation and $\mathbf{r}_{i}^\prime$ and $\mathbf{r}_j^\prime$ the positions of the same beads after the 
deformation, then:
\begin{align}
   \mathbf{r}_i^\prime = \mathbb{F} \mathbf{r}_i \tag{C11} \label{eq:ri_deformed} \\
   \mathbf{r}_j^\prime = \mathbb{F} \mathbf{r}_j \tag{C12} \label{eq:rj_deformed}
\end{align}
By subtracting eq \ref{eq:rj_deformed} from eq \ref{eq:ri_deformed}, we get:
\begin{equation}
   \mathbf{r}_{ij}^\prime = \mathbb{F} \mathbf{r}_{ij} 
   \tag{C13}
\end{equation}
Thus, one component of the deformed vector is connected to the components of the undeformed through the relation:
\begin{equation}
   r_{ij, \beta}^\prime = \sum_{\gamma = 1}^3 \mathbb{F}_{\beta \gamma} r_{ij, \gamma}
   \tag{C14}
\end{equation}
and the derivative appearing in eq \ref{eq:bonded_stress} can now be calculated:
\begin{equation}
   \frac{\partial r_{ij,\beta}}{\partial \mathbb{F}_{\beta \gamma}} = r_{ij, \gamma}
   \tag{C15}
\end{equation}
Eq \ref{eq:bonded_stress} takes the form:
\begin{align}
   \tau_{\alpha \beta}^{\rm b}\left(\left\{r_{ij} \right\}, T \right)
   & = \frac{1}{V_\mathscr{R}} \sum_{\left( i, j\right)} 
   \frac{\partial A_{\rm pair} \left(r_{ij}, T\right)}{\partial r_{ij}} 
   \frac{r_{ij,\beta}}{r_{ij}}
   \sum_{\gamma = 1}^{3} r_{ij,\gamma} \mathbb{F}_{\alpha \gamma} \nonumber \\
   & = 
   \frac{1}{V_\mathscr{R}} \sum_{\left( i, j\right)} \frac{\partial A_{\rm pair} \left(r_{ij}, T\right)}{\partial r_{ij}}
   \; \frac{r_{ij,\beta} r_{ij, \alpha}}{r_{ij}}
   \tag{C16}
   \label{eq:bonded_stress_affine}
\end{align}
where the virial theorem is recovered.\cite{MacromolTheorySimul_2_191}

We now move to the estimation of the nonbonded contribution to the stress tensor, $\tau_{\alpha \beta}^{\rm nb}$.
We calculate the stress by taking the derivative of the equation-of-state free energy density, 
$a_{\rm vol}\left(\rho_{{\rm cell},k}, T \right)$, with respect to the deformation gradient tensor, $\mathbb{F}$,
considering the set of space discretization voxels (introduced in Appendix A) as a permanent scaffold into which beads 
would be reassigned upon deformation. In that approach, a tensile deformation in a periodic system would amount to 
extending the system such that it occupies an additional layer of voxels, for example. This would indeed be 
infinitesimal for a very large model system. There are practical problems with this, however, so we consider our 
voxels as a means of spatial discretization in integrating the free energy density over 3-dimensional space, allowing them to 
deform affinely following the macroscopically applied deformation. This may not be too bad if the voxel size exceeds 
the correlation length of density fluctuations in the polymer. 
Following our assumptions, the nonbonded contribution to the stress tensor can be cast as:
\begingroup
\allowdisplaybreaks
\begin{align}
   \tau_{\alpha \beta}^{\rm nb}\left(\left\{\rho_{\rm cell} \right\}, T \right) 
   & =  \frac{1}{V_\mathscr{R}} \sum_{\gamma = 1}^3 \mathbb{F}_{\alpha \gamma} \left\{
   \frac{\partial}{\partial \mathbb{F}_{\beta \gamma}}\left[ \sum_{k\; \in\;{\rm cells}} V_{{\rm cell},k}^{\rm acc} 
   a_{\rm vol}\left(\rho_{{\rm cell},k}, T \right) \right]  \right \} \nonumber \\
   & = \frac{1}{V_\mathscr{R}} \sum_{\gamma = 1}^{3} \mathbb{F}_{\alpha \gamma} 
   \sum_{k \; \in \; {\rm cells}} \left[ \frac{\partial V_{{\rm cell},k}^{\rm acc}}{\partial \mathbb{F}_{\beta \gamma}} 
   \; a_{\rm vol}\left(\rho_{{\rm cell},k}, T \right) \right] \nonumber \\
   & + \frac{1}{V_\mathscr{R}} \sum_{\gamma = 1}^{3} \mathbb{F}_{\alpha \gamma} 
   \sum_{k \; \in \; {\rm cells}} \left[ V_{{\rm cell},k}^{\rm acc} 
   \frac{\partial a_{\rm vol}\left(\rho_{{\rm cell},k}, T \right)}{\partial \mathbb{F}_{\beta \gamma}}\right]
   \nonumber \\
   & = \frac{1}{V_\mathscr{R}} \sum_{k \; \in \; {\rm cells}}  
   a_{\rm vol}\left(\rho_{{\rm cell},k},T \right) \sum_{\gamma = 1}^{3}  \mathbb{F}_{\alpha \gamma} 
   \frac{\partial V_{{\rm cell},k}^{\rm acc}}{\partial \mathbb{F}_{\beta \gamma}} \nonumber \\
   & + \frac{1}{V_\mathscr{R}} \sum_{k \; \in \; {\rm cells}} V_{{\rm cell},k}^{\rm acc} 
   \sum_{\gamma = 1}^{3} \mathbb{F}_{\alpha \gamma} 
   \frac{\partial a_{\rm vol}\left(\rho_{{\rm cell},k}, T \right)}{\partial \mathbb{F}_{\beta \gamma}} 
   \nonumber \\
   & = \frac{1}{V_\mathscr{R}} \sum_{k \; \in \; {\rm cells}}  
   a_{\rm vol}\left(\rho_{{\rm cell},k},T \right) \sum_{\gamma = 1}^{3}  \mathbb{F}_{\alpha \gamma} 
   \frac{\partial V_{{\rm cell},k}^{\rm acc}}{\partial \mathbb{F}_{\beta \gamma}} \nonumber \\
   & + \frac{1}{V_\mathscr{R}} \sum_{k \; \in \; {\rm cells}} V_{{\rm cell},k}^{\rm acc} 
   \sum_{\gamma = 1}^{3} \mathbb{F}_{\alpha \gamma} 
   \frac{\partial a_{\rm vol}\left(\rho_{{\rm cell},k}, T \right)}{\partial \rho_{{\rm cell},k}}
   \frac{\partial \rho_{{\rm cell},k}}{\partial \mathbb{F}_{\beta \gamma}} \nonumber \\
   & = \frac{1}{V_\mathscr{R}} \sum_{k \; \in \; {\rm cells}}  
   a_{\rm vol}\left(\rho_{{\rm cell},k},T \right) \sum_{\gamma = 1}^{3}  \mathbb{F}_{\alpha \gamma} 
   \frac{\partial V_{{\rm cell},k}^{\rm acc}}{\partial \mathbb{F}_{\beta \gamma}} \nonumber \\
   & - \frac{1}{V_\mathscr{R}} \sum_{k \; \in \; {\rm cells}} \rho_{{\rm cell},k} 
   \frac{\partial a_{\rm vol}\left(\rho_{{\rm cell},k}, T \right)}{\partial \rho_{{\rm cell},k}}
   \sum_{\gamma = 1}^{3} \mathbb{F}_{\alpha \gamma} 
   \frac{\partial V_{{\rm cell},k}^{\rm acc}}{\partial \mathbb{F}_{\beta \gamma}} \nonumber \\
   & = \frac{1}{V_\mathscr{R}} \sum_{k \; \in \; {\rm cells}}\Big[
   a_{\rm vol}\left(\rho_{{\rm cell},k}, T \right)
   - \rho_{{\rm cell},k} \frac{\partial a_{\rm vol}\left(\rho_{{\rm cell},k}, T \right)}{\partial \rho_{{\rm cell},k}}
   \Big] \sum_{\gamma = 1}^{3} \mathbb{F}_{\alpha \gamma} 
   \frac{\partial V_{{\rm cell},k}^{\rm acc}}{\partial \mathbb{F}_{\beta \gamma}}
   \tag{C17}
   \label{eq:b3DkMC_stress_non_bonded_pre_final}
\end{align}%
\endgroup
where $a_{\rm vol}\parnths{\rho_{{\rm cell},k},T }$ is the nonbonded Helmholtz energy density in cell $k$. 
The term in brackets is the negative of the pressure, $- p\parnths{\rho_{{\rm cell},k},T}$ as that is predicted by 
the equation of state 
(eq \ref{eq:method_polymer_sl_equation_of_state}) under given density $\rho_{{\rm cell},k}$ and temperature, $T$.
At this point we have to calculate the derivatives expressing the variation of the volume of a cell with respect to
an element of the deformation gradient tensor, $\partial V_{{\rm cell},k}^{\rm acc}/\partial \mathbb{F}_{\beta \gamma}$.

The determinant of the deformation gradient tensor is the ratio of volumes or densities of the deformed and initial 
configurations:
\begin{equation}
   \det{\parnths{\mathbb{F}}} = \frac{V^\prime}{V_\mathscr{R}} 
   = \frac{V_{{\rm cell},k}^{\rm acc \;\;\prime}}{V_{{\rm cell},k}^{\rm acc}}
   = \frac{\rho}{\rho^\prime} 
   = \frac{\rho_{{\rm cell},k}}{\rho_{{\rm cell},k}^\prime}
   \tag{C18}
\end{equation}
where the use of primes denotes the deformed configuration. 
The derivative of the determinant of $\mathbb{F}$ with 
respect to the tensor $\mathbb{F}$ itself is calculated by the following equation:\cite{JElast_16_221}
\begin{equation}
   \frac{\partial \left( \det{\left( \mathbb{F} \right)} \right)}{\partial \mathbb{F}} = 
   \det{\left(\mathbb{F} \right)} \left(\mathbb{F}^{-1}\right)^{\rm T}
   \tag{C19}
\end{equation}
Finally, the nonbonded contribution to the stress tensor, substituting the above terms in 
eq \ref{eq:b3DkMC_stress_non_bonded_pre_final} is:
\begin{align}
   \tau_\albe^{\rm nb}\parnths{\cbraces{\rho_{\rm cell}}, T}
   & = - \frac{1}{V_\mathscr{R}} \sum_{k \; \in \; {\rm cells}}p\parnths{\rho_{{\rm cell},k},T}
   \sum_{\gamma = 1}^{3} \mathbb{F}_{\alpha \gamma} 
   \frac{\partial V_{{\rm cell},k}^{\rm acc}}{\partial \mathbb{F}_{\beta \gamma}} \nonumber \\
   & = - \frac{1}{V_\mathscr{R}} \sum_{k \; \in \; {\rm cells}}p\parnths{\rho_{{\rm cell},k},T}
   \sum_{\gamma = 1}^{3} \mathbb{F}_{\alpha \gamma} \pderv{V_{{\rm cell},k}^{\rm acc}}{\det{\parnths{\mathbb{F}}}}
   \pderv{\parnths{\det{\parnths{\mathbb{F}}}}}{\mathbb{F}_{\beta \gamma}} \nonumber \\
   & = - \delta_\albe \frac{1}{V_\mathscr{R}} \sum_{k \; \in \; {\rm cells}}p\parnths{\rho_{{\rm cell},k},T}
    V_{{\rm cell},k}^{\rm acc}
    \tag{C20}
   \label{eq:b3DkMC_stress_non_bonded_final}
\end{align}
where all volumes, $V_\mathscr{R}$ and $V_{{\rm cell},k}^{\rm acc}$ refer to the undeformed (reference) state of the
system. Eq \ref{eq:b3DkMC_stress_non_bonded_final} could be fully anticipated. The contribution of the equation of 
state to the stress tensor of the system is limited to the diagonal components and its magnitude is the negative of 
the weighted average pressure over the cells of the grid, with volumes $V_{{\rm cell},k}^{\rm acc}$ being the 
weights multiplying the individual contributions. 

\section{Appendix D: Initial Estimation of the Hopping Frequency Factor}
The mean square displacement of a slip-spring in time $\Delta t$ is $k_{\rm hop}\Delta t \parnths{n_{\rm Kuhns/bead} b^2}$,
where the mean square displacement of the chain along the primitive path due to the slip-springs is:
\begin{align}
   \angles{\Delta r_{\rm cm}^2} & = \frac{\text{mean square displacement of slip-springs}}{n_{\rm ss/chain}} \nonumber \\
   & = k_{\rm hop}\Delta t \frac{n_{\rm Kuhns/bead}b^2}{n_{\rm ss/chain}} \nonumber \\
   & = k_{\rm hop}\Delta t \frac{n_{\rm Kuhns/bead}b^2}{\frac{2N_{\rm ss}}{n}} \nonumber \\
   & = \frac{k_{\rm hop}\Delta t \parnths{n_{\rm Kuhns/bead} b^2}}{2 \frac{N_{\rm ss}}{N_{\rm beads}} N}
   = 2 D_{{\rm cm}_{\rm along \:contour}} \Delta t
   \tag{D1}
\end{align} 
with $N_{\rm beads}$ and $N_{\rm ss}$ being the total number of beads and slip-springs in the system, respectively,
and $n$ the number of chains. The factor 2 is included because a slip-spring is attached to two chains. $N$ is 
the chain length measured in number of Kuhn segments.
The chain diffusivity by means of slip-spring jumps is:
\begin{equation}
   D_{{\rm cm}_{\rm along \: contour}} = \frac{k_{\rm hop}\parnths{n_{\rm Kuhns/bead}b^2}}{4\frac{N_{\rm ss}}{N_{\rm beads}}N}
   \tag{D2}
\end{equation}
while the corresponding diffusivity, according to the Rouse model:
\begin{equation}
   D_{\rm cm, Rouse} = \frac{k_{\rm B}T}{N \zeta}
   \tag{D3}
\end{equation}
for the above two to become equal, it should hold:
\begin{equation}
   k_{\rm hop} = \frac{4 k_{\rm B}T}{\zeta} \frac{N_{\rm ss}}{N_{\rm beads}}
   \frac{1}{\parnths{n_{\rm Kuhns/bead} b^2}}
   \tag{D4}
   \label{eq:hop_rate_estimate}
\end{equation}
The last expression provides an upper limit to the hopping rate, in the case the diffusion of the polymer chains 
was based only on their motion along their primitive paths. Following the discussion of the main text, an estimate
of the average hopping rate, in the case of Gaussian slip-springs, is:
\begin{equation}
   \angles{k_{\rm hop}} \simeq 38 \nu_{\rm hop}
   \tag{D5}
\end{equation}
which, in view of eq \ref{eq:hop_rate_estimate}, allows us for an upper limit for $\nu_{\rm hop}$:
\begin{equation}
   \nu_{\rm hop} < \frac{2}{19} \frac{k_{\rm B}T}{\zeta} \frac{N_{\rm ss}}{N_{\rm beads}}
   \frac{1}{\parnths{n_{\rm Kuhns/bead} b^2}}
   \tag{D6}
\end{equation}
The exact value of $\nu_{\rm hop}$ to be used during the simulation is the one ensuring conservation of the 
average number of slip-springs when implementing the fluctuating number of slip-springs scheme (Appendix E).

\section{Appendix E: Microscopic Reversibility in Slip-spring Formation and Destruction 
         (fluctuating number of slip-\\springs scheme)}

Consider a specific configuration bearing a slip-spring connecting an end $b_0$ of a chain contained in it with
a bead $a_0$ of the same or another chain (which may be either an interior or an end bead). Let us call this 
configuration $a_0 \land b_0$, for simplicity. Consider also a configuration which is identical to $a_0 \land b_0$,
the only difference being that the slip-spring connecting $a_0$ and $b_0$ is missing. Let us call that configuration
$a_0 \land b_\times$ (see Figure \ref{fig:kMC_hopping_destruction}).

If $a_0$ is not a chain end, the only way in which one can go from $a_0 \land b_0$ to $a_0 \land b_\times$ is slippage
of the slip-spring past the chain end $b_0$. The probability per unit time for observing the transition is:
\begin{equation}
   \mathcal{Q}_{a_0 \land b_0 \to a_0 \land b_\times} = k_{{\rm hop},a_0\land b_0} P_{a_0 \land b_0}
   \tag{E1}
\end{equation}
with $P_{a_0 \land b_0}$ being the a priori probability of state $a_0 \land b_0$ and $k_{{\rm hop}, a_0 \land b_0}$ being
the rate constant for hopping of the slip-spring along the chain in one of the two directions away from chain end $b_0$
in configuration $a_0 \land b_0$. In the reverse transition, the only way in which one can go from $a_0 \land b_\times$
to $a_0 \land b_0$ is formation of a slip-spring off of the chain end $b_0$. The probability per unit time for observing 
a transition from $a_0 \land b_\times$ to $a_0 \land b_0$ is:
\begin{equation}
   \mathcal{Q}_{a_0 \land b_\times \to a_0 \land b_0} = k_{{\rm form}, a_0\land b_\times \to a_0 \land b_0}\;\;
   P_{a_0 \land b_\times} \;\;\; .
   \tag{E2}
\end{equation}

At equilibrium, we demand detailed balance:
\begin{equation}
   \mathcal{Q}_{a_0 \land b_0 \to a_0 \land b_\times} = \mathcal{Q}_{a_0 \land b_\times \to a_0 \land b_0}
\end{equation}
which enables us to define $k_{{\rm form}}$ in terms of $k_{\rm hop}$:
\begin{equation}
   k_{{\rm form}, a_0\land b_\times \to a_0 \land b_0} = k_{{\rm hop}, a_0 \land b_0} 
   \frac{P_{a_0 \land b_0}}{P_{a_0 \land b_\times}} \;\;.
   \tag{E3}
   \label{eq:appE_kform_definition}
\end{equation}
In the fluctuating number of slip-springs scheme the probability distribution of configurations is dictated by an
ensemble which is canonical with respect to the chains and grand canonical with respect to the slip-springs. Multiple
slip-springs connecting the same pair of beads are, in principle, possible and are considered indistinguishable.
\cite{PhysRevLett_109_148302} The probabilities $P_{a_0 \land b_0}$ and $P_{a_0 \land b_\times}$ satisfy the following
proportionalities with the same proportionality constant:
\begin{align}
   P_{a_0 \land b_0} \propto &
   \exp{\parnths{-\frac{A_{a_0 \land b_0}^{\parnths{N_{\rm ss}}}}{k_{\rm B}T}}}
   z^{N_{\rm ss}} \frac{1}{n_{{\rm ss}, a_0 \land b_0}!}
   \tag{E4} \\
   P_{a_0 \land b_\times} \propto & \exp{\parnths{-\frac{A^{\parnths{N_{\rm ss}-1}}}{k_{\rm B}T}}}
   z^{N_{\rm ss}-1} \frac{1}{\parnths{n_{{\rm ss},a_0 \land b_0}-1}!}
   \tag{E5}
\end{align}
where $N_{\rm ss}$ is the total number of slip-springs in the starting configuration, $a_0 \land b_0$ and 
$n_{{\rm ss},{a_0 \land b_0}}$ is the total number of slip-springs (usually 1) connecting beads $a_0$ and $b_0$ in that
configuration. $A_{a_0 \land b_0}^{\parnths{N_{\rm ss}}}$ and $A^{\parnths{N_{\rm ss}-1}}$ are the total Helmholtz 
energies of the configuration including and not including the slip-spring, respectively 
 (c.f. Figure \ref{fig:free_energy_levels}). $z$ is the fugacity of the slip-springs, connected to the chemical 
potential of the slip-springs by:
\begin{equation}
   z = \exp{\parnths{\frac{\mu}{k_{\rm B}T}}} \;\;.
   \tag{E6}
   \label{eq:appE_fugacity_definition}
\end{equation}
From eqs \ref{eq:appE_kform_definition} to \ref{eq:appE_fugacity_definition} one obtains
\begin{equation}
   k_{{\rm form},a_0\land b_\times \to a_0 \land b_0} = k_{\rm hop, a_0 \land b_0} 
   \exp{\parnths{-\frac{A_{a_0 \land b_0}^{N_{\rm ss}} - A^{\parnths{N_{\rm ss}-1}}}{k_{\rm B}T}}}
   \frac{z}{n_{{\rm ss},a_0\land b_0}}
   \tag{E7}
\end{equation}

According to the approach we have introduced in the main text, all hopping moves take 
place with the same frequency factor, $\nu_0$, surpassing a constant free energy barrier, $A_{\mathscr{O} \to \mathscr{N}}^\ddagger$, for 
all slip-springs. This ensures microscopic reversibility during slip-springs moves. In other words, 
we assume:
\begin{equation}
   k_{{\rm hop}, a_0 \land b_0} = \nu_{0} 
   \exp{\parnths{- \frac{A_{\mathscr{O} \to \mathscr{N}}^\ddagger - A_{a_0 \land b_0}}{k_{\rm B} T}}}
   \tag{E8}
\end{equation}
with $A_{a_0 \land b_0}$ being the free energy stored in the considered slip-spring connecting $a_0$ and
$b_0$ (\textit{not of} the entire configuration). With this approach, the imposition of detailed balance,
eq \ref{eq:appE_kform_definition}, gives:
\begin{equation}
   k_{{\rm form},a_0 \land b_\times \to a_0 \land b_0}  = \nu_0 
   \exp{\parnths{-\frac{A_{\mathscr{O} \to \mathscr{N}}^\ddagger - A_{a_0 \land b_0}}{k_{\rm B}T}}}
   \exp{\parnths{-\frac{A_{a_0 \land b_0}^{\parnths{N_{\rm ss}}} - A^{\parnths{N_{\rm ss}-1}}}{k_{\rm B}T}}}
   \frac{z}{n_{{\rm ss},a_0\land b_0}}
   \tag{E9}
\end{equation}
or
\begin{equation}
   k_{{\rm form}, a_0 \land b_\times \to a_0 \land b_0} = \nu_0 
   \exp{\parnths{-\frac{A_{\mathscr{O} \to \mathscr{N}}^\ddagger}{k_{\rm B}T}}}
   \frac{z}{n_{{\rm ss},a_0\land b_0}}
   = \nu_{\rm hop}
   \frac{z}{n_{{\rm ss},a_0\land b_0}}
   \tag{E10}
\end{equation}
since 
\begin{equation}
   A_{a_0 \land b_0}^{\parnths{N_{\rm ss}}} = A^{\parnths{N_{\rm ss}-1}} + A_{a_0 \land b_0}
   \tag{E11}
\end{equation}
For the slip-spring chemical potentials considered here, multiple connections between the same two beads are extremely
improbable; the quantity $n_{{\rm ss}, a_0 \land b_0}$ is practically equal to 1 in all cases.
Thus, $k_{{\rm form}, a_0 \land b_\times \to a_0 \land b_0}$ amounts to a configuration-independent constant.

The total rate of slip-spring formation off of the chain end $b_0$ is:
\begin{equation}
   k_{{\rm form}, a_0 \land b_\times \to} = \nu_{\rm hop} n_{\rm cands}\parnths{a_0 \land b_\times}
   \frac{z}{n_{{\rm ss},a_0\land b_0}}
   \tag{E12}
\end{equation}
and is proportional to the number of candidate segments, $n_{\rm cands}\parnths{a_0 \land b_\times}$, with which
end $b_0$ can be bridged through a new slip-spring. In our scheme, since the candidate bridging sites are selected
among all sites within a radius $\alpha_{\rm attempt}$ from $b_0$ (Figure \ref{fig:kMC_hopping_creation}), no 
slip-springs longer than $\alpha_{\rm attempt}$ should  be allowed to slip past an end in the reverse, slip-spring 
destruction, move.

The slip-spring creation move can proceed as follows. For each chain end, we maintain a list of all bridgeable 
beads (either belonging to same or different chains) lying within distance $\alpha_{\rm attempt}$ from it, including other
chain ends. For each end $b_0$ and candidate partner $a_0$, we attempt construction of a slip-spring between them at
a constant rate $\parnths{\nu_{\rm hop} z}/n_{{\rm ss},a_0\land b_0}$. This could be implemented by considering 
each chain end and deciding whether a new slip-spring will be constructed off of it with probability:
\begin{equation}
   P_{{\rm form},a_0 \land b_0} = \nu_{\rm hop} 
   \frac{z}{n_{{\rm ss},a_0\land b_0}} n_{\rm cands}\parnths{a_0 \land b_0} \Delta t_{\rm kMC}
   \tag{E13}
\end{equation}%
where $n_{\rm cands} \parnths{a_0 \land b_0}$ is the number of bridgeable beads around $b_0$ and $\Delta t_{\rm kMC}$
is the time span between successive implementations of kMC events. For each picked end $b_0$, pick one of its 
bridgeable $n_{\rm cands}\parnths{a_0 \land b_0}$ beads at random and construct a slip-spring. Clearly,
$\Delta t_{\rm kMC}$ should be small enough for the probability $P_{{\rm form},a_0 \land b_0}$ to be considerably
smaller than $1$. This would also make the construction of double slip-spring bridges between chain ends very
unlikely.

\section{Appendix F: Microscopic Reversibility in Slip-spring Formation and Destruction 
         (constant number of slip-springs scheme)}

Following the main text and the preceding Appendix E, we consider a specific configuration bearing a slip-spring 
connecting an end $b_0$ of a chain contained in it with a bead $a_0$ of the same or another chain (which may be 
either an interior or an end bead). We have denoted this configuration with $\mathscr{O} \equiv a_0 \land b_0$, 
for simplicity. Moreover, we consider also a configuration which is identical to $\mathscr{O}$,
the only difference being that the slip-spring connecting $a_0$ and $b_0$ has been replaced by 
a slip-spring connecting beads $a_0^\prime$ and $b_0^\prime$. Let us call that configuration 
$\mathscr{N} \equiv a_0^\prime \land b_0^\prime$. The process driving us from the former to the latter configuration
is by destroying the slip-spring $a_0 \land b_0$ and subsequently creating the slip-spring $a_0^\prime \land b_0^\prime$.
The coupled destruction/formation process ensures the conservation of the number of slip-springs during the 
simulation. 

In the following we will consider only the case where the one end of the slip-spring is a chain end, for clarity.
However, the same procedure applies to the case where both ends of the slip-spring are chain ends.
If $a_0$ is not a chain end, the only way in which one can go from $a_0 \land b_0$ to $a^\prime_0 \land b^\prime_0$ is 
slippage of the slip-spring past the chain end $b_0$. The probability per unit time for observing the transition is:
\begin{equation}
   \mathcal{Q}_{a_0 \land b_0 \to a^\prime_0 \land b^\prime_0} = P_{a_0 \land b_0} k_{{\rm hop},a_0\land b_0} 
   P_{a^\prime_0 \land b^\prime_0}^{\rm sel} P_{\mathscr{O} \to \mathscr{N}}^{\rm accept}
   \label{eq:f1}
   \tag{F1}
\end{equation}
with $P_{a_0 \land b_0} = \exp{\parnths{-\beta A^{\parnths{\rm N_{\rm ss}}}_{a_0 \land b_0}}}$ being the a priori 
probability of state $a_0 \land b_0$ and $k_{{\rm hop}, a_0 \land b_0} = \nu_{\rm hop} \exp{\parnths{\beta A_{a_0 \land b_0}}}$ being
the rate constant for hopping of the slip-spring along the chain in one of the two directions away from chain end $b_0$
in configuration $a_0 \land b_0$.
At this point we should recall the distinction between $A^{\parnths{\rm N_{\rm ss}}}_{a_0 \land b_0}$ which is 
the total free energy of a configuration whose $N_{\rm ss}$-th slip-spring finds itself connecting beads $a_0$, and
$b_0$ and $A_{a_0 \land b_0}$ that is the free energy stored in the considered slip-spring connecting $a_0$ and
$b_0$ (not of the entire configuration).
Following the constant number of slip-springs scheme introduced in the main text, we should also multiply by the 
probability of selecting the pair $a^\prime_0 \land b^\prime_0$ as the new configuration, 
$P_{a^\prime_0 \land b^\prime_0}^{\rm sel}$, once the slip-spring has passed through the chain end and is 
considered for destruction. 
The final term in eq \ref{eq:f1}, $P_{\mathscr{O} \to \mathscr{N}}^{\rm accept}$, is the acceptance probability of
the combined destruction/creation move. Equivalently, the probability
of destroying a slip-spring anchored at $a^\prime_0$ and $b^\prime_0$ and subsequently creating a slip-spring 
anchored at $a_0$ and $b_0$ is:
\begin{equation}
   \mathcal{Q}_{a^\prime_0 \land b^\prime_0 \to a_0 \land b_0} = P_{a^\prime_0 \land b^\prime_0}
    k_{{\rm hop},a^\prime_0\land b^\prime_0}  P_{a_0 \land b_0}^{\rm sel}
   P_{\mathscr{N} \to \mathscr{O}}^{\rm accept}
   \tag{F1}
\end{equation}

In order to create a new slip-spring, we randomly select one of the $2n$ end beads (with $n$ being the number of
chains) available in the system. We then try to create a new slip-spring emanating from the selected chain end, e.g. 
$a_0^\prime$. This is accomplished with probability:
\begin{equation}
   P_{{\rm sel} \: a^\prime_0} = \frac{1}{2n}
    \tag{F2}
\end{equation}
After searching for candidates $b^\prime$ inside a sphere of radius $\alpha_{\rm attempt}$ centered at $a^\prime_0$, 
one of them, e.g. $b_0^\prime$, is selected with probability:
\begin{equation}
   P_{{\rm sel} \: b^\prime_0} = \frac{\exp{\parnths{-\beta A_{a^\prime_0 \land b^\prime_0}}}}{W_{\mathscr{N}}}
   \tag{F3}
\end{equation}
with $W_{\mathscr{N}}$ being the corresponding Rosenbluth weight:
\begin{equation}
   W_{\mathscr{N}} = \sum_{b^\prime = 1}^{n_{{\rm cands}\:a^\prime_0}} \exp{\parnths{-\beta A_{a^\prime_0  \land b^\prime}}}
    \tag{F4}
\end{equation}
with $b^\prime$ running over all possible candidates lying in the vicinity of $a^\prime_0$, $n_{{\rm cands}\:a^\prime_0}$. 
The overall probability of choosing to create a new slip-spring connecting $a^\prime_0$ with $b^\prime_0$ is:
\begin{equation}
   P_{a^\prime_0 \land b^\prime_0}^{\rm sel} = \frac{1}{2n} \frac{\exp{\parnths{-\beta A_{a^\prime_0 \land b^\prime_0}}}}
   {\sum_{b^\prime = 1}^{n_{{\rm cands}\:a^\prime_0}} \exp{\parnths{-\beta A_{a^\prime_0 \land b^\prime}}}}
   \tag{F5}
\end{equation}
In a completely analogous way, the probability of creating a slip-spring anchored at $a_0$ and $b_0$, once a slip-spring
anchored at $a^\prime_0$ and $b^\prime_0$ is considered for destruction, is:
\begin{equation}
   P_{a_0 \land b_0}^{\rm sel} = \frac{1}{2n} \frac{\exp{\parnths{-\beta A_{a_0 \land b_0}}}}
   {W_\mathscr{O}} =
   \frac{1}{2n} \frac{\exp{\parnths{-\beta A_{a_0 \land b_0}}}}
   {\sum_{b = 1}^{n_{{\rm cands}\:a_0}} \exp{\parnths{-\beta A_{a_0 \land b}}}}
   \tag{F6}
\end{equation}
with $b$ running over all anchoring candidates of $a_0$.

At equilibrium, we demand detailed balance:
\begin{equation}
   \mathcal{Q}_{a_0 \land b_0 \to a^\prime_0 \land b^\prime_0} = \mathcal{Q}_{a^\prime_0 \land b^\prime_0 \to a_0 \land b_0}
   \tag{F7}
   \label{eq:appendix_f_detailed_balance_a}
\end{equation}
which, after replacing all factors yields:
\begin{align}
   & \exp{\parnths{-\beta A^{\parnths{ N_{\rm ss}}}_{a_0 \land b_0}}} \;\; \nu_{\rm hop} 
   \exp{\parnths{\beta A_{a_0 \land b_0}}} \;\; \frac{1}{2n} 
   \frac{\exp{\parnths{-\beta A_{a^\prime_0 \land b^\prime_0}}}}{W_{\mathscr{N}}} 
   P_{\mathscr{O} \to \mathscr{N}}^{\rm accept}
   \nonumber \\
   = & \exp{\parnths{-\beta A^{\parnths{N_{\rm ss}}}_{a^\prime_0 \land b^\prime_0}}} \;\; \nu_{\rm hop} 
   \exp{\parnths{\beta A_{a^\prime_0 \land b^\prime_0}}}  \;\;
   \frac{1}{2n} \frac{\exp{\parnths{-\beta A_{a_0 \land b_0}}}} {W_\mathscr{O}}
   P_{\mathscr{N} \to \mathscr{O}}^{\rm accept}
   \label{eq:appendix_f_detailed_balance_b}
   \tag{F8}
\end{align}
At this point, by recalling the definitions of the free energy levels introduced in Figure \ref{fig:free_energy_levels}, we can replace 
$A^{\parnths{N_{\rm ss}}}_{a_0 \land b_0} = A^{\parnths{N_{\rm ss}-1}} + A_{a_0 \land b_0}$ and
$A^{\parnths{N_{\rm ss}}}_{a^\prime_0 \land b^\prime_0} = A^{\parnths{N_{\rm ss}-1}} + A_{a^\prime_0 \land b^\prime_0}$.
Thus, eq \ref{eq:appendix_f_detailed_balance_b} can be simplified to:
\begin{equation}
   \frac{\exp{\parnths{-\beta A_{a^\prime_0 \land b^\prime_0}}}} {W_\mathscr{N}}
   P_{\mathscr{O} \to \mathscr{N}}^{\rm accept} = 
   \frac{\exp{\parnths{-\beta A_{a_0 \land b_0}}}}{W_{\mathscr{O}}}
   P_{\mathscr{N} \to \mathscr{O}}^{\rm accept}
   \tag{F9}
   \label{eq:appendix_f_detailed_balance_c}
\end{equation} 
which is the necessary requirement for microscopic reversibility. For the detailed balance condition to hold, eq 
\ref{eq:appendix_f_detailed_balance_c} implies that a combined destuction/creation move leading from configuration 
$\mathscr{O} = a_0 \land b_0$ to a configuration $\mathscr{N} = a^\prime_0 \land b^\prime_0$, should be accepted 
with probability:
\begin{equation}
   P^{\rm accept}_{\mathscr{O}\to \mathscr{N}} = \min{\bracks{1,\exp{\parnths{-\frac{A_{a_0 \land b_0} - A_{a^\prime_0 \land b^\prime_0}}{k_{\rm B}T}}}
   \frac{W_{\mathscr{N}}}{W_{\mathscr{O}}}}} 
   \tag{F10}
\end{equation}
which is eq \ref{eq:kMC_creation_acceptance_criterion} of the main text.

\begin{acknowledgement} 
This work was funded by the European Union through the project COMPNANOCOMP
under grant number 295355. G.G.V. thanks the Alexander S. Onassis Public Benefit
Foundation for a doctoral scholarship.
During the course of this research G.M. and D.N.T. have been funded by the Volkswagen Foundation 
in the context of the project ``Mesoscopic Simulations of Viscoelastic Properties of Networks''. 
These authors also thank the Limmat Foundation for giving them the opportunity to extend the present research 
to polymer networks (under the project entitled ``Multiscale Simulations of Complex Polymer Systems'').
The authors thank Mr. Aris Sgouros (National Technical University of Athens) for his help with improving 
parts of the methodology and the associated computer code.
Fruitful and stimulating discussions with Prof. Dr. Marcus M\"uller (Georg-August-Universit\"at G\"ottingen),
and Mr. Ludwig Schneider (Georg-August-Universit\"at G\"ottingen) are gratefully acknowledged.
\end{acknowledgement} 

\bibliography{refs}    

\end{document}